\documentclass[aps,prd,reprint,twocolumn,superscriptaddress,longbibliography,nofootinbib,floatfix,showpacs]{revtex4-1}
\usepackage{epsfig}
\usepackage{amsmath,amssymb,amsfonts}
\usepackage{hyperref}
\usepackage{mathrsfs}
\usepackage{bbm}
\usepackage{slashed}
\usepackage{graphicx}
\usepackage{verbatim}

\usepackage{bm}

\usepackage[usenames]{xcolor}

\definecolor{darkgreen}{rgb}{0.2,0.6,0}

\newcommand{\be}{\begin{equation}}
\newcommand{\ee}{\end{equation}}
\newcommand{\bw}{\begin{widetext}}
\newcommand{\ew}{\end{widetext}}
\newcommand{\bi}{\begin{itemize}}
\newcommand{\ei}{\end{itemize}}
\newcommand{\ud}{\mathrm{d}}

\newcommand{\LCm}{{\scriptscriptstyle -}} \newcommand{\LCp}{{\scriptscriptstyle +}}
\newcommand{\LCpm}{{\scriptscriptstyle \pm}}

\newcommand{\LCperp}{{\scriptscriptstyle \perp}}

\newcommand{\LCpara}{{\scriptscriptstyle \parallel}}

\newcommand{\pa}{\partial}
\newcommand{\bo}{\textbf}
\renewcommand{\Re}{\text{Re} \,}
\renewcommand{\Im}{\text{Im} \,}
\newcommand{\tr}{\mathrm{T}}

\usepackage[T1]{fontenc} \usepackage[latin1]{inputenc}

\begin{document}

\title{Momentum spectrum of nonlinear Breit-Wheeler pair production in space-time fields}

\author{Gianluca Degli Esposti}
\email{g.degli-esposti@hzdr.de}
\affiliation{Helmholtz-Zentrum Dresden-Rossendorf, Bautzner Landstra{\ss}e 400, 01328 Dresden, Germany}

\affiliation{Institut f\"ur Theoretische Physik, 
Technische Universit\"at Dresden, 01062 Dresden, Germany}

\author{Greger Torgrimsson}
\email{greger.torgrimsson@umu.se}
\affiliation{Department of Physics, Ume{\aa} University, SE-901 87 Ume{\aa}, Sweden}

\begin{abstract}

We show how to use a worldline-instanton formalism to calculate, to leading order in the weak-field expansion, the momentum spectrum of nonlinear Breit-Wheeler pair production in fields that depend on time and one spatial coordinate. We find a nontrivial dependence on the width, $\lambda$, of the photon wave packet, and the existence of a critical point $\lambda_c$. For $\lambda<\lambda_c$ and a field with one peak, the spectrum has one peak where the electron and positron have the same energy. For $\lambda>\lambda_c$ this splits into two peaks. We calculate a high-energy ($\Omega\gg1$) expansion, which to leading order agrees with the results obtained by replacing the space-time field with a plane wave and using the well-known Volkov solutions. We also calculate an expansion for $\Omega\sim a_0\gg1$, where the field is strong enough to significantly bend the trajectories of the fermions despite $\Omega\gg1$.      

\end{abstract}

\maketitle

\section{Introduction}

Nonlinear Breit-Wheeler pair production, $\gamma\to e^\LCp e^\LCm$, in a strong electromagnetic background field, such as a high-intensity laser, is an important strong-field-QED process. See~\cite{DiPiazza:2011tq,Gonoskov:2021hwf,Fedotov:2022ely} for reviews. In this paper when we refer to the ``photon'', we mean the single incoherent photon, not the photons that make up the background field.
Most of the existing calculations have been done with the field as a plane wave, which only depends on one lightlike coordinate, e.g. $t+x$. The reason for this is that the solution to the Dirac equation is very simple for a plane wave, but complicated for other fields. One can justify this by considering photons with high frequency, for which one can expect a rather general field to behave as if it were a plane wave. The Volkov solutions allow one to study various pulsed plane waves, which has been done in many papers~\cite{DiPiazza:2011tq,Fedotov:2022ely}. 
To go beyond plane waves to fields depending on more than one space-time coordinate, one could try to solve the Dirac equation numerically, but this is very difficult and has not been done.
However, one also realizes that one might not need to calculate the probability exactly. Indeed, the thing that makes nonlinear Breit-Wheeler pair production experimentally relevant even with existing lasers, while spontaneous Schwinger pair production is not, is the fact that the photon can have frequencies, $\Omega$, well above the electron mass\footnote{We use units with $m_e=1$ in addition to $c=\hbar=1$. We also rescale the background field as $eF_{\mu\nu}\to F_{\mu\nu}$}. This energy scale thus gives a large parameter which allows one to approximate the probability. By assuming the frequency of the photon to be the largest parameter in the system, such an approximation has been calculated in \cite{BaierKatkov,BaierBook,DiPiazza:2016maj,DiPiazza:2016tdf,Raicher:2018cih}.
The resulting approximation shares some similarities with the result obtained for plane waves. 

Another parameter that one also often wants to be large is $a_0=E/\omega$, where $E$ is the maximum field strength and $\omega$ a characteristic frequency scale of the background. $a_0$ is common notation in papers considering plane waves. In the following we will also use $\gamma=1/a_0$, which is more common in studies on Schwinger pair production.
As explained in~\cite{DiPiazza:2016maj,DiPiazza:2016tdf,Raicher:2018cih}, the plane-wave-like/high-energy approximation assumes that $\Omega$ is large compared to $a_0$. This is well justified if e.g. the photon has been produced using conventional high-energy accelerators. 

Another regime that could potentially be interesting is $\Omega\sim a_0\gg1$. This could be relevant e.g. if the photon has been produced by nonlinear Compton scattering by an electron or positron which have been accelerated by the background, because then $a_0$ gives the typical scale of the Lorentz-force momentum. When $\Omega\gg\{1,a_0\}$ the particles follow almost straight lines, but when $\Omega\sim a_0\gg1$ the field is strong enough to significantly bend the trajectories of the fermions, which means the fermions will ``feel'' a field that is more complicated than a plane wave.

In this paper our starting assumption is not $\Omega\gg1$ but rather that the field is weak, so the probability has an exponential form. To be able to calculate the probability we have developed a worldline instanton formalism\footnote{For processes in certain simple classes of fields, such as plane waves, it is possible to calculate the worldline integrals without using saddle-point approximations, see~\cite{Edwards:2021vhg,Schubert:2023gsl,Copinger:2023ctz}.}. 
There have been several papers using closed worldline instantons to study Schwinger pair production~\cite{Affleck:1981bma,Dunne:2005sx,Dunne:2006st,Dunne:2006ur,Dumlu:2015paa,Schneider:2018huk}. These studies have shown that worldline instantons allow one to obtain the probability for general multidimensional fields. Closed worldline instantons have also been used to study the total/integrated probability of nonlinear Breit-Wheeler, but so far only for constant fields~\cite{Monin:2010qj,Satunin:2013an} and time-dependent fields~\cite{Torgrimsson:2016ant}. To obtain the dependence of the probability on the momenta and spins of the produced particles, we consider instead open worldlines\footnote{See also~\cite{Barut:1989mc,Rajeev:2021zae} for open worldlines for Schwinger pair production by a constant electric field.}.
We started in~\cite{DegliEsposti:2021its} with fields that only depend on time. However, there are important simplifications that work for a time dependent field but not if the field also depends on space. In this paper we will show how to use the worldline formalism for fields that depend on time and one spatial coordinate, $z$. This allows us to investigate the role of the transverse shape of the field. We will for simplicity consider fields with only one maximum.  

When one calculates the probability for a plane wave or a field that only depends on time, the photon wave packet is usually not considered\footnote{Effects of electron wave packets have been considered in photon emission in~\cite{wavePacketPeatross,Angioi:2016vir,Angioi:2017ygv}.}. And even in papers where the initial state is at the beginning of the calculations written explicitly in terms of a wave packet, it is common to assume early in the calculations that the wave packet is very sharply peaked, and then the wave packet drops out as in~\eqref{sharplyPeaked} below.    
In this paper we will assume that the wave packet is peaked in the directions perpendicular to the field inhomogeneity, i.e. in $x$ and $y$, but not in $z$. By allowing the wave packet to be wide, we have found that there is a critical point for the width of the wave packet, $\lambda=\lambda_c$. For $\lambda<\lambda_c$ we find that the electron-positron momentum spectrum has one peak, but for $\lambda>\lambda_c$ it splits into two peaks. The fact that there are two peaks can be understood as follows. We assume for simplicity that the wave packet is maximized for a photon travelling transverse to the field inhomogeneity, i.e. for photon momentum $k_\mu=l_\mu=\Omega(1,1,0,0)$, where $l_\mu$ is the position of the peak. For large $\lambda$ the wave packet is wide in momentum space, which means momentum components with large $k_3$ is less suppressed. The result from~\cite{Dunne:2009gi} for a constant field suggests that increasing the $k_3$ component of the absorbed photon would not actually help to make it easier to produce a pair. However, in the opposite limit, i.e. for a rapidly varying field, we can expect pair production via a perturbative mechanism, and in that regime the direction of the photon momentum with respect to the field polarization is less relevant compared to the overall increase in the photon energy when any of its components increases. When $\gamma$ is neither very small nor very large, we have an interplay between tunneling/nonperturbative and perturbative pair production and it may not be a priori obvious which mechanism is more important.   

However, this argument alone does not explain why the split into two peaks happens at a finite critical point $\lambda_c\ne0$. Mathematically the criticality comes from saddle-point equations which are similar to Landau's theory of second-order phase transitions. We find such criticality both using the instanton formalism and by treating the field to leading order in perturbation theory.

As an aside, a different type of criticality has been studied in Schwinger pair production by purely spatially inhomogeneous electric fields in~\cite{Gies:2015hia,Gies:2016coz,Pimentel:2018nkl}, where the order parameter is the integrated probability rather than the position of peaks in the momentum spectrum. In that case, the probability is exactly zero for $\gamma_z>1$ (with a suitable normalization of $\gamma_z$) and becomes nonzero at $\gamma_z<1$.

\section{Notation and starting point}

The starting point is the worldline representation of the fermion propagator,
\be\label{propagatorWorldline}
\begin{split}
&S(x_\LCp,x_\LCm)=(i\slashed{\partial}_{x_\LCp}-\slashed{A}(x_\LCp)+1)\int_0^\infty\frac{\ud T}{2}\int\limits_{q(0)=x_\LCm}^{q(1)=x_\LCp}\mathcal{D}q\,\mathcal{P}\\
&\times\exp\left\{-i\left[\frac{T}{2}+\int_0^1\!\ud\tau\left(\frac{\dot{q}^2}{2T}+A\dot{q}+\frac{T}{4}\sigma^{\mu\nu}F_{\mu\nu}\right)\right]\right\} \;,
\end{split}
\ee
where $\tau$ is proper time divided by the total proper time $T$, $\mathcal{P}$ indicates proper-time ordering of the the spin term with $\sigma^{\mu\nu}=\frac{i}{2}[\gamma^\mu,\gamma^\nu]$.  
From~\eqref{propagatorWorldline} we obtain the pair-production amplitude using the LSZ (Lehmann-Symanzik-Zimmermann) reduction formula~\cite{Barut:1989mc,ItzyksonZuber},
\be\label{LSZ3pair}
\begin{split}
M=\lim_{t_\LCpm\to\infty}\int \ud^3x_\LCp\ud^3 x_\LCm e^{ipx_\LCp+ip'x_\LCm}
\bar{u}\gamma^0S(x_\LCp,x_\LCm)\gamma^0 v \;.
\end{split}
\ee
\eqref{propagatorWorldline} gives the fermion propagator for an arbitrary coherent background field $A_\mu$. We can also use it for pair production by a coherent field and an incoherent (and high-frequency) photon with momentum $k_\mu$ and polarization vector $\epsilon_\mu$ by replacing 
\be\label{replacingAepsilon}
A_\mu\to A_\mu+\epsilon_\mu e^{-ikx}
\ee
and then expanding and selecting the term in $M$ that is linear in $\epsilon_\mu$~\cite{Strassler:1992zr,McKeon:1994hd,Shaisultanov:1995tm,Dittrich:2000wz,Schubert:2000yt,Schubert:2001he,Ahmad:2016vvw,Gies:2011he}. We denote $\sigma$ for the point in $\tau$ when the photon is absorbed. 
For the 1D fields which we considered in~\cite{DegliEsposti:2021its}, one has to be more careful with the asymptotic states, because, even for a field with $\lim_{t\to\pm\infty}E(t)=0$, one can still have $A(-\infty)\ne A(\infty)$. This is the case for fields with one maximum, such as a Sauter pulse, $A(t)=(1/\gamma)\tanh(\omega t)$. But now when we turn to 2D fields the asymptotic states are actually simpler. We consider for example
\be
A_3(t,z)=\frac{E}{\omega_t}\tanh(\omega_t t)\text{sech}^2(\omega_z z) \;.
\ee
Even though we have the same pulse shape in $t$, the potential nevertheless vanishes asymptotically because the worldline starts and ends at $|z|\to\infty$. The electron and positron states, $u_s(p)$ and $v_{s'}(p')$, are therefore free states, and $p$ and $p'$ are the physical momenta. 
We denote $\gamma_t=\omega_t/E$ and $\gamma_z=\omega_z/E$, and, unless otherwise stated, consider $\gamma_t$ and $\gamma_z$ as $E$-independent parameters.

\section{Wave packet}

We use the same notation and conventions as in Appendix A of~\cite{DegliEsposti:2021its}. When dealing with a background field that only depends on one space-time coordinate, e.g. a purely time-dependent electric field $E_3(t)$ or a plane wave $a_\LCperp(t+z)$, one does not need to choose a specific wave packet if one assumes that it is so sharply peaked that one can approximate
\be\label{sharplyPeaked}
\int\frac{\ud^3k}{(2\pi)^32k_0}|f(k)|^2 h(k)\approx h(l) \;,
\ee
where $l$ is the position of the peak, for all relevant functions $h$. However, when considering fields that depend on two or more space-time coordinates, we need to take the wave packet into account. We will not assume~\eqref{sharplyPeaked}. We choose a Gaussian wave packet
\be\label{fWavePacket}
f(k)=\rho(k)\exp\left\{-\sum_{j=1}^3\frac{(k_j-l_j)^2}{2\lambda_j^2}+ib^jk_j\right\} \;,
\ee
where we allow the widths $\lambda_j$ to be different, and where $b^j$ gives the impact parameter.  
The normalization
\be
\int\frac{\ud^3k}{(2\pi)^32k_0}|f(k)|^2=1 \;,
\ee
is satisfied with 
\be\label{rhoWavePacket}
\rho(k)=\sqrt{\frac{(2\pi)^3 2k_0}{\pi^{3/2}\lambda_1\lambda_2\lambda_3}} \;.
\ee

In this paper we will focus on fields that depend on $t$ and $z$. The $t$ dependence gives a nontrivial pulse shape even in the $\Omega\gg1$ limit, and the $z$ dependence allows us to investigate the role of the shape of the field transverse to its propagation direction, and the role of the wave packet.

The $x^\LCperp=\{x,y\}$ directions are trivial. We change variables from
\be\label{varphithetadef}
x_\LCp^\LCperp=\varphi^\LCperp+\frac{\theta^\LCperp}{2}
\qquad
x_\LCm^\LCperp=\varphi^\LCperp-\frac{\theta^\LCperp}{2}
\ee
to $\varphi$ and $\theta$. After a constant shift of the transverse components of the worldline, $q^\LCperp(\tau)\to\varphi^\LCperp+q^\LCperp(\tau)$, the $\varphi$ integral gives momentum conservation for the transverse components
\be
\int\ud^2\varphi \, e^{i(p+p'-k)_\LCperp\varphi^\LCperp}=(2\pi)^2\delta^2_\LCperp(p+p'-k) \;.
\ee
We use the delta function to perform the $k_\LCperp$ integral. Then we assume that $\lambda_1$ and $\lambda_2$ are sufficiently small that
\be
\begin{split}
&\int\frac{\ud^2 p'_\LCperp}{(2\pi)^2}\exp\left\{-\sum_{j=1}^2\frac{(p+p'-l)_j^2}{\lambda_j^2}\right\}F(p'_\LCperp) \\
&\approx\frac{\lambda_1\lambda_2}{4\pi}F(l_\LCperp-p_\LCperp) \;.
\end{split}
\ee
The factors of $\lambda_1$ and $\lambda_2$ drop out and we are left with only $\lambda_3=:\lambda$. We thus arrive at the starting point for the more nontrivial calculations,
\be\label{general2Dstart}
\begin{split}
\mathbb{P}=&\frac{e^2}{4\pi^{3/2}\lambda}\int\frac{\ud^2 p_\LCperp}{(2\pi)^2}\frac{\ud p_3}{2\pi}\frac{\ud p'_3}{2\pi}\\
&\times\left|\int\frac{\ud k_3}{\sqrt{k_0}}\exp\left[-\frac{(k-l)_3^2}{2\lambda^2}+ibk_3\right]\mathbb{M}\right|^2 \;,
\end{split}
\ee
where $k_\LCperp=l_\LCperp$.
The normalization of the amplitude is such that for a field that is independent of $z$ we have $\mathbb{M}=2\pi\delta(p_3+p'_3-k_3)M$, with $M$ as in Eq.~(A5) in~\cite{DegliEsposti:2021its}.

If $\lambda$ is smaller than any other scale, then we find
\be\label{PwidePhoton}
\mathbb{P}\approx\frac{e^2}{2\Omega}\int\frac{\ud^2 p_\LCperp}{(2\pi)^2}\frac{\ud p_3}{2\pi}\frac{\ud p'_3}{2\pi}\frac{\lambda}{\sqrt{\pi}}|\mathbb{M}(k_3=l_3)|^2 \;.
\ee
In this limit we could simplify the calculation of $\mathbb{M}$ by setting $l_3=0$. However, the limit $\lambda\to0$ means a wave packet that is very spread out in space. The probability goes to zero as $\lambda\to0$ because if the photon wave packet is much wider than the field, then there is less probability that the photon will be at the field maximum.

To allow for different sizes of the wave packet compared to the size of the field, we will instead perform the $k_3$ integral with the saddle-point method, which is of course natural since all the other integrals are anyway performed with this method. 
In this paper we will focus on $l_3=0$.

\section{Amplitude}

To compute the remaining integrals in~\eqref{general2Dstart} we use the saddle-point method. For the exponent, we simply need to evaluate it at all the saddle points. The result is given by
\be\label{expPsiStart}
-\frac{(k_3 -l_3)^2}{2\lambda^2} + ibk_3 +\psi \; , \qquad \psi = i\int \ud u \, q^\mu \partial_\mu A_\nu \frac{\ud q^\nu}{\ud u}
\ee
where we have changed the variable to $u=T(\tau-\sigma)$ and $q(u)$ is the saddle point of the path integral, i.e. the solution to the Lorentz force equation~\eqref{eq:LFE}. The calculations are shown in Appendix~\ref{app:exponent}.

For the prefactor, we start with the path integral using the Gelfand-Yaglom method to compute the determinant of the kinetic operator as in Appendix~\ref{app:PI}, followed by the ordinary integrals over $(T,\sigma, z_\LCpm)$ in~\ref{app:OI} (we compute separately the one over $k_3$ at the end). Since the individual terms at the prefactor are proportional to some power of the asymptotic times $t_\LCpm$, which in the limit go to $\infty$, it is convenient to extract such dependence analytically to cancel them out. Finally, we deal with the spin contribution in Appendix~\ref{app:spin}.

\section{Instanton}

The instanton is determined by the Lorentz-force equation, which for $u\ne0$ is given by
\be
t''=E(t,z)z'
\qquad
z''=E(t,z)t' \;,
\ee
with asymptotic boundary condition
\be\label{asympNotSaddle}
z'(\infty)=-p_3
\qquad
z'(-\infty)=p'_3 \;. 
\ee
The derivatives are discontinuous at $u=0$ due to photon absorption,
\be\label{discDtDz}
t'(0+)-t'(0-)=k_0
\qquad
z'(0+)-z'(0-)=-k_3 \;.
\ee
The transverse velocities also change discontinuously at $u=0$, but are otherwise constant,
\be
x_\LCperp'(u>0)=p_\LCperp
\qquad
x_\LCperp'(u<0)=-p'_\LCperp=-(k-p)_\LCperp \;.
\ee
From $q^{\prime2}=1$ we have
\be\label{dtFromSqrt}
\begin{split}
t'(u>0)=&\sqrt{m_\LCperp^2+z^{\prime2}(u)} \\
t'(u<0)=&-\sqrt{m_\LCperp^{\prime2}+z^{\prime2}(u)} \;,
\end{split}
\ee
where $m_\LCperp=\sqrt{1+p_\LCperp^2}$ and $m_\LCperp'=\sqrt{1+p_\LCperp^{\prime2}}$. Since the field, $E_3(t,z)$ is trivial in $x_\LCperp$, we can without loss of generality choose $k_2=0$ and $k_1>0$.  From~\eqref{discDtDz} and~\eqref{dtFromSqrt} we find (we assume $0<p_1<k_1$, which is not particularly restrictive since the most important point is $p_1\sim k_1/2$)
\be\label{dz0Fromk}
\begin{split}
z'(0+)=&i\frac{m_2k_0}{k_1}-\frac{p_1}{k_1}k_3 \\
z'(0-)=&i\frac{m_2k_0}{k_1}+\frac{k_1-p_1}{k_1}k_3 \;,
\end{split}
\ee
where $m_2=\sqrt{1+p_2^2}$, while $t'(0\pm)$ is obtained from~\eqref{dtFromSqrt}. For $p_1=k_1/2$ we have 
\be\label{dt0Fromk}
\begin{split}
t'(0+)&=\frac{k_0}{2}-i\frac{k_3}{k_1} \\
t'(0-)&=-\frac{k_0}{2}-i\frac{k_3}{k_1} \;.
\end{split}
\ee
Thus, the velocities at $u=0$ are already determined. 
We can find $t(0)$ and $z(0)$ by varying them until we find the instanton with some particular asymptotic momenta~\eqref{asympNotSaddle}. 

The small parameter that justifies the use of the saddle-point method is the maximum field strength, $E$. Terms in the exponent scale as $e^{.../E}$, and terms in the prefactor scale as some power of $E$. There is no nontrivial dependence on $E$ anywhere. To see this we rescale as 
\be\label{Erescaling}
q^\mu\to\frac{q^\mu}{E}
\qquad
u\to\frac{u}{E} \;,
\ee
after which $E$ drops out from the Lorentz-force equation. In other words, if we change the value of $E$, we do not need to find new instantons etc.

\section{Momentum saddle points}\label{sec:MomentumSP}

We can deal with the momentum integrals in a way similar to how we obtained the $p$ and $p'$ widths in~\cite{DegliEsposti:2022yqw,DegliEsposti:2023qqu} for Schwinger pair production. After we have performed the path integral and the ordinary integrals over $T$, $\sigma$, $x_\LCp^j$ and $x_\LCm^j$, the exponent $e^{\psi}$ depends in a nontrivial way on the momenta, $p$, $p'$ and $k_3$. The prefactor, on the other hand, is simply evaluated by setting all variables, including $p$ and $p'$, to their saddle-point values, which is justified since we consider the leading order in the saddle-point expansion. 
The nontrivial dependence of $\psi$ on the momenta comes from the momentum dependence of the saddle-point values of $q^\mu$, $T$, $\sigma$, $x_\LCp^j$ and $x_\LCm^j$. 
However, the partial derivatives of $\psi$ with respect to these integration variables vanish at their saddle points, so the first momentum derivatives of $\psi$ are simply obtained by differentiating as if the other variables did not depend on the momenta. This gives 
\be\label{dpsidk}
\frac{\partial\psi}{\partial k_3}=-i\left[z(0)+\frac{k_3}{k_0}t(0)\right] \;,
\ee
where the parametrization of the instanton has been chosen such that the photon is absorbed at $u=0$ regardless of the values $k$, $p$ and $p'$,
\be\label{dpsidp}
\frac{\partial\psi}{\partial p_3}=i\left[z(\infty)+\frac{p_3}{p_0}t(\infty)\right] \;,
\ee
\be\label{dpsidpp}
\frac{\partial\psi}{\partial p'_3}=i\left[z(-\infty)+\frac{p'_3}{p'_0}t(-\infty)\right] 
\ee
and (remember $p'_\LCperp=k_\LCperp-p_\LCperp$)
\be\label{dpsidpperp}
\begin{split}
\frac{\partial\psi}{\partial p_\LCperp}&=i\left(x^\LCperp+\frac{p_\LCperp}{p_0}t\right)(u_1)-i\left(x^\LCperp+\frac{p'_\LCperp}{p'_0}t\right)(u_0) \\
&=ip_\LCperp\left(\frac{t(u_1)}{p_0}-u_1\right)-ip'_\LCperp\left(\frac{t(u_0)}{p'_0}+u_0\right)\;.
\end{split}
\ee
The saddle point, $k_s$, is hence determined by
\be\label{saddleEqk3}
-\frac{k_3-l_3}{\lambda^2}+ib+\frac{\partial\psi}{\partial k_3}=0 \;.
\ee
Writing the exponential part on the probability level as 
\be
\mathbb{P}\propto e^{-\mathcal{A}} \;,
\ee
we have
\be\label{dAdp}
\frac{\partial\mathcal{A}}{\partial p_3}=2\lim_{u\to\infty}\text{Im}\left(z+\frac{p_3}{p_0}t\right)
=2\lim_{u\to\infty}\text{Im}\left(z-\frac{z'}{t'}t\right)
\ee
and
\be\label{dAdpp}
\frac{\partial\mathcal{A}}{\partial p'_3}=2\lim_{u\to-\infty}\text{Im}\left(z+\frac{p'_3}{p'_0}t\right)
=2\lim_{u\to-\infty}\text{Im}\left(z-\frac{z'}{t'}t\right)\;,
\ee
so the saddle points, $p_{3s}$ and $p'_{3s}$, are determined by
\be\label{saddleFromIm}
\text{Im}\left(z-\frac{z'}{t'}t\right)(u=\pm\infty)=0 \;.
\ee 
For the transverse components we have
\be\label{dAdpperp}
\frac{\partial\mathcal{A}}{\partial p_\LCperp}=2p_\LCperp\text{Im}\left(\frac{t(u_1)}{p_0}-u_1\right)-2p'_\LCperp\text{Im}\left(\frac{t(u_0)}{p'_0}+u_0\right)\;.
\ee

Since $k_2=0$ we see immediately from~\eqref{dAdpperp} that there is always a saddle point at $p_2=p'_2=0$. This momentum component decouples from the others, i.e. differentiating $\partial\mathcal{A}/\partial p_2$ with respect to the other momentum components gives zero when evaluated at the saddle points. The spectrum therefore scales as 
\be
\mathbb{P}\propto\exp\left\{-\frac{p_2^2}{d_2^2}\right\} \;,
\ee
where the width is given by
\be\label{eq:d2Width}
d_2^{-2}=\lim_{u_1\to\infty}\text{Im}\left[\frac{t(u_1)}{p_0}-u_1\right]+
\lim_{u_0\to-\infty}\text{Im}\left[\frac{t(u_0)}{p'_0}+u_0\right]\;,
\ee
which for $b=0$, as we will show below, reduces to 
\be\label{d2k0}
d_2^{-2}=\lim_{u_1\to\infty}2\text{Im}\left[\frac{t(u_1)}{p_0}-u_1\right] \;,
\ee
where $p_2=0$ in $p_0$ and $p_0'$. Significantly more work is needed to find the spectrum for the other momentum components.   

When the wave packet size is below some critical value $\lambda_c$, the saddle point of the $k_3$ integral is zero, thus, from the initial conditions~\eqref{dz0Fromk}, \eqref{dt0Fromk} and the symmetry of the field, the $(t,z)$ components of the instanton are respectively even and odd on the saddle point of the momenta.
The (anti-)symmetry of~\eqref{dz0Fromk} suggests that there is a saddle point at $p_1=k_1/2$, and it is straightforward to check that it indeed solves $\eqref{dAdpperp}=0$ below the critical point and when $b=0$.
However, above the critical point, the instantons are no longer symmetric under $u \to -u$.
Nonetheless, if we restrict to zero impact parameter, the saddle point is still $p_1=k_1/2$ due to a different symmetry of the instantons. If we substitute $p_1 = p'_1 = k_1/2$ in~\eqref{dAdpperp} we find
\be
\frac{\pa \mathcal A}{\pa p_1} = \frac{k_1}{2} \Im \left( \frac{t(u_1)}{p_0} + \frac{t(u_0)}{p'_0} -u_1 -u_0 \right)
\ee
then, choosing a contour in the complex $u$ plane as in Sec.~\ref{app:Complex}, the first two terms drop and we are left with
\be
\frac{\pa \mathcal A}{\pa p_1} = -\frac{k_1}{2} \Im \left(u_1 +u_0 \right) \; .
\ee
In the region where the contour is below the real axis the instanton corresponds to the solution to the Lorentz force equation with $u=0+$ initial conditions, while above the real axis it corresponds to the solution with $u=0-$ initial conditions. When $p_1 = k_1/2$ and $b=0$ the saddle point of $k_3$ is real, therefore $t'(0+) = -t'(0-)^*$ and $z'(0+) = -z'(0-)^*$. Furthermore, when $b=0$, $t(0)$ and $z(0)$ are always purely imaginary, therefore the instantons satisfy $t(u^*) = -t(u)^*$ and $z(u^*) = -z(u)^*$. 
From this it follows that the contour, as shown in Fig.~\ref{fig:Complex}, is of the form
\be
\begin{split}
    u_1 &= u_c +r \\
    u_0 &= u_c^*-r \;,
\end{split}
\ee
where $r \in \mathbb R$, which means
\be
\frac{\pa \mathcal A}{\pa p_1} = -\frac{k_1}{2} \Im (u_c + u_c^*) = 0 \;,
\ee
i.e. $p_1 = k_1/2$ is a saddle point when $b=0$. Unless stated explicitly, we will focus on zero impact parameter throughout the present work.

The saddle points for the longitudinal momenta, $p_{3s}$ and $p'_{3s}$, are nontrivial. An inefficient way to find them, and the corresponding instanton, would be to use~\eqref{asympNotSaddle} and vary $p$ and $p'$ until one finds a maximum of the exponent. One would then need to use the shooting method to find different $t(0)$ and $z(0)$ for every value of $p$ and $p'$. A much faster way is to express $\partial\mathcal{A}/\partial p_3=0$ and $\partial\mathcal{A}/\partial p'_3=0$ as in the second expressions in~\eqref{dAdp} and~\eqref{dAdpp}, so that $p_3$ and $p'_3$ do not explicitly enter the equations that determine $p_{3s}$ and $p'_{3s}$.
If we evaluate the above at one point $u=u_1$, where $u_1$ is where we stop the numerical integration, then we have two real conditions. To determine the two complex constants $t(0)$ and $z(0)$ we need four real conditions. So in addition to~\eqref{saddleFromIm} we also take the imaginary parts of~\eqref{asympNotSaddle},
\be\label{realMomentum}
\text{Im }z'(\pm\infty)=0 \;.
\ee
This allows us to find the instanton without having to first find $p_s$ and $p_s'$. Afterwards we can find $p_{3s}$ and $p'_{3s}$ by simply evaluating $\mp z'(\pm\infty)$.
\section{Instantons on the complex plane}\label{app:Complex}

In the study of the Schwinger effect in~\cite{DegliEsposti:2023qqu} we consider a contour made up of two distinct sections. One of them, the ``formation region'', runs along the imaginary axis from $u = 0$ to $u = u_c \in i \mathbb R$ where $t(u_c) = 0$, and describes particles tunneling through the field. The other section, the ``acceleration region'', describes particles after creation when they are accelerated by the field.

\begin{figure*}
    \centering
    \includegraphics[width=\linewidth]{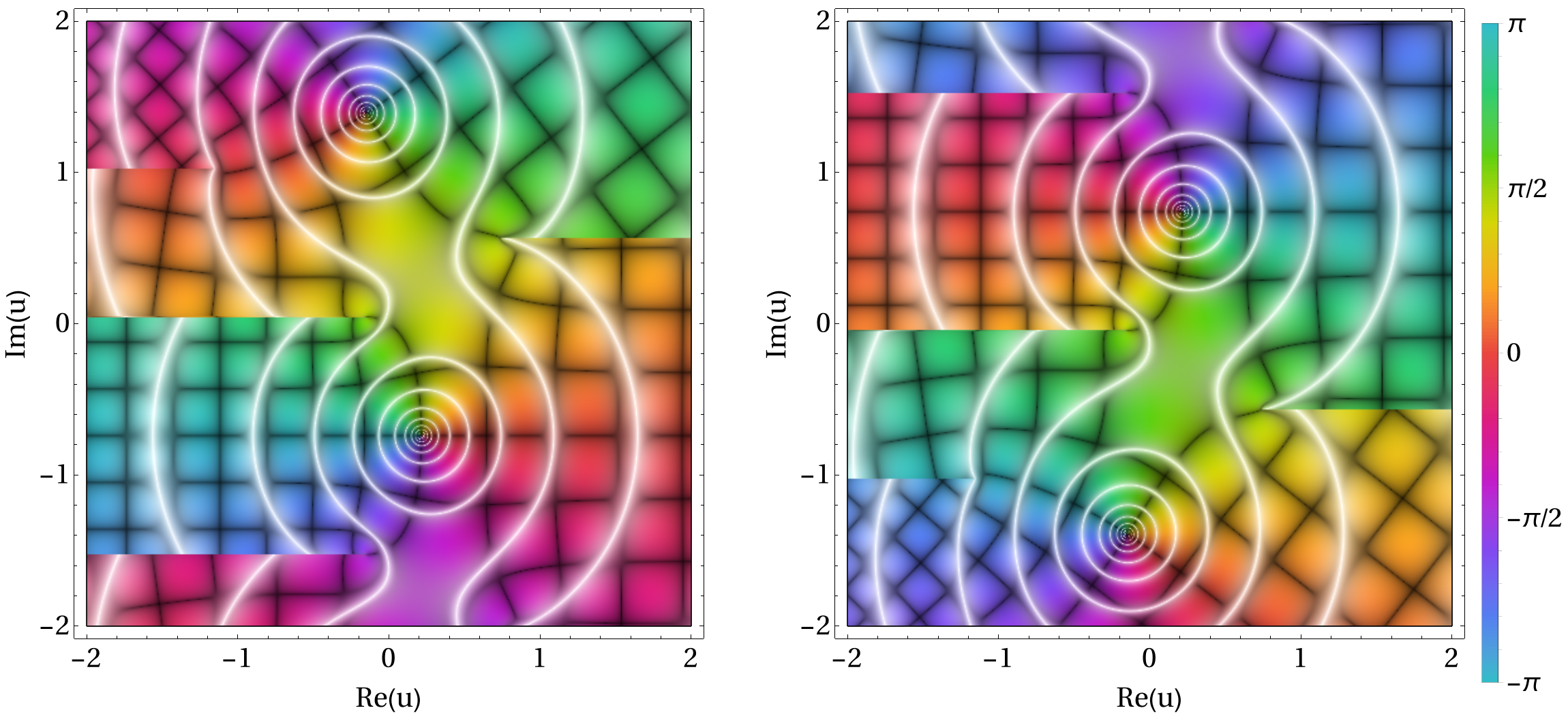}
    \caption{$t_{(\LCp)}(u)$ (left) and $t_{(\LCm)}(u)$ (right) with $\Omega = \gamma_t = \gamma_z = 1$, $k_3 \approx 1.15$, and $\lambda = 3.2 \sqrt{E}$ (i.e. larger than the critical value $\lambda_c \approx 2.23 \sqrt{E}$). The white lines represent contour lines of $|t(u)|$, the black ones contour lines with constant real/imaginary parts, and the color represents the phase. Gluing together these two functions along the real axis with the prescription~\eqref{eq:InstGlued} gives the plot shown in Fig.~\ref{fig:Complex}.}
    \label{fig:tPlusMinus}
\end{figure*}

For Breit-Wheeler we have something similar. In this case, however, we have to be careful due to the delta function in the Lorentz-force equation representing the photon absorption. As mentioned in Appendix~\ref{app:exponent}, we can consider the instanton components as effectively being two distinct functions, $q_{(\LCp)}(u)$ and $q_{(\LCm)}(u)$, defined as the solutions to the Lorentz force equation with initial conditions at $u=0$ given by
\be
\begin{split}
   t'_{(\LCp)}(0) &= \frac{k_0}{2} -i\frac{k_3}{k_1} \\
   z'_{(\LCp)}(0) &= i\frac{k_0}{k_1} -\frac{k_3}{2}
\end{split}
\ee
and
\be
\begin{split}
   t'_{(\LCm)}(0) &= -\frac{k_0}{2} -i\frac{k_3}{k_1} \\
   z'_{(\LCm)}(0) &= i\frac{k_0}{k_1} +\frac{k_3}{2} \;.
\end{split}
\ee
In principle, both functions can be defined on the whole complex plane obtaining the plots in Fig.~\ref{fig:tPlusMinus}, but since we want to find an analogue to the contour used for dipole fields, we want the asymptotic region $u \to +\infty$ to represent the solution with $q'(0+)$ initial conditions and the asymptotic region $u \to -\infty$ to represent the solution with $q'(0-)$.

\begin{figure}
    \centering
    \includegraphics[width=\linewidth]{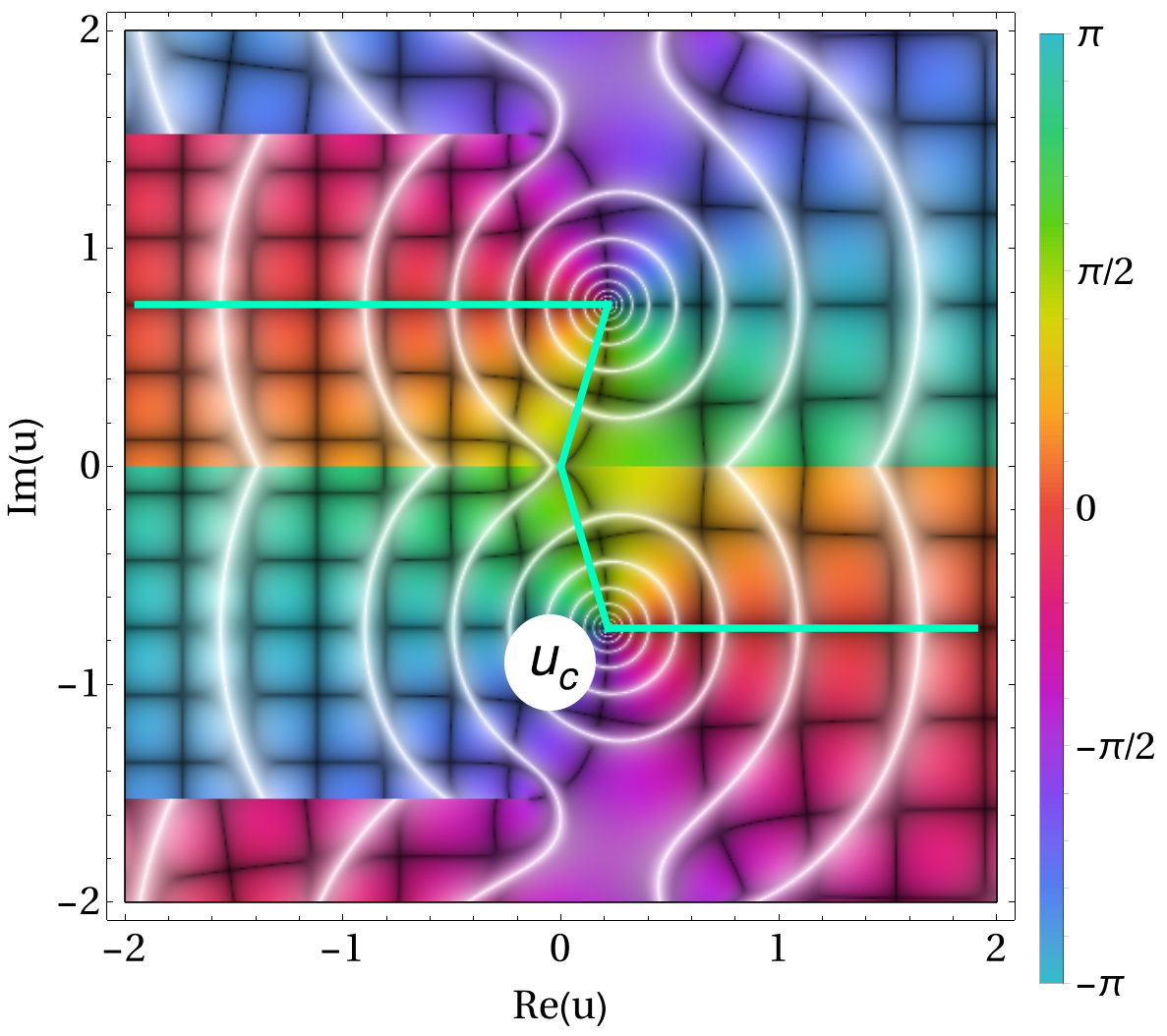}
    \caption{Time component of the instantons defined in~\eqref{eq:InstGlued} obtained gluing together the two plots in Fig.~\ref{fig:tPlusMinus}. The discontinuity along the real axis is due to the discontinuity in~\eqref{eq:InstGlued}, i.e. they are different functions joined along the real line $\Im(u) = 0$.}
    \label{fig:Complex}
\end{figure}

Furthermore, when $p_1 = k_1/2$, one can see that the two distinct solutions satisfy
\be
q_{(\LCp)}(u^*) = -q_{(\LCm)}(u)^*
\ee
therefore it is convenient to define single-valued instanton components as follows
\be\label{eq:InstGlued}
q(u) :=
\begin{cases}
    &q_{(\LCp)}(u) \qquad \Im(u)<0 \\
    &q_{(\LCm)}(u) \qquad \Im(u)>0
\end{cases}
\ee
obtaining Fig.~\ref{fig:Complex}.
In this case, since the initial velocities are generic complex numbers, the zero of the time component $u_c$ is not purely imaginary.

Complex contours made up of different segments, e.g. parallel to the real or the imaginary axis, also appear in~\cite{Mclaughlin:1972ws,DollGeorgeMiller} for nonrelativistic tunneling through potential barriers (where ordinary time $t$ is used to parameterize the trajectory instead of proper time $u$), and in~\cite{Lavrelashvili:1989he,Dumlu:2011cc} for particle production.

\section{Widths}

To calculate the contribution from the $k_3$ integral to the prefactor and the widths of the spectrum, we need to expand the exponent to quadratic order around $k_{3s}$, $p_{js}$ and $p'_{3s}$. Since we have already found simple expressions for the first-order derivatives, i.e.~\eqref{dpsidk}, \eqref{dpsidp}, \eqref{dpsidpp} and~\eqref{dpsidpperp}, we start with these equations and expand them to linear order in $\delta p=p-p_s$ etc. Thus, even though we are calculating $\mathcal{O}(\delta^2)$ terms in the exponent, we only need to find $\mathcal{O}(\delta)$ corrections to the instanton, $q^\mu\to q^\mu+\delta q^\mu+\mathcal{O}(\delta^2)$.  

We change variables from
\be\label{pToPDeltap}
p_3=-P+\frac{\Delta}{2}
\qquad
p'_3=P+\frac{\Delta}{2}
\ee
to $P$ and $\Delta$.

$\delta t$ and $\delta z$ are determined by expanding the Lorentz-force equation to linear order, which can in general be written as
\be\label{dtdzeq}
\begin{split}
\delta t''&=E\delta z'+\nabla E\cdot\{\delta t,\delta z\}z' \\
\delta z''&=E\delta t'+\nabla E\cdot\{\delta t,\delta z\}t' \;,
\end{split}
\ee
where $\nabla E=\{\partial_t E,\partial_z E\}$. \eqref{dtdzeq} holds in general for either $u<0$ or $u>0$. At $u=0$ the velocities are in general discontinuous due to the photon absorption. 
Any solution can be expressed as a superposition,
\be\label{qjsum}
\delta q(u)=\sum_{j=1}^4 a_j\delta q_{[j]}(u) \;,
\ee
where the basis $\delta q_{[j]}$ have simple initial conditions,
\be
\delta t_{[1]}(0)=
\delta z_{[2]}(0)=
\delta t_{[3]}'(0)=
\delta z_{[4]}'(0)=1 
\ee
and all the others are zero. The coefficients $a_3$ and $a_4$ can be different for $u<0$ and $u>0$. Before expanding, we have $t^{\prime2}-z^{\prime2}=m_\LCperp^2$, so $\nu=t'\delta t'-z'\delta z'$ is a constant of motion and gives a relation between $a_3$ and $a_4$.  

We consider first the variation of the instanton due to $\delta k=k_3-k_{3s}$ with $p$ and $p'$ kept constant. We have 
\be
\delta t_k'(\pm\infty)=\delta z_k'(\pm\infty)=0 \;.
\ee
Setting $p_1=k_1/2$ and expanding~\eqref{dz0Fromk} gives
\be\label{dz0deltak}
\delta z'_k(0\pm)=i\frac{k_3}{k_1k_0}\mp\frac{1}{2}
\ee
Since the $\delta k$ variation leaves $m_\LCperp$ constant, we have
\be\label{dt0deltak}
\delta t'_k(0\pm)=\frac{z'}{t'}\delta z'_k\bigg|_{u=0\pm} \;.
\ee
\eqref{dz0deltak} and~\eqref{dt0deltak} give $a_3$ and $a_4$ in~\eqref{qjsum}.

Next we consider the change in the instanton due to either $\delta\Delta$, $\delta P$ or $\delta p_1$. We define these variations with $k_3$ kept constant, and then we take care of the cross terms, e.g. $\delta k\delta\Delta$, afterwards.  From~\eqref{asympNotSaddle}, \eqref{pToPDeltap} and~\eqref{dz0Fromk} we have
\be\label{deltaqDeltaDer}
\delta z_\Delta'(\pm\infty)=\mp\frac{1}{2}
\qquad
\delta z_\Delta'(0)=\delta t_\Delta'(0)=0 \;,
\ee
\be\label{deltaqPDer}
\delta z_P'(\pm\infty)=1
\qquad
\delta z_P'(0)=\delta t_P'(0)=0 
\ee
and
\be\label{deltaqp1Der}
\begin{split}
&\delta z_{p_1}'(\pm\infty)=0 
\qquad
\delta z_{p_1}'(0\pm)=-\frac{k_3}{k_1} \\
&\delta t_{p_1}'(0\pm)=\frac{z'\delta z_{p_1}'\pm p_1}{t'}\bigg|_{u=0\pm}\;.
\end{split}
\ee
$a_3$ and $a_4$ in~\eqref{qjsum} are determined directly from the $u=0\pm$ conditions in~\eqref{deltaqDeltaDer}, \eqref{deltaqPDer} and~\eqref{deltaqp1Der}: e.g. $a_3=a_4=0$ for $\delta q_\Delta$ and $\delta q_P$. $a_1$ and $a_2$ are determined algebraically from the conditions on $\delta z'(\pm\infty)$.

We begin with the last remaining integral on the amplitude level, $k_3$. We need the derivative of~\eqref{dpsidk} with respect to $k_3$ with $p$ and $p'$ kept constant. Since we only need this derivative evaluated for all integration variables at their saddle points, we have
\be\label{eq:dkWidth}
d_k^{-2}=\frac{1}{2}\left(\frac{1}{\lambda^2}-X_{kk}\right) \;,
\ee
where
\be
X_{kk}=-i\left[\delta z_k(0)+\frac{k_3}{k_0}\delta t_k(0)+\frac{k_1^2}{k_0^3}t(0)\right] \;.
\ee
To simplify notation, rather than keep writing a subscript ``s'' on e.g. $k_{3s}$, we will instead simply write $k_3$, i.e. we change integration variable from $k_3=k_{3s}+\delta k$ to $\delta k$ and afterwards rename $k_{3s}\to k_3$.

On the amplitude level, we use $X_{k\Delta}$ to denote the term in the exponent proportional to $\delta k\delta\Delta$, and similarly for the other combinations. We find
\be
\begin{split}
&X_{k\Delta}=-i\left[\delta z_\Delta(0)+\frac{k_3}{k_0}\delta t_\Delta(0)\right]\\
&=\frac{i}{2}\left[\delta z_k(\infty)+\frac{p_3}{p_0}\delta t_k(\infty)+\delta z_k(-\infty)+\frac{p'_3}{p'_0}\delta t_k(-\infty)\right]
\end{split}
\ee
\be
\begin{split}
&X_{kP}=-i\left[\delta z_P(0)+\frac{k_3}{k_0}\delta t_P(0)\right]\\
&=i\left[-\delta z_k(\infty)-\frac{p_3}{p_0}\delta t_k(\infty)+\delta z_k(-\infty)+\frac{p'_3}{p'_0}\delta t_k(-\infty)\right]
\end{split}
\ee
\be
\begin{split}
X_{kp_1}&=-i\left[\delta z_{p_1}(0)+\frac{k_3}{k_0}\delta t_{p_1}(0)\right]\\
&=ip_1\left[\frac{\delta t_k(\infty)}{p_0}-\frac{\delta t_k(-\infty)}{p'_0}\right]
\end{split}
\ee
\be
\begin{split}
X_{\Delta P}=&\frac{i}{2}\bigg[\delta z_P(\infty)+\frac{p_3}{p_0}\delta t_P(\infty)-\frac{m_\LCperp^2}{p_0^3}t(\infty)\\
&+\delta z_P(-\infty)+\frac{p'_3}{p'_0}\delta t_P(-\infty)+\frac{m_\LCperp^2}{p_0^{\prime3}}t(-\infty)\bigg] \\
=&i\bigg[-\delta z_\Delta(\infty)-\frac{p_3}{p_0}\delta t_\Delta(\infty)-\frac{m_\LCperp^2}{2p_0^3}t(\infty)\\
&+\delta z_\Delta(-\infty)+\frac{p'_3}{p'_0}\delta t_\Delta(-\infty)+\frac{m_\LCperp^2}{2p_0^{\prime3}}t(-\infty)\bigg]
\end{split}
\ee 
\be
\begin{split}
&X_{\Delta p_1}=\frac{i}{2}\bigg[\delta z_{p_1}(\infty)+\frac{p_3}{p_0}\delta t_{p_1}(\infty)-\frac{p_1p_3}{p_0^3}t(\infty)\\
&\hspace{1.2cm}+\delta z_{p_1}(-\infty)+\frac{p'_3}{p'_0}\delta t_{p_1}(-\infty)+\frac{p_1p'_3}{p_0^{\prime3}}t(-\infty)\bigg]\\
=&ip_1\bigg[\frac{\delta t_\Delta(\infty)}{p_0}-\frac{p_3}{2p_0^3}t(\infty)
-\frac{\delta t_\Delta(-\infty)}{p'_0}+\frac{p'_3}{2p_0^{\prime3}}t(-\infty)\bigg]
\end{split}
\ee
\be
\begin{split}
&X_{Pp_1}=i\bigg[-\delta z_{p_1}(\infty)-\frac{p_3}{p_0}\delta t_{p_1}(\infty)+\frac{p_1p_3}{p_0^3}t(\infty)\\
&\hspace{1.2cm}+\delta z_{p_1}(-\infty)+\frac{p'_3}{p'_0}\delta t_{p_1}(-\infty)+\frac{p_1p'_3}{p_0^{\prime3}}t(-\infty)\bigg]\\
&=ip_1\bigg[\frac{\delta t_P(\infty)}{p_0}+\frac{p_3}{p_0^3}t(\infty)-\frac{\delta t_P(-\infty)}{p'_0}+\frac{p'_3}{p_0^{\prime3}}t(-\infty)\bigg]
\end{split}
\ee
\be
\begin{split}
X_{\Delta\Delta}=\frac{i}{2}&\bigg[\delta z_\Delta(\infty)+\frac{p_3}{p_0}\delta t_\Delta(\infty)+\frac{m_\LCperp^2}{2p_0^3}t(\infty)\\
+&\delta z_\Delta(-\infty)+\frac{p'_3}{p'_0}\delta t_\Delta(-\infty)+\frac{m_\LCperp^2}{2p_0^{\prime3}}t(-\infty)\bigg]    
\end{split}
\ee
\be
\begin{split}
X_{PP}=i&\bigg[-\delta z_P(\infty)-\frac{p_3}{p_0}\delta t_P(\infty)+\frac{m_\LCperp^2}{p_0^3}t(\infty)\\
&+\delta z_P(-\infty)+\frac{p'_3}{p'_0}\delta t_P(-\infty)+\frac{m_\LCperp^2}{p_0^{\prime3}}t(-\infty)\bigg]    
\end{split}
\ee
\be
\begin{split}
X_{p_1p_1}=&ip_1\bigg[\frac{\delta t_{p_1}(u_1)}{p_0}-\frac{p_1}{p_0^3}t(u_1)\\
&-\frac{\delta t_{p_1}(u_0)}{p'_0}-\frac{p_1}{p_0^{\prime3}}t(u_0)\bigg]\\
+&i\bigg[\frac{t(u_1)}{p_0}-u_1+\frac{t(u_0)}{p'_0}+u_0\bigg]
\end{split}
\ee

With these expressions we are now ready to perform the $k_3$ integral, which is given by
\be
\begin{split}
\int\ud\delta k\exp(&-d_k^{-2}\delta k^2\\
&+[X_{kp_1}\delta p_1+X_{k\Delta}\delta\Delta+X_{kP}\delta P]\delta k) \;.
\end{split}
\ee

\subsection{$k_s=0$}\label{ks0section}

When $k_s=0$ the solutions are either symmetric or antisymmetric in $u$, and $\Delta_s=0$. $t$, $\delta z_k$, $\delta z_{p_1}$, $\delta t_P$ and $\delta z_\Delta$ are symmetric, while $z$, $\delta t_k$, $\delta t_{p_1}$, $\delta z_P$ and $\delta t_\Delta$ are antisymmetric. The $X$'s simplify
\be\label{Xkk0}
X_{kk}=-i\left[\delta z_k(0)+\frac{t(0)}{k_0}\right] \;,
\ee
\be\label{XkDelta0}
X_{k\Delta}=-i\delta z_\Delta(0) 
=i\left[\delta z_k(\infty)-\frac{P}{p_0}\delta t_k(\infty)\right] \;,
\ee
\be\label{Xkp10}
X_{kp_1}=-i\delta z_{p_1}(0)=\frac{ik_1}{p_0}\delta t_k(\infty) \;,
\ee
\be\label{Xp1Delta0}
\begin{split}
X_{p_1\Delta}&=i\left[\delta z_{p_1}-\frac{P}{p_0}\delta t_{p_1}+\frac{p_1P}{p_0^3}t\right](\infty)\\
&=ik_1\left[\frac{\delta t_\Delta}{p_0}+\frac{Pt}{2p_0^3}\right](\infty) \;,
\end{split}
\ee
\be\label{XDeltaDeltak0}
X_{\Delta\Delta}=i\left[\delta z_\Delta-\frac{P}{p_0}\delta t_\Delta+\frac{m_\LCperp^2t}{2p_0^3}\right](\infty) \;,
\ee
\be\label{XPPk0}
X_{PP}=2i\left[-\delta z_P+\frac{P}{p_0}\delta t_P+\frac{m_\LCperp^2t}{p_0^3}\right](\infty) \;,
\ee
\be\label{Xp1p1k0}
X_{p_1p_1}=ik_1\left[\frac{\delta t_{p_1}}{p_0}-\frac{p_1t}{p_0^3}\right]+2i\left[\frac{t}{p_0}-u_1\right](u_1\to\infty) \;,
\ee
and $X_{kP}=X_{\Delta P}=X_{p_1P}=0$. The absence of cross terms with $\delta P$ implies that $d_{P,z}$ is not affected by the wave packet.  

In terms of the basis solutions we have
\be\label{deltaqDeltaP21}
\delta q_\Delta^\mu(u)=-\frac{\delta q_{[2]}^\mu(u)}{2\delta z_{[2]}'(\infty)}
\qquad
\delta q_P^\mu(u)=\frac{\delta q_{[1]}^\mu(u)}{\delta z_{[1]}'(\infty)} \;,
\ee
\be\label{deltaqp1From32}
\delta q_{p_1}^\mu(u)=\delta q_{[3]}^\mu(u)-\frac{\delta z_{[3]}'(\infty)}{\delta z_{[2]}'(\infty)}\delta q_{[2]}^\mu(u)
\ee
and, for $u>0$,
\be\label{deltakFrom234}
\begin{split}
&\delta q_k^\mu(u)=-\frac{i}{k_1}\delta q_{[3]}^\mu(u)-\frac{1}{2}\delta q_{[4]}^\mu(u) \\
&+\frac{1}{\delta z_{[2]}'(\infty)}\left(\frac{i}{k_1}\delta z_{[3]}'(\infty)+\frac{1}{2}\delta z_{[4]}'(\infty)\right)\delta q_{[2]}^\mu(u) \;.
\end{split}
\ee
Thus, all four basis solutions are needed.

For the $k$ integral we have
\be
\begin{split}
&\int\ud\delta k\exp\bigg\{\left(-\frac{1}{\lambda^2}+X_{kk}\right)\frac{\delta k^2}{2}\\
&\hspace{2cm}+(X_{k\Delta}\delta\Delta+X_{kp_1}\delta p_1)\delta k\bigg\}\\
&=\sqrt{2\pi}\left(\frac{1}{\lambda^2}-X_{kk}\right)^{-1/2}\\
&\times\exp\bigg\{\frac{\lambda^2}{2}\frac{(X_{k\Delta}\delta\Delta+X_{kp_1}\delta p_1)^2}{1-X_{kk}\lambda^2}\bigg\} \;.
\end{split}
\ee
The widths for $\delta\Delta$ and $\delta p_1$ are finally given by
\be\label{widthsDelta1}
\begin{split}
\exp\bigg\{\text{Re}\bigg[&\left(X_{\Delta\Delta}+\frac{X_{k\Delta}^2\lambda^2}{1-X_{kk}\lambda^2}\right)\delta\Delta^2\\
+&\left(X_{p_1p_1}+\frac{X_{kp_1}^2\lambda^2}{1-X_{kk}\lambda^2}\right)\delta p_1^2\\
+&2\left(X_{\Delta p_1}+\frac{X_{k\Delta}X_{kp_1}\lambda^2}{1-X_{kk}\lambda^2}\right)\delta\Delta\delta p_1
\bigg]\bigg\} \;.
\end{split}
\ee
In contrast to the widths for $\delta P$ and $p_2$, the widths in~\eqref{widthsDelta1} depend in general on the width of the wave packet.

\section{Phase transition}

The saddle point of the $k_3$ integral, given by~\eqref{saddleEqk3}, might be $k_3 \neq 0$ even if the wave packet is peaked at $l_3 = 0$. We begin for simplicity with zero impact parameter $b = 0$.

Letting
\be
\psi_a = \psi - \frac{k_3^2}{2\lambda^2} \, \qquad \psi_r = 2 \, \Re \psi_a
\ee
be the full exponent respectively before/after taking the modulus squared, the saddle point of the $k_3$ and momentum integrals is defined as the  solution to
\be
\begin{split}
    & \frac{\pa \psi}{\pa k_3}(k_s, \Pi_s) - \frac{k_s}{\lambda^2} = 0 \\
    & \frac{\pa \psi_r}{\pa \Pi_\alpha}(k_s, \Pi_s) = 0 \;,
\end{split}
\ee
where $\Pi_\alpha$ denotes all momentum components (in this case $\alpha \in \{p_1,\Delta, P \}$ because there is no mixing with the other transversal component, $p_2$). The implicit function theorem allows us to find a solution $\Pi(k_3)$ to the second equation in a neighborhood of $k_s$ such that $\Pi(k_s) = \Pi_s$. Taking the total derivative of the second equation with respect to $k_3$ we then find
\be
    R_{\alpha \beta}\frac{\ud \Pi_\beta}{\ud k_3} + R_{\alpha k} = 0 \; \to \; \frac{\ud \Pi_\alpha}{\ud k_3} = -R_{\alpha \beta}^{-1}R_{\beta k}
\ee
where
\be\label{eq:defR}
R_{\alpha \beta} = 2 \, \Re X_{\alpha \beta} \, \qquad R_{\alpha k} = 2 \, \Re X_{\alpha k}\; .
\ee
Plugging $\Pi(k)$ in the first equation and defining
\be
\chi(k_3) := \frac{\pa \psi}{\pa k_3}(k_3, \Pi(k_3)) = -i\left[z(0)+\frac{k_3}{k_0}t(0)\right]
\ee
we obtain a single variable equation
\be\label{saddleEqk3new}
\chi(k_3) = \frac{k_3}{\lambda^2} \; .
\ee
If $\lambda$ is sufficiently small, the trivial solution $k_3 = 0$ is the only intersection point between the two sides of~\eqref{saddleEqk3new} because
\be
\left|\frac{k_3}{\lambda^2}\right| > \left|\chi(k_3)\right|
\ee
for all $k_3 \neq 0$. However, above some critical size $\lambda > \lambda_c$, the right-hand side grows faster near $k_3 \sim 0$, so there are two new nonzero intersection points $k_3 \neq 0$ besides $k_3 = 0$. 

The critical size is therefore the value $\lambda_c$ such that the two sides have the same slope near to $k_3 = 0$, i.e.
\be\label{eq:defCrit}
\frac{1}{\lambda_c^2} = \chi'(0) \;,
\ee
therefore
\be\label{criticalLambda}
\lambda_c = \frac{1}{\sqrt{X_{kk} -X_{\alpha k} R_{\alpha \beta}^{-1}R_{\beta k}}} \; .
\ee
We stress that $\lambda_c$ can be finite even for a purely time-dependent field, showing that the phase transition is not a special feature of the field being spacetime dependent. Since we only need to evaluate the coefficients at $k_3 = 0$, the computation of $\lambda_c$ is very simple. Furthermore, from~\eqref{criticalLambda} we immediately see that $\lambda_c \sim \sqrt{E}$.

Even though there are two new saddle points, this does not necessarily imply that $k_3 = 0$ no longer corresponds to a maximum of the spectrum. To see this for a generic value of $\lambda$, we need to look at the prefactor. If $k_3 = 0$ is no longer a maximum, the determinant of the Hessian matrix of the spectrum becomes negative, thus evaluating blindly at such a saddle point gives an imaginary prefactor.

\begin{figure}
    \centering
    \includegraphics[width=\linewidth]{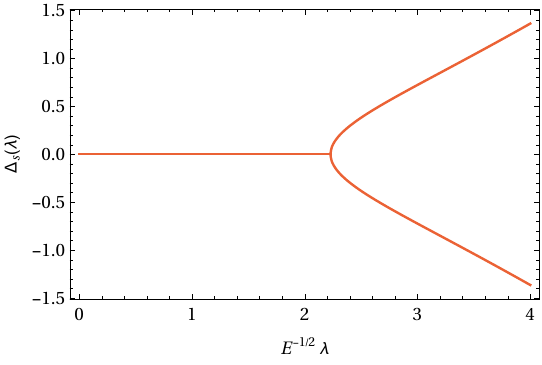}
    \caption{Pitchfork bifurcation of the order parameter $\Delta_s$ as a function of $\lambda$. With parameters $\Omega = \gamma_t = \gamma_z = 1$, the critical width is $\lambda_c \approx 2.23 \sqrt{E}$. Since $\Delta_s \sim k_s$ near $\lambda_c$, we take $\Delta_s$ as the order parameter because, unlike the saddle points $k_s$, it is measurable.}
    \label{fig:transition}
\end{figure}

The high-energy behavior of $\lambda_c$ is obtained expanding the $X$ coefficients as in Sec.~\ref{sec:highOmega}. The terms at the denominator of~\eqref{criticalLambda} contain powers of $\Omega$, $1/\Omega$, and so on with higher negative odd powers. However, using the relations~\eqref{XkkXkDelta} and~\eqref{eq:Xfirstorder}, both the leading and next-to-leading-order terms cancel, so the first nonzero order is actually
\be
X_{kk} -X_{\alpha k} R_{\alpha \beta}^{-1}R_{\beta k} \sim 1/\Omega^3 \;,
\ee
which means
\be\label{lambdacLargeOmega}
\lambda_c \sim \Omega^{3/2} \; .
\ee
\begin{figure*}
    \centering
    \includegraphics[width=.48\linewidth]{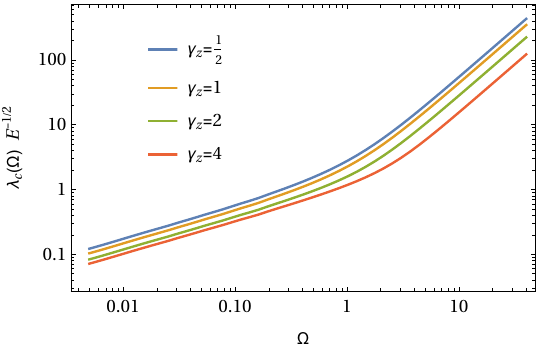}
    \hfill
    \includegraphics[width=.48\linewidth]{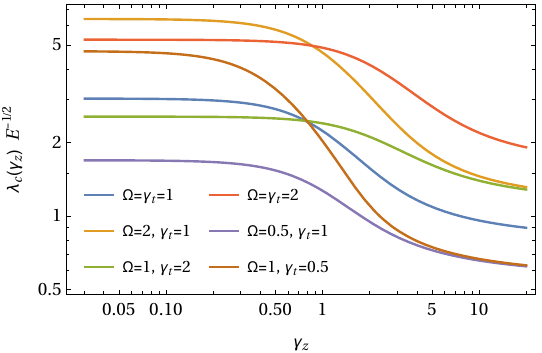}
    \caption{Critical wave packet size~\eqref{criticalLambda} as a function of the photon energy at $\gamma_t = 1$ (left) and as a function of the inverse size of the field $\gamma_z$ for some values of $\Omega$ and $\gamma_t$. We see that, even in the time-dependent limit, the phase transition is still present, but as the size of the field decreases the critical wave packet size becomes smaller.}
    \label{fig:criticals}
\end{figure*}

\subsection{Instantons}

As to the nontrivial saddle point above the critical size, we can find it using a method similar to the one outlined in Sec.~\ref{sec:MomentumSP}. Instead of computing the instantons for several values of $k_3$ and asymptotic momenta, we vary the initial conditions only once in order to find the instanton at the saddle point. From~\eqref{saddleEqk3new} we find
\be\label{z0Saddle}
z(0) = k_3 \left( \frac{i}{\lambda^2} -\frac{t(0)}{k_0} \right) \;,
\ee
which we can use as a new initial condition, varying $t(0)$ and $k_3$ such that~\eqref{saddleFromIm} and~\eqref{realMomentum} are satisfied. While the last two give us the instantons at the saddle point of the momentum, the constraint~\eqref{z0Saddle} ensures that we are on the saddle point of $k_3$ as well. For this field and for $b=0$, the turning point $t(0)$ is always imaginary and $k_3$ is real, so we only need two real conditions to fix these parameters. It turns out that it suffices to consider the two conditions~\eqref{saddleFromIm} and~\eqref{realMomentum} at $+\infty$.

Since $\chi(k_3)$ is odd, if $k_s$ is a solution then so is $-k_s$, which means that the saddle point effectively splits from $k_3 = 0$ when $\lambda < \lambda_c$ to $k_3 = \pm k_s$ when $\lambda > \lambda_c$ as in Fig.~\ref{fig:transition}. At $\lambda = \lambda_c$, the two saddle points merge. The instanton at $-k_s$ can be obtained from the $k_s$ one by applying $\{z(u), t(u)\} \to \{-z(-u),  t(-u)\}$. In fact, when $k_s = 0$ the transformation does nothing because the components are respectively odd and even.

This in turn implies that $\{p_{3s},p'_{3s}\} \to \{-p'_{3s},-p_{3s}\}$ or $\Delta_s \to -\Delta_s$ and $P_s \to P_s$. In other words, at $\lambda_c$ the single peak splits into two peaks symmetric with respect to $\pm \Delta_s$. Note that the two saddle points, $k_3 = \pm k_s$, correspond to two different saddle points for $p$ and $p'$, so there is no interference, i.e. the amplitude only has a single term. 

\subsection{Widths}

To obtain a general expression for the widths, we begin by performing the integral over $k_3$. Letting $k_s(\Pi)$ be the saddle point in a neighborhood of $\Pi_s$, we have
\be
\frac{\pa \psi_a}{\pa k} \bigr(k_s(\Pi), \Pi \bigr) \overset{!}{=} 0
\ee
for all $\Pi$, therefore taking the derivative with respect to any component $\Pi_\alpha$ we find
\be
\frac{\pa k_s}{\pa \Pi_\alpha} = -\frac{\pa^2 \psi_a}{\pa k \pa \Pi_\alpha} \frac{1}{\frac{\pa^2 \psi_a}{\pa k^2}} \; .
\ee
Now we compute the momentum integrals, whose saddle points are given by
\be
0 \overset{!}{=} \frac{\ud\psi_r}{\ud\Pi_\alpha}(k_s(\Pi_s), \Pi_s) = \frac{\pa \psi_r}{\pa \Pi_\alpha} + 2\, \Re \frac{\pa k_s}{\pa \Pi_\alpha} \frac{\pa \psi_a}{\pa k} = \frac{\pa \psi_r}{\pa \Pi_\alpha} \;.
\ee
Thus, we only need to consider the explicit dependence on the momenta to find the saddle points. From the second derivative we obtain the widths as follows
\be\label{eq:WidthsFull}
\begin{split}
        2 d_{\alpha \beta}^{-2} &:= -\frac{\ud^2\psi_r}{\ud\Pi_\alpha \ud \Pi_\beta} \\
        &= -\frac{\pa^2 \psi_r}{\pa \Pi_\alpha \pa \Pi_\beta} - 2\, \Re \frac{\pa k_s}{\pa \Pi_\alpha} X_{\beta k} \\
    &= -R_{\alpha \beta} - 2\lambda^2 \, \Re \frac{ X_{\alpha k} \, X_{\beta k}}{1- X_{kk} \lambda^2} \;.
\end{split}
\ee
Thus, we get the usual widths $\pa_\alpha\pa_\beta \psi_r$ plus a correction proportional to the mixed partial derivatives. If $k_s = 0$, most terms cancel and we obtain~\eqref{widthsDelta1}.

Since there is no mixing with the integration over $p_2$ and the corresponding width is simple, we focus on the rest of the spectrum. Because there is no mixing with the $P$ component, in the subcritical regime we have a simple structure (the power of $-2$ is just part of the name of the matrix)
\be\label{eq:WidthsMatrix}
\bo d^{-2} :=
\begin{pmatrix}
    d^{-2}_{p_1 p_1} & d^{-2}_{p_1 \Delta} & 0 \\
    d^{-2}_{p_1 \Delta} & d^{-2}_{\Delta \Delta} & 0 \\
    0 & 0 & d^{-2}_{P P}
\end{pmatrix} \; .
\ee

\subsection{Spectrum at $k_3\ne 0$}

From Eq.~\eqref{eq:defCrit} we see that when $\lambda > \lambda_c$ we have new saddle points $k_3 = \pm k_s \neq 0$ besides $k_3 = 0$, but this does give any information about the spectrum. In fact, Eq.~\eqref{eq:defCrit} only tells us which values are saddle points of the $k_3$ integral. To see whether a point is a maximum or not, we have to compute the eigenvalues of the Hessian matrix of the spectrum. If the determinant vanishes at $\lambda = \lambda_c$ and becomes negative when $\lambda > \lambda_c$, then we know that the point is no longer a maximum.

To do this analytically we can compare the prefactors obtained integrating over $k_3$ first or over the momenta first. However, it is more convenient to study both at the same time, integrating over all variables together and using a linear algebra property of block matrices. 

Since the $k_3$ integral is inside the modulus squared, we first rewrite
\be
\left| \int \ud k_3 \, e^{\psi_a(k_3,\Pi)} \right|^2 = \int \ud k_3 \ud \tilde k_3 \, e^{\psi_a(k_3,\Pi) + \psi_a^*(\tilde k_3,\Pi)}
\ee
and then perform the integrals over $(k_3, \tilde k_3, p_1, \Delta, P)$ together with the saddle-point method. The saddle points are the same as before, with the extra $k_s = \tilde k_s$. To find the prefactor we need to compute the global Hessian matrix of coefficients
\be
H_{ij} = -\frac{\pa^2}{\pa \xi_i \pa \xi_j}\bigr( \psi_a(k_3, \Pi) + \psi_a(\tilde k_3, \Pi)^* \bigr)
\ee
with $\xi_i \in \{k_3, \tilde k_3, p_1, \Delta, P\}$. 

The resulting matrix can be seen as a block matrix with a 2x2 block made out of the $(k_3, \tilde k_3)$ derivatives, a 3x3 block with momentum derivatives, an two mixed blocks. Since the saddle point is $k_s = \tilde k_s$, we have
\be\label{eq:PartialHessian}
\begin{split}
    H_{kk} &= \frac{1}{\lambda^2} - X_{kk} \\
    H_{\tilde k \tilde k} &= \frac{1}{\lambda^2} - X_{kk}^* \\
    H_{k \tilde k} &= H_{\tilde k k} = 0
\end{split}
\ee
for the $k_3$-block, then (letting $\alpha, \beta \in \{p_1, \Delta,P\})$ 
\be
H_{\alpha \beta} = -(X_{\alpha \beta} + X_{\alpha \beta}^*) = -R_{\alpha \beta}
\ee
with $R$ defined as in~\eqref{eq:defR}, and finally
\be
    H_{\alpha k} = -X_{\alpha k} \qquad H_{\alpha \tilde k} = -X_{\alpha k}^* \; .
\ee
If we write $H_{\alpha k} = -X_{\alpha k}$ as a column vector $\bo v$ and $R_{\alpha \beta}$ as a matrix $\bo R$, we have
\be\label{eq:FullHessian}
H_{ij}=
\begin{pmatrix}
    \frac{1}{\lambda^2} - X_{kk} & 0 & \bo v^\tr \\
    0 & \frac{1}{\lambda^2} - X_{kk}^* & (\bo v^\tr)^* \\
    \bo v & \bo v^* & - \bo R \;.
\end{pmatrix}
\ee
Thus, the contribution to the prefactor from the $k_3$ and momentum integrals is then given by
\be
\sqrt{\frac{(2\pi)^5}{\det \bo H}} \; .
\ee

However, since $H_{ij}$ has a natural block structure
\be\label{eq:Block}
\begin{pmatrix}
    A & B \\
    C & D
\end{pmatrix}
\ee
we can write the determinant in two equivalent ways
\be
\begin{split}
\det \eqref{eq:Block} &= \det (A) \det(D -C A^{-1} B) \\
&= \det (D) \det (A - B D^{-1} C)
\end{split}
\ee
which correspond to, respectively,
\be\label{eq:kFirst}
\left| \frac{1}{\lambda^2} - X_{kk} \right| \det \left( -R_{\alpha \beta} - 2 \lambda^2\Re \frac{X_{\alpha k} X_{\beta k}}{1-X_{kk}\lambda^2}  \right)
\ee
and
\be\label{eq:PiFirst}
\begin{split}
    \det&(-R_{\alpha \beta}) \times \\
    &\left(\left| \frac{1}{\lambda^2} -X_{kk} + X_{\alpha k}R^{-1}_{\alpha \beta} X_{\beta k}\right|^2 - \left| X_{\alpha k}R^{-1}_{\alpha \beta} X_{\beta k}^* \right|^2 \right)\; .
\end{split}
\ee
\eqref{eq:kFirst} and~\eqref{eq:PiFirst} are, respectively, what we would obtain if we integrated with respect to $k_3$ or over the momenta first, since the blocks on the diagonal of~\eqref{eq:FullHessian} correspond to the partial derivatives~\eqref{eq:PartialHessian}. In fact, the second factor in Eq.~\eqref{eq:kFirst} is exactly the matrix of coefficients of the spectrum~\eqref{eq:WidthsFull}.

From~\eqref{eq:WidthsFull} we see that $-R_{\alpha \beta}$ is the spectrum at $\lambda = 0$, therefore we must have $\det(-R_{\alpha \beta} ) > 0$. This is because when $\lambda = 0$ the wave packet is a delta function around $l_3 = 0$, which means that only $k_3 = 0$ can contribute, so it can only be a maximum. Furthermore, since $X_{kk}$ is complex, $\left| 1/\lambda^2 - X_{kk} \right| \neq 0$.

From these considerations and the fact that $\eqref{eq:kFirst} = \eqref{eq:PiFirst}$ we see that the determinant of the spectrum $d^{-2}_{\alpha \beta}$ vanishes if and only if
\be\label{eq:LcritSolutions}
\left| \frac{1}{\lambda^2} -X_{kk} + X_{\alpha k}R^{-1}_{\alpha \beta} X_{\beta k}\right|^2 = \left| X_{\alpha k}R^{-1}_{\alpha \beta} X_{\beta k}^* \right|^2 \;.
\ee
Thus, from~\eqref{eq:defCrit} we immediately see that one solution is precisely $\lambda = \lambda_c$, which means that $\det d^{-2}_{\alpha \beta} = 0$ exactly at $\lambda_c$ and $\det \bo d^{-2} < 0$ at $\lambda \gtrsim \lambda_c$, i.e. the saddle point $k_3 = 0$ no longer corresponds to a maximum. 

However, for some values of the parameters, there is a second solution $\widetilde \lambda_c > \lambda_c$ to~\eqref{eq:LcritSolutions} given by
\be
\widetilde \lambda_c = \frac{1}{\sqrt{X_{kk} - X_{\alpha k}R^{-1}_{\alpha \beta} I_{\beta k}}} \; ,
\ee
where $I_{\beta k} = 2i \, \Im  X_{\beta k}$. Since $\det \bo d^{-2} > 0$ and all eigenvalues are positive at $\lambda > \widetilde \lambda_c$, the saddle point $k_3 = 0$ becomes a maximum again. 

Furthermore, as $\lambda \to \lambda_c$ from below we have $d^{-2}_{p_1 \Delta}, d^{-2}_{\Delta \Delta} \to 0$, so the spectrum becomes very spread out in the $\Delta$ direction. At $\lambda = \lambda_c$ the two coefficients are exactly zero and for $\lambda \gtrsim \lambda_c$ we have $d^{-2}_{\Delta \Delta} < 0$ at $k_3 = 0$, but this is to be expected because $k_3 = 0$ is no longer a maximum.

\begin{figure}
    \centering
    \includegraphics[width=\linewidth]{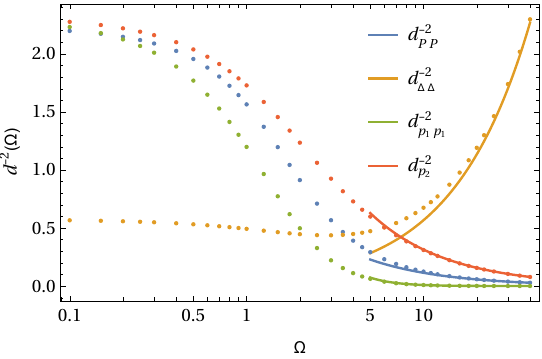}
    \caption{Widths~\eqref{eq:d2Width}, \eqref{eq:WidthsMatrix} (without the overall factor of $1/E$) as functions of the photon energy $\Omega$ for $\gamma_t = \gamma_z = 1$ and $\lambda = 0$ (where the widths are simply $d^{-2}_{\alpha \beta} = -\Re \, X_{\alpha \beta}$) (dotted) and their high-energy approximations~\eqref{XPP}, \eqref{XDeltaDeltaPW}, \eqref{Xp1p1PW}  (solid lines).}
    \label{fig:widthssub}
\end{figure}

For some value of the parameters, the second critical wave packet size $\widetilde \lambda_c$ becomes imaginary, indicating that $k_3 = 0$ never returns to be a maximum. From now on, we focus on this case. In the supercritical regime $\lambda > \lambda_c$, all the coefficients become nonzero. Although the individual elements are different, the eigenvalues are positive and equal for the two saddle points, thus the determinant and the prefactor are the same.

\begin{figure}
    \centering
    \includegraphics[width=\linewidth]{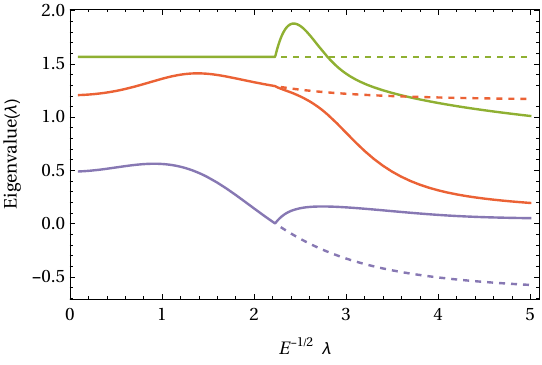}
    \caption{Eigenvalues of the matrix of coefficients $d^{-2}_{\alpha \beta}$ (without the overall factor of $1/E$) as a function of $\lambda$ for $\Omega = \gamma_t = \gamma_z = 1$ (hence critical size $\lambda_c \approx 2.23 \sqrt{E}$). The dashed lines represent the eigenvalues at $k_3 = 0$ after the phase transition, i.e. when it is no longer a maximum. We see that the lilac dashed line becomes negative at $\lambda_c$ and never returns to be positive, indicating that for these parameters $k_3 = 0$ never returns to be a maximum.}
    \label{fig:widths}
\end{figure}

In Fig.~\ref{fig:widths} we can see the eigenvalues of $\bo d^{-2}$ before and after the split. We immediately see that, in the subcritical regime, the green line corresponds to $d_{PP}^{-2}$ and it is independent of $\lambda$. Furthermore, also in the subcritical regime, the red line is approximately equal to $d_{p_1 p_1}^{-2}$ and the lilac one to $d_{\Delta \Delta}^{-2}$ since the mixing term $d_{p_1 \Delta}^{-2}$ is smaller compared to the other two. At the critical size, the coefficient $d_{\Delta \Delta}^{-2}$ is exactly zero. The reason why the coefficients are not smooth is that the saddle point changes at once at $\lambda_c$, since if we look at the eigenvalues at $k_3 = 0$ (dashed lines in Fig.~\ref{fig:widths}) they are smooth and the one corresponding to $d_{\Delta \Delta}^{-2}$ becomes negative when $\lambda > \lambda_c$, indicating that $k_3 = 0$ is no longer a maximum. In the supercritical regime, all the off-diagonal coefficients become nonzero and the contour surfaces of the spectrum are ellipsoids with principal axes determined by the eigenvectors and relative lengths proportional to the inverses of the eigenvalues.

The fact that $\det \bo d^{-2} = 0$ at the critical point tells us that the saddle-point approximation breaks down close to and at $\lambda = \lambda_c$ because the saddle point is degenerate and the prefactor would be infinite. The spectrum is no longer gaussian in the direction $\Delta$ because $d^{-2}_{\Delta \Delta} = 0$, so we need to consider higher order terms. To obtain a bounded spectrum, we must consider the term $\mathcal O (\Delta^4)$.

\begin{figure*}
    \centering
    \includegraphics[width=0.48\linewidth]{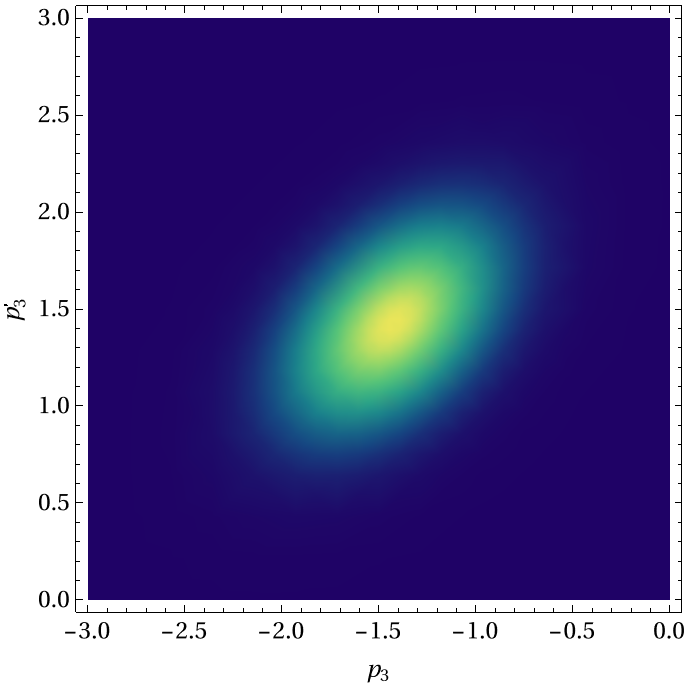}
    \hfill
    \includegraphics[width=0.48\linewidth]{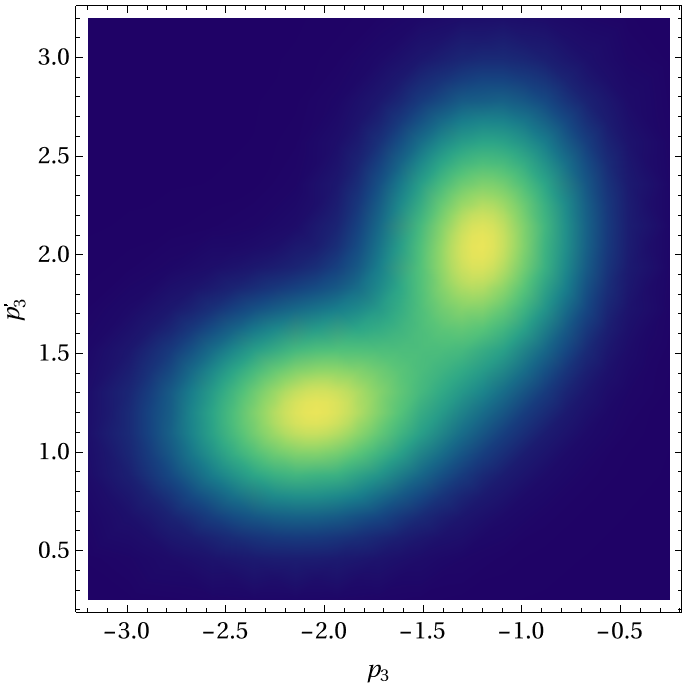}
    \caption{Longitudinal spectrum before and after phase transition ($p_\LCperp = p_{\LCperp s})$. The parameters are $E= 1/10$ and $\Omega = \gamma_t = \gamma_z = 1$ for both, while $\lambda = 2 \sqrt{E}$ on the left and $\lambda = 3.2 \sqrt{E}$ on the right. With these parameters, the critical size is $\lambda_c \approx 2.23\sqrt{E}$. The principal axes of the elliptical shapes in the spectrum are given by the eigenvectors of the matrix $d^{-2}_{\alpha \beta}$.}
    \label{fig:splitting}
\end{figure*}

Focusing on the longitudinal part of the spectrum, i.e. considering the transverse momenta at their saddle point $p_\LCperp = k_\LCperp/2$, we see the transition in Fig.~\ref{fig:splitting} with the splitting of the peak into two.

\subsection{Critical exponents}

We note that the total exponent
\be
\psi_a(\lambda,b,k_3) = -\frac{k_3^2}{2\lambda^2} +\psi(k_3) + ibk_3
\ee
has an expansion at small $k_3$ and $b = 0$
\be
\psi_a(\lambda,k_3) \sim a_2(\lambda) k^2_3 + a_4 k^4_3
\ee
with
\be
\begin{split}
    a_2(\lambda) &= -\frac{1}{2\lambda^2} +\frac{1}{2}\frac{\ud^2 \psi}{\ud k_3^2} \sim \lambda - \lambda_c \\ 
    a_4 &= \frac{1}{4!}\frac{\ud^4 \psi}{\ud k_3^4} \;,
\end{split}
\ee
therefore it has the same form as the free energy in Landau theory of second-order phase transitions.
From this and the fact that $\Delta_s \sim k_s$ when $k \ll 1$ we can already conclude that the critical exponents defined as
\be 
\begin{split}
    &\Delta_s(b = 0) \sim \bar \lambda^\beta \qquad \Delta_s(\lambda = \lambda_c) \sim b^{\frac{1}{\delta}} \\
    &\frac{\pa \Delta_s}{\pa b} \Bigr|_{b=0} \sim 
\begin{cases}
    \bar \lambda^{-\gamma} \quad \; \lambda < \lambda_c\\
    \bar \lambda^{-\gamma'} \quad \lambda > \lambda_c
\end{cases}
    \\
    &C = -\lambda \frac{\ud^2 g}{\ud \lambda^2} \sim 
\begin{cases}
    (-\bar \lambda)^{-\alpha} \quad \; \lambda < \lambda_c\\
   (-\bar \lambda)^{-\alpha'} \quad \lambda > \lambda_c
\end{cases}
\end{split}
\ee
where $g(\lambda) = \psi_a \bigr(\lambda,k_s(\lambda) \bigr)$ and $\bar \lambda = (\lambda - \lambda_c)/\lambda_c$, are given by
\be
\alpha = \alpha' = 0 \qquad \beta = \frac{1}{2} \qquad \gamma = \gamma' = 1 \qquad \delta = 3
\ee
and as a result, in the saddle-point approximation, this phase transition falls in the universality class of mean field theory.

\section{Integrated probability}

To obtain the integrated probability we evaluate the exponent at the saddle points and collect all the contributions to the prefactor. We write
\be
\mathbb{P}=\int\ud^2p_\LCperp\ud p_3\ud p'_3\mathbb{P}(p_\LCperp,p_3,p'_3)
\ee
for the integrated probability and the spectrum. In this paper we sum over the spins of the produced pair. The dependence on the photon polarization is included by writing the spectrum as $\mathbb{P}(p,p')={\bf N}\cdot{\bf M}(p,p')$, where ${\bf N}$ is the Stokes vector for the photon~\eqref{stokesVector}. 
We have
\be
\begin{split}
{\bf M}(p,p') =& \frac{\alpha}{\sqrt{\pi}\lambda(2\pi)^4}\frac{p_0 p'_0{\bf m}}{m_\perp^2} \\
&\left|\int\frac{\ud k_3}{\sqrt{k_0}}\frac{e^{-\frac{k_3^2}{2\lambda^2}+ibk_3+\psi}}{2\pi T} \frac{T}{\sqrt{\det \Lambda}} \frac{\pi^2}{\sqrt{\det \bo H}}\right|^2 \;,
\end{split}
\ee
where ${\bf m}$ comes from the spin and polarization terms, which we calculate in Appendix~\ref{app:spin}; $\det\Lambda$ is the functional determinant from the worldline path integral, which we calculate in Appendix~\ref{app:PI}; $\det{\bf H}$ is the determinant of the Hessian matrix coming from the expansion around the saddle point of the $\{T,\sigma,z_\LCpm\}$ integrals, which we calculate in Appendix~\ref{app:OI}. Collecting everything we finally find 
\be\label{generalMueller}
{\bf M}(p,p')=\frac{\alpha\{1+3p_1^2,0,0,1-p_1^2\}}{2\pi^{3/2}k_0p_0p'_0m_\LCperp^2\lambda|D''/2|}\frac{e^{-\mathcal{A}(p,p')}}{\left|\frac{1}{\lambda^2}-X_{kk}\right|} \;,
\ee
where we have used~\eqref{DetGYFinal} and \eqref{eq:HessianOrdinary}. 
Note that for $\lambda>\lambda_c$, the spectrum has more than one peak. The total probability is the sum of the contributions of the individual peaks
\be
\mathbb P = \sum_n \mathbb P_n \; .
\ee
Integrating over momenta we obtain
\be
\begin{split}
\bo M =& \frac{\alpha \{1+3p_1^2,0,0,1-p_1^2\}  \sqrt \pi}{2\lambda k_0 m_\LCperp^2 p_0 p'_0 |D''/2|}\\
&\frac{1}{\left|X_{kk} - \frac{1}{\lambda^2}\right|} \frac{e^{-\mathcal A}}{\sqrt{d_2^{-2}\det \bo d^{-2}}} \;,
\end{split}
\ee
where $d_2^{-2}$ comes from the integration over $p_2$ and $\det \bo d^{-2}$ from $(p_1, \Delta, P)$.
When $\lambda \to 0$, we recover Eq.~\eqref{PwidePhoton}.

\begin{figure}
    \centering    \includegraphics[width=\linewidth]{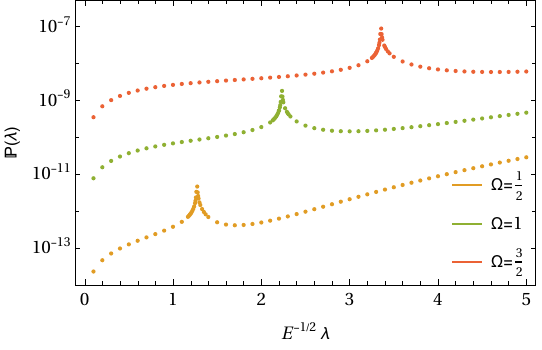}
    \caption{Total probability as a function of $\lambda$ for $\gamma_t = \gamma_z = 1$, $E = 1/10$, perpendicular polarization, and different values of the photon energy $\Omega$.}
    \label{fig:total}
\end{figure}

In Fig.~\ref{fig:total} we see the total probability as a function of $\lambda$ for fixed field parameters and different photon energies. At the phase transition, we clearly see a peak in the probability. In the saddle-point approximation, the prefactor goes to infinity, but this is simply due to the fact that we neglect higher order terms, which are actually needed when the saddle point is degenerate $\det \bo d^{-2} = 0$. To obtain the value of the probability at the critical point, to know how high the peak is, we would need to consider terms up to order four in $\Delta$. 

Higher-order stationary points, in the language of catastrophe theory, have been studied in~\cite{Kharin:2018dxa} for nonlinear Compton scattering by chirped plane-wave fields. Merging saddle points have been studied in nonlinear Compton and Breit-Wheeler in~\cite{Seipt:2015rda,Nousch:2015pja}.  

Note that we can in principle use the instanton formalism to calculate the value of the spectrum at some generic point $({\bf p},{\bf p}')$ in the momentum phase space without expanding around the saddle points at all. So, for example, at the critical point we could ``just'' set up a grid of points in $({\bf p},{\bf p}')$, find the instantons at each of these points and then use them to compute the exponent etc. This, though, might be relatively expensive numerically, as one can understand by imagining how many grid points, i.e. how many different instantons, would be needed to obtain a smooth plot of the spectrum, compared to what we have focused on, where the gaussian shapes of the peaks are described by their widths, which we obtain by finding a single instanton plus the four solutions $\delta q_{[j]}$ which describe the variation around the instanton. It is usually also easier to find the instantons that correspond to the saddle-point values of the momenta, since away from the saddle points there is less symmetry. However, it should be noted that, at least in most of the cases we have considered, it actually does not take much time to find the instantons.

\section{High-energy limit}\label{sec:highOmega}

In this section we explain how to obtain the high-energy $\Omega\gg1$ expansion. Since the critical point increases as $\lambda_c\propto\Omega^{3/2}$ (see~\eqref{lambdacLargeOmega}), we will focus on $\lambda<\lambda_c$.

\subsection{Instanton expansion}

To obtain the high-energy limit we change variable from $u$ to $\phi=t(u)-\tilde{t}$, where $\tilde{t}=t(u=0)$. We can expand the instanton as
\be\label{tzPWser}
\tilde{t}=\sum_{n=0}^\infty\frac{\tilde{t}_{(n)}}{\rho^{2n}} 
\qquad
z(\phi)=\frac{1}{\rho}\sum_{n=0}^\infty\frac{z_{(n)}(\phi)}{\rho^{2n}} \;,
\ee
where $\rho=\Omega/2$, while $\phi=\mathcal{O}(\rho^0)$.
The Jacobian, $d(\phi)=\ud\phi/\ud u$ is obtained from $t^{\prime2}(u)-z^{\prime2}(u)=m_\LCperp^2$, which gives, assuming $u>0$,
\be\label{dphidu}
\begin{split}
d(\phi)&=\frac{m_\LCperp}{\sqrt{1-z^{\prime2}(\phi)}}=\rho\sum_{n=0}^\infty\frac{d_{(n)}(\phi)}{\rho^{2n}}\\
&=\rho+\frac{1+z_{(0)}^{\prime2}(\phi)}{2\rho}+\mathcal{O}(\rho^{-3}) \;.
\end{split}
\ee
From $z'(\phi)=z'(u)/t'(u)$ we have $z'(\phi=0)=i/\rho$, so $z_{(0)}'(0)=i$ and $z_{(n>0)}'(0)=0$. We have divided out the trivial $E$ scaling as in~\eqref{Erescaling}. We consider a field on the form $E_3(t,z)=F_t(\gamma_t t)F_z(\gamma_zz)$. $F_z$ has a maximum at $z=0$, $F_z'(0)=0$, and we normalize $F_z$ and $\gamma_z$ such that
\be\label{eq:FieldNorm}
F_z(0)=1
\qquad
F_z''(0)=-2 \;.
\ee
For the temporal pulse shape we instead write $F_t(x)=f'(x)$ and normalize as $F_t(0)=1$, $f(0)=0$ and $f(\infty)=1$. For a double Sauter pulse we have $F_t(x)=F_z(x)=\text{sech}^2(x)$ and $f(x)=\tanh(x)$.
Expanding the Lorentz-force equation to leading order gives $z_{(0)}''(\phi)=F_t(\gamma_t[\tilde{t}_0+\phi])$, so
\be
z_0'(\phi)=i+\frac{1}{\gamma_t}f(\gamma_t[\tilde{t}_0+\phi])-\frac{1}{\gamma_t}f(\gamma_t\tilde{t}_0) \;.
\ee
$z'$ should be real at asymptotic times, which implies
\be\label{PWsolLead}
\tilde{t}_0=\frac{iv}{\gamma_t}
\qquad
z_0'(\phi)=\frac{1}{\gamma_t}f(\gamma_t[\tilde{t}_0+\phi])
\;,
\ee
with $v$ given by
\be\label{vDef}
v=\tilde{f}^{-1}\left(\frac{1}{a_0}\right)>0 \;,
\ee
where $f(ix)=i\tilde{f}(x)$ and $a_0=1/\gamma_t$. Expanding the Lorentz-force equation to the next order gives
\be
\begin{split}
&z_{(1)}''-\gamma_t\tilde{t}_{1} F_t'(\gamma_t[\tilde{t}_0+\phi])\\
&=-\left[\frac{1}{2}+\gamma_z^2z_{(0)}^2+\frac{3}{2}z_{(0)}^{\prime2}\right]F_z=:-h(\gamma_t[\tilde{t}_0+\phi]) \;,
\end{split}
\ee
where $h$ is known from the zeroth-order solution. So
\be
z_{(1)}'(\phi)=\tilde{t}_1\{F_t(\gamma_t[\tilde{t}_0+\phi])-F_t(\gamma_t\tilde{t}_0)\}
-\int_0^\phi\ud\phi' h \;.
\ee
Demanding $\text{Im }z_{(1)}'(\infty)=0$ gives
\be
\tilde{t}_1=\frac{1}{\gamma_t F_t(\gamma_t\tilde{t}_0)}\int_0^{\gamma_t\tilde{t}_0}\ud x\, h(x) \;.
\ee
Both $\text{Re }\tilde{t}_0=\text{Re }\tilde{t}_1=0$.

Expanding~\eqref{expPsiStart} using~\eqref{PWsolLead}, we find to leading order $2\text{Re }\psi=-\mathcal{A}_{PW}$, where $\mathcal{A}_{PW}$ is the plane wave result in~\eqref{ds} with $\chi=\sqrt{-(F^{\nu\nu}k_\nu)^2}=E\Omega$.

For a Sauter pulse, $F_t(x)=\text{sech}^2(x)$, we have $\tilde{t}_0=(i/\gamma_t)\text{arctan}(\gamma_t)$,
\be
\begin{split}
z_{(0)}&=\frac{1}{\gamma_t^2}\ln\left[\frac{\cosh(\gamma_t[\tilde{t}_0+\phi])}{\cosh(\gamma_t\tilde{t}_0)}\right] \\
&=\frac{1}{\gamma_t^2}\ln[\cosh(\gamma_t\phi)+i\gamma_t\sinh(\gamma_t\phi)] \;,
\end{split}
\ee
\be
\begin{split}
\tilde{t}_1=&\frac{i\gamma_z^2}{\gamma_t^5(1+\gamma_t^2)}\bigg[2\gamma_t-\text{Im Li}_2(e^{i\theta})\\
&+2\text{arctan}(\gamma_t)\left(\ln\left[\frac{2}{\sqrt{1+\gamma_t^2}}\right]-1\right)\bigg] \;,
\end{split}
\ee
where $\text{Li}_2$ is the dilogarithm~\cite{PolyLogDLMF} and
\be
\theta=\pi-2\text{arctan}(\gamma_t) \;.
\ee

From~\eqref{dphidu} we find 
\be\label{uTophiSer}
u(\phi)=\frac{1}{\rho}\sum_{n=0}^\infty\frac{u_{(n)}(\phi)}{\rho^{2n}}=
\frac{\phi}{\rho}+\mathcal{O}(\rho^{-3}) \;,
\ee
which, together with~\eqref{PWsolLead} and~\eqref{d2k0}, gives
\be
d_2^{-2}=\frac{4v}{E\Omega\gamma_t} \;,
\ee
which agrees with~\eqref{d23PW} and~\eqref{highEnergydPz}. For the other widths we need $\delta q_{[j]}$. 

\subsection{$\delta q_{[1]}$}

We find
\be\label{delta1ser}
\begin{split}
\delta t_{[1]}(\phi)&=\sum_{n=0}^{\infty}\frac{\delta t_{[1]}^{(n)}(\phi)}{\rho^{2n}}=1+\mathcal{O}(\rho^{-2})\\
\delta z_{[1]}(\phi)&=\sum_{n=1}^\infty\frac{\delta z_{[1]}^{n}(\phi)}{\rho^{2n-1}}=\frac{\delta z_{[1]}^{(1)}(\phi)}{\rho}+\mathcal{O}(\rho^{-3}) \;.
\end{split}
\ee
Expanding~\eqref{dtdzeq} gives to leading order
\be\label{deltaz10}
\delta z_{[1]}^{(1)\prime\prime}(\phi)=\frac{\ud}{\ud\phi} F_t(\gamma_t[\tilde{t}_0+\phi]) \;,
\ee
and hence
\be
\delta z_{[1]}^{(1)}(\phi)=\frac{1}{\gamma_t}f(\gamma_t[\tilde{t}_0+\phi])-i-F_t(\gamma_t\tilde{t}_0)\phi \;.
\ee
From $\nu_{[1]}=0$ we have
\be
\delta t_{[1]}^{(1)\prime}(\phi)=z_{(0)}'(\phi)\delta z_{[1]}^{(1)\prime}(\phi) \;,
\ee
which gives
\be\label{deltat11}
\begin{split}
\delta t_{[1]}^{(1)}(\phi)&=\frac{1}{2}\left[1+\frac{1}{\gamma_t^2}f^2(\gamma_t[\tilde{t}_0+\phi])\right]\\
&-\frac{F_t(\gamma_t\tilde{t}_0)}{\gamma_t^2}\int_{\gamma_t\tilde{t}_0}^{\gamma_t(\tilde{t}_0+\phi)}\ud x\, f(x) \;.
\end{split}
\ee
Since $\delta q_P$ only involves this basis solution, see~\eqref{deltaqDeltaP21}, we can now approximate~\eqref{XPPk0}. We find
\be\label{XPP}
X_{PP}=\frac{1}{\rho}\sum_{n=0}\frac{X_{PP}^{(n)}}{\rho^{2n}}=-\frac{4a_0}{E\Omega}\left(v-\frac{1}{a_0\tilde{f}'(v)}\right)+\mathcal{O}(\Omega^{-3}) \;.
\ee
The leading order agrees with~\eqref{d23PW} and~\eqref{highEnergydPz}.

\subsection{$\delta q_{[2]}$}

The second solution can be expanded as
\be\label{delta2ser}
\begin{split}
\delta t_{[2]}(\phi)&=\sum_{n=1}^\infty\frac{\delta t_{[2]}^{(n)}(\phi)}{\rho^{2n+1}}=\frac{\delta t_{[2]}^{(1)}(\phi)}{\rho^3}+\mathcal{O}(\rho^{-5}) \\
\delta z_{[2]}(\phi)&=\sum_{n=0}^\infty\frac{\delta z_{[2]}^{(n)}(\phi)}{\rho^{2n}}=1+\mathcal{O}(\rho^{-2}) \;.
\end{split}
\ee
Expanding~\eqref{dtdzeq} gives
\be\label{ddz21}
\delta z_{[2]}^{(1)\prime\prime}(\phi)=-2\gamma_z^2z_{(0)}(\phi)F_t(\gamma_t[\tilde{t}_0+\phi]) \;,
\ee
so
\be\label{deltaz21prime}
\begin{split}
\delta z_{[2]}^{(1)\prime}(\phi)=&\frac{2\gamma_z^2}{\gamma_t^2}\int_0^\phi\ud\phi' f(\gamma_t[\tilde{t}_0+\phi'])\\
&\times\{f(\gamma_t[\tilde{t}_0+\phi'])-f(\gamma_t[\tilde{t}_0+\phi])\} \;.
\end{split}
\ee
$\delta q_\Delta$ only involves $\delta q_{[2]}$, see~\eqref{deltaqDeltaP21}. So with $\delta q_{[2]}$ we can obtain $X_{\Delta\Delta}$ in~\eqref{XDeltaDeltak0}. We find
\be\label{XDeltaDeltaPW}
X_{\Delta\Delta}=\rho\sum_{n=0}^\infty\frac{X_{\Delta\Delta}^{(n)}}{\rho^{2n}}=-\frac{i\Omega}{4E\delta z_{[2]}^{(1)\prime}(\infty)}+\mathcal{O}(\Omega^{-1}) \;,
\ee
where
\be\label{dz21}
\delta z_{[2]}^{(1)\prime}(\infty)=\frac{2\gamma_z^2}{\gamma_t^3}\int_{iv}^\infty\ud x\, f(x)[f(x)-1] \;.
\ee

\subsection{$\delta q_{[3]}$}

For the third basis solution we have
\be\label{delta3ser}
\begin{split}
\delta t_{[3]}(\phi)&=\sum_{n=0}^\infty\frac{\delta t_{[3]}^{(n)}(\phi)}{\rho^{2n+1}}=\frac{\phi}{\rho}+\mathcal{O}(\rho^{-3}) \\
\delta z_{[3]}(\phi)&=\sum_{n=1}^\infty\frac{\delta z_{[3]}^{(n)}(\phi)}{\rho^{2n}}=\frac{\delta z_{[3]}^{(1)}(\phi)}{\rho^2}+\mathcal{O}(\rho^{-4}) \;.
\end{split}
\ee
Expanding~\eqref{dtdzeq} to leading order and integrating once gives
\be
\delta z_{[3]}^{(1)\prime}=\phi F_t(\gamma_t[\tilde{t}_0+\phi]) \;.
\ee
From $\nu_{[3]}(u>0)=\rho$ we find
\be
\delta t_{[3]}^{(1)\prime}(\phi)=-1-z_{(0)}^{\prime2}(\phi)+z_{(0)}'(\phi)\delta z_{[3]}^{(1)\prime}(\phi) \;.
\ee
\eqref{deltaqp1From32} shows that in general we need both $\delta q_{[3]}$ and $\delta q_{[2]}$ to obtain $\delta q_{p_1}$. But to leading order we have
\be
\delta t_{p_1}(\phi)\approx\frac{\phi}{\rho}+\frac{\delta t_{[3]}^{(1)}(\phi)}{\rho^3} \;,
\ee
so $\delta t_{p_1}(\phi)\approx \delta t_{[3]}(\phi)$.
To calculate $X_{p_1p_1}$ in~\eqref{Xp1p1k0} we need to find $u_{(1)}$ in~\eqref{uTophiSer}. We find
\be
u_{(1)}(\phi\to\infty)=-\frac{1}{2}\left[1+\frac{1}{\gamma_t^2}\right]\phi+\frac{L}{2} \;,
\ee
where
\be
L=\frac{1}{\gamma_t^3}\int_{iv}^\infty\ud x[f^2(\infty)-f^2(x)] \;.
\ee
We also need
\be
\delta t_{[3]}^{(1)}(\phi\to\infty)=-\left[1+\frac{1}{\gamma_t^2}\right]\phi+\frac{3}{2}L \;.
\ee
With this we find
\be\label{Xp1p1PW}
\begin{split}
X_{p_1p_1}&=\frac{1}{\rho^3}\sum_{n=0}^\infty\frac{X_{p_1p_1}^{(n)}}{\rho^{2n}}\\
&=-\frac{16a_0}{E\Omega^3}\left(v-a_0^2\mathcal{J}-\frac{i}{a_0}\text{Re }L\right)+\mathcal{O}(\Omega^{-5}) \;.
\end{split}
\ee
The leading order agrees with~\eqref{ds} and~\eqref{timed1HighEnergy}.

\subsection{$\delta q_{[4]}$}

For the fourth basis solution we have
\be\label{delta4ser}
\begin{split}
\delta t_{[4]}(\phi)&=\sum_{n=1}^\infty\frac{\delta t_{[4]}^{(n)}(\phi)}{\rho^{2n}}=\frac{\delta t_{[4]}^{(1)}(\phi)}{\rho^2}+\mathcal{O}(\rho^{-4}) \\
\delta z_{[4]}(\phi)&=\sum_{n=0}^\infty\frac{\delta z_{[4]}^{(n)}(\phi)}{\rho^{2n+1}}=\frac{\phi}{\rho}+\mathcal{O}(\rho^{-3}) \;.
\end{split}
\ee
Expanding~\eqref{dtdzeq} gives
\be
\delta t_{[4]}^{(1)\prime\prime}(\phi)=F_t(\gamma_t[\tilde{t}_0+\phi]) \;,
\ee
so 
\be\label{ddeltat41}
\delta t_{[4]}^{(1)}(\phi)=z_0(\phi)-i\phi \;.
\ee
Using~\eqref{ddeltat41}, we can integrate the equation for $\delta z_{[4]}^{(1)}$ once to obtain
\be\label{deltaz41}
\begin{split}
\delta z_{[4]}^{(1)\prime}=&[z_0(\phi)-i\phi]z_0''(\phi)-\frac{1}{2}[1+z_0^{\prime2}(\phi)]\\
&+\phi\delta z_{[2]}^{(1)\prime}(\phi)-\delta z_{[2]}^{(1)}(\phi) \;.
\end{split}
\ee 

From~\eqref{deltakFrom234} we find $\delta t_k=\mathcal{O}(\rho^{-2})$ and $\delta z_k\approx\rho/(2\delta z_{[2]}^{(1)\prime}(\infty))$. From~\eqref{Xkk0} and~\eqref{XkDelta0} we find
\be
X_{kk}=\rho\sum_{n=0}^\infty\frac{X_{kk}^{(n)}}{\rho^{2n}}
\qquad
X_{k\Delta}=\rho\sum_{n=0}^\infty\frac{X_{k\Delta}^{(n)}}{\rho^{2n}} \;.
\ee
To leading order we have
\be\label{XkkXkDelta}
X_{kk}^{(0)}=-X_{k\Delta}^{(0)}=X_{\Delta\Delta}^{(0)} \;,
\ee
where $X_{\Delta\Delta}^{(0)}$ is given by~\eqref{XDeltaDeltaPW}. 

At next-to-leading order we also have
\be\label{NLOzero}
\begin{split}
&2i\delta z_{[2]}^{(1)\prime}(\phi)[X_{kk}^{(1)}+2X_{k\Delta}^{(1)}+X_{\Delta\Delta}^{(1)}]\\
&=\frac{1}{2}[1+z_0^{\prime2}(\phi)]-\phi\delta z_{[2]}^{(1)\prime}+\delta z_{[2]}^{(1)}+\delta z_{[4]}^{(1)\prime} \;,
\end{split}
\ee
where $\phi\to\infty$. Using~\eqref{deltaz41} we find
\be
\eqref{NLOzero}=[-i\phi+z_0(\phi)]z_0''(\phi) \;,
\ee
which vanishes as $\phi\to\infty$. Thus,
\be\label{eq:Xfirstorder}
X_{kk}^{(1)}+2X_{k\Delta}^{(1)}+X_{\Delta\Delta}^{(1)} = 0 \; .
\ee

\subsection{Off-diagonal $X$}

From~\eqref{Xkp10} and~\eqref{Xp1Delta0} we find 
\be
X_{kp_1}=\frac{1}{\rho^2}\sum_{n=0}^\infty\frac{X_{kp_1}^{(n)}}{\rho^{2n}} 
\qquad
X_{p_1\Delta}=\frac{1}{\rho^2}\sum_{n=0}^\infty\frac{X_{p_1\Delta}^{(n)}}{\rho^{2n}} \;.
\ee
To leading order we have
\be
X_{kp_1}^{(0)}=\lim_{\phi\to\infty}i\left[-i\phi-\delta t_{[4]}^{(1)}+\frac{\delta t_{[2]}^{(1)}}{\delta z_{[2]}^{(1)\prime}}\right] \;.
\ee
From $\nu_{[2]}=0$ we find $\delta t_{[2]}^{(1)\prime}=z_{(0)}'\delta z_{[2]}^{(1)\prime}$, so after a partial integration
\be
\delta t_{[2]}^{(1)}=z_{(0)}\delta z_{[2]}^{(1)\prime}-\int_0^\phi\ud\phi'\, z_{(0)}\delta z_{[2]}^{(1)\prime\prime} \;,
\ee
where $\delta z_{[2]}^{(1)\prime\prime}$ is given by~\eqref{ddz21}. From~\eqref{ddeltat41} we have $\delta t_{[4]}^{(1)}=z_{(0)}-i\phi$. Thus,
\be
X_{kp_1}^{(0)}=\frac{i2\gamma_z^2}{\delta z_{[2]}^{(1)\prime}(\infty)}\int_0^\infty\ud\phi\, z_{(0)}^2(\phi)F_t[\gamma_t(\tilde{t}_0+\phi)] \;,
\ee
where $\delta z_{[2]}^{(1)\prime}(\infty)$ is given by~\eqref{dz21} (so the factors of $2\gamma_z^2$ cancel).

We also have 
\be
X_{p_1\Delta}^{(0)}=\lim_{\phi\to\infty}i\left[(\tilde{t}_0+\phi)z_{(0)}'-\frac{\delta t_{[2]}^{(1)}}{\delta z_{[2]}^{(1)\prime}}\right] \;,
\ee
which simplifies using
\be
\begin{split}
&(\tilde{t}_0+\phi)z_{(0)}'-z_{(0)}\\
&\to
-\frac{1}{\gamma_t^2}\left[\int_0^v\ud x\tilde{f}(x)
+\int_0^\infty\ud x[f(x)-1]\right] \;,
\end{split}
\ee
which gives a purely imaginary contribution to $X_{p_1\Delta}^{(0)}$. While both the real and imaginary parts of $X_{kp_1}$ contribute to the probability, only the real part of $X_{p_1\Delta}$ contributes, see~\eqref{widthsDelta1}. Thus,
\be\label{Xp1Deltakp1}
X_{p_1\Delta}^{(0)}=-X_{kp_1}^{(0)}+i... \;.
\ee

When expanding the widths in~\eqref{widthsDelta1} we need to choose how we want $\lambda$ to scale with $\Omega$. Consider first $\lambda$ as independent of $\Omega$, then
\be
\begin{split}
&X_{\Delta p_1}+\frac{X_{k\Delta}X_{kp_1}\lambda^2}{1-X_{kk}\lambda^2}\\
&=\frac{1}{\rho^2}\left(X_{\Delta p_1}^{(0)}-\frac{X_{k\Delta}^{(0)}}{X_{kk}^{(0)}}X_{kp_1}^{(0)}\right)+\mathcal{O}(\rho^{-3})
=\mathcal{O}(\rho^{-3}) \;,
\end{split}
\ee
where the terms that would otherwise contribute to leading order cancel due to~\eqref{XkkXkDelta} and~\eqref{Xp1Deltakp1}. A similar cancellation of the leading-order terms happens for the coefficient of $\delta\Delta^2$, so that, instead of being $\mathcal{O}(\Omega)$, it is $\mathcal{O}(\Omega^0)$. This shows the importance of obtaining higher order terms, or at least knowing that the expansions of $X$ are in terms of $1/\Omega^2$ rather than just $1/\Omega$. After these cancellations we find
\be\label{largeOmegalambda0}
\exp\left\{-\frac{\delta\Delta^2}{\lambda^2}+\frac{X_{p_1p_1}^{(0)}}{\rho^3}\delta p_1^2+\frac{2X_{kp_1}^{(0)}}{X_{\Delta\Delta}^{(0)}\lambda^2\rho^3}\delta\Delta\delta p_1\right\} \;.
\ee
To be able to tell whether one component of $X_{ij}$ is negligible to leading order we have to also take the natural scaling of $\delta_i\delta_j$ into account, i.e. we should rescale the momentum variables so that the exponent becomes independent of $\Omega$ to leading order (but still depends on all the momentum variables).  
From~\eqref{largeOmegalambda0} we see that we should rescale $\delta p_1=\rho^{3/2}\delta\hat{p}_1$ and consider $\delta\hat{p}_1$ as $\mathcal{O}(\Omega^0)$. The cross term is then $\delta\Delta\delta p_1/\rho^3=\mathcal{O}(\Omega^{-3/2})$ and can therefore be neglected compared to the diagonal terms.

$X_{kk}=\mathcal{O}(\Omega)$, so if we want to keep a nontrivial dependence on $\lambda$ even at leading order, we should rescale $\lambda$ so that the combination $X_{kk}\lambda^2$ in $1/(1-X_{kk}\lambda^2)$ is not very small or large. We therefore consider $\lambda=\mathcal{O}(\Omega^{-1/2})$. We then have, schematically,
\be
\exp\left\{\mathcal{O}(\Omega)\delta\Delta^2+\mathcal{O}(\Omega^{-3})\delta p_1^2+\mathcal{O}(\Omega^{-2})\delta\Delta\delta p_1\right\} \;.
\ee
To determine the relative importance of these terms, the natural rescaling is $\delta p_1=\rho^{3/2}\delta\hat{p}_1$, as before, and now also $\delta\Delta=\delta\hat{\Delta}/\sqrt{\rho}$. This makes the diagonal terms $\mathcal{O}(\Omega^0)$, while the cross term is $\delta\Delta\delta p_1/\rho^2=\mathcal{O}(\Omega^{-1})$. The cross term is still small for sufficiently large $\Omega$, though only by $\mathcal{O}(\Omega^{-1})$. In the next subsections we consider the width for $\delta\Delta$ in more detail.

\subsection{$\lambda=\mathcal{O}(\Omega^{-1/2})$}

We have just explained that the cross term in~\eqref{widthsDelta1} can be neglected to leading order. 
From~\eqref{XDeltaDeltaPW} we see that we can write
\be\label{XhatDef}
X_{kk}=-\frac{\Omega}{\gamma_z^2}\hat{X}(\gamma_t) \;,
\ee
where $\text{Re }\hat{X}>0$. We thus find
\be
\text{Re }\left\{X_{kk}\delta\Delta^2+\frac{X_{kk}^2\lambda^2\delta\Delta^2}{1-\lambda^2X_{kk}}\right\}=-\text{Re}\frac{\beta\hat{X}}{1+\beta\hat{X}}\frac{\delta\Delta^2}{\lambda^2} \;,
\ee
where
\be\label{betaDef}
\beta=\Omega\frac{\lambda^2}{\gamma_z^2} \;.
\ee
We can interpret $\beta$ by noting that $\lambda/\gamma_z$ is essentially the size of field divided by the size of photon wave packet.
Thus, 
\be\label{dDeltaRe}
d_\Delta^{-2}=\frac{1}{\lambda^2}\text{Re}\frac{\beta\hat{X}}{1+\beta\hat{X}} \;.
\ee
It is straightforward to check that $d_\Delta^{-2}>0$. For a Sauter pulse we have
\be
\hat{X}=\frac{\gamma_t^3}{2}\left(\gamma_t-\text{arctan}(\gamma_t)+i+i\ln\left[\frac{1}{2}\sqrt{1+\gamma_t^2}\right]\right)^{-1} \;.
\ee
In deriving~\eqref{dDeltaRe} we have assumed that $\Omega$ is large compared to some function of $\gamma_t$ and $\gamma_z$.  

In the limit where $\Omega$ is much larger than everything else, in particular $\beta\gg1$, we have
\be\label{dDeltaBetaLarge}
d_\Delta\approx\lambda \;,
\ee
so $d_\Delta$ becomes field independent and is equal to the width of the photon wave packet.

In the limit where $\lambda$ is smaller than everything else, in particular $\beta\ll1$, we instead have
\be
\delta_\Delta^{-2}\approx\frac{\Omega}{\gamma_z^2}\text{Re }\hat{X} \;,
\ee
which is independent of the size of the wave packet.

\subsection{$\lambda=\mathcal{O}(\Omega^{3/2})$}

Expanding~\eqref{widthsDelta1} with $\lambda=\mathcal{O}(\Omega^{3/2})$ using~\eqref{XkkXkDelta}, \eqref{eq:Xfirstorder} and~\eqref{Xp1Deltakp1}, we find
\be
\begin{split}
&\text{Re}\left[-\frac{\rho^3}{\lambda^2}-\frac{(X_{kk}^{(1)}-X_{\Delta\Delta}^{(1)})^2}{4 X_{\Delta\Delta}^{(0)}}+X_{kk}^{(2)}+2X_{k\Delta}^{(2)}+X_{\Delta\Delta}^{(2)}\right]\\
&\times\frac{\delta\Delta^2}{\rho^3}+\frac{\text{Re }X_{p_1p_1}^{(0)}}{\rho^3}\delta p_1^2+\mathcal{O}(\rho^{-4})\delta \Delta\delta p_1 \;.
\end{split}
\ee
With both $\delta\Delta=\mathcal{O}(\rho^{3/2})$ and $\delta p_1=\mathcal{O}(\rho^{3/2})$, we see that the cross term is $\mathcal{O}(1/\rho)$ and can hence be neglected to leading order. We change time variable from $\phi$ to
\be
x=\gamma_t(\tilde{t}_0+\phi) \;,
\ee
and choose a contour that starts at $x=iv$, follows the imaginary axis down to $x=0$, and then turn and follows the real $x$ axis to $x\to\infty$. 
Using the expansions above we find
\be\label{firsTermLargeOmegagammaz}
-\frac{(X_{kk}^{(1)}-X_{\Delta\Delta}^{(1)})^2}{4 X_{\Delta\Delta}^{(0)}}
=\frac{i}{2\rho^3\gamma_t}\frac{W^2}{\delta z_{[2]}^{(1)\prime}(x)} \;,
\ee
where
\be
W=x\delta z_{[2]}^{(1)\prime}(x)-\delta z_{[2]}^{(1)}(x) \;.
\ee
We also find
\be\label{secondTermLargeOmegagammaz}
\begin{split}
&\text{Re}\left[X_{kk}^{(2)}+2X_{k\Delta}^{(2)}+X_{\Delta\Delta}^{(2)}\right]=X_{0r}\\
&+\text{Re}\frac{i}{2\rho^3\gamma_t}\left(\frac{\delta z_{[2]}^{(1)}}{\delta z_{[2]}^{(1)\prime}(x)}W-\int_{iv}^x\ud y\,W(y)\right) \;,
\end{split}
\ee
where $X_{0r}$ is given by~\eqref{X0rlargeOmega}, i.e. it is independent of $\gamma_z$ and agrees with the $\Omega\gg1$ limit of the corresponding result for a purely time dependent field. Combining~\eqref{firsTermLargeOmegagammaz} and~\eqref{secondTermLargeOmegagammaz} gives
\be
\left(-\frac{\rho^3}{\lambda^2}+X_{0r}+X_z\right)\frac{\delta\Delta^2}{\rho^3}\;,
\ee
where 
\be
X_z=\text{Re}\frac{ia_0}{2\rho^3}\left(xW-\int_{iv}^x\ud y\, W(y)\right) \;,
\ee
where $a_0=1/\gamma_t$. From the symmetry of the field, $f(iu)=i\tilde{f}(u)$, we find
\be
X_z=-\frac{a_0}{2\rho^3}\int_0^v\ud u\,W(iu) \;.
\ee
Using~\eqref{deltaz21prime} we finally find
\be
X_z=\frac{2\gamma_z^2a_0^5}{\rho^3}\int_0^v\ud u\,u\int_u^v\ud w\,\tilde{f}(w)[\tilde{f}(w)-\tilde{f}(u)] \;.
\ee
Both $X_{0r}$ and $X_z$ are positive. $X_z$ depends on the spatial width of the field via the quadratic factor $\gamma_z^2$. Thus, in this $\Omega\gg1$ limit where $\lambda=\mathcal{O}(\Omega^{3/2})$, the width, $d_\Delta$, depends on two different adiabaticity parameters, $\gamma_t=1/a_0$ and $\gamma_z$. This is therefore something that one cannot obtain if one starts with Volkov solutions for plane waves.

\subsection{$\det\Lambda$}

To obtain the prefactor we also need to find $D''/2$ in~\eqref{ddD}. We write $\bar{\phi}^{(1)}$ and $\bar{\phi}^{(2)}$ as sums of $\delta q_{[j]}$ as in~\eqref{qjsum}. For the coefficients for $\bar{\phi}^{(1)}$ we find
\be
\begin{split}
a_1&=1+\frac{a_1^{(1)}}{\rho^2}+\mathcal{O}(\rho^{-4}) \\
a_2&=\frac{a_2^{(1)}}{\rho}+\mathcal{O}(\rho^{-3})
\qquad
a_3=\frac{a_3^{(1)}}{\rho}+\mathcal{O}(\rho^{-3}) \;,
\end{split}
\ee
and $a_4=i\rho a_3$, which follows from $\nu(u<0)=0$. This together with the expansions of $\delta z_{[j]}$ gives to leading order
\be
\begin{split}
\bar{\phi}^{(1)}_1&=1+\frac{1}{\rho^2}[a_1^{(1)}+\delta t_{[1]}^{(1)}+a_3^{(1)}(\phi+\delta t_{[4]}^{(1)})]+\mathcal{O}(\rho^{-4}) \\
\bar{\phi}^{(1)}_2&=\frac{1}{\rho}[a_2^{(1)}+i\phi a_3^{(1)}+\delta z_{[1]}^{(1)}(\phi)]+\mathcal{O}(\rho^{-3}) \;.
\end{split}
\ee
Demanding $\bar{\phi}^{(1)\prime}_2\to0$ as $u\to-\infty$ gives, using~\eqref{deltaz10}, $a_3^{(1)}=-iE(\gamma_t\tilde{t}_0)$. We can also obtain $a_2^{(1)}$ by demanding $\bar{\phi}^{(1)}_2\to0$, but we do not need it. Since $\delta z_{[1]}'$ is symmetric, we have $\bar{\phi}^{(1)\prime}_2\to0$ is negligible at $u\to+\infty$.
From~\eqref{deltat11} and~\eqref{ddeltat41} we have
\be
\delta t_{[1]}^{(1)\prime}(\phi\to\infty)=-\frac{E(\gamma_t\tilde{t}_0)}{\gamma_t}
\ee
and
\be
\delta t_{[4]}^{(1)\prime}(\phi\to\infty)=\frac{1}{\gamma_t}-i \;.
\ee
With $F'(u)\approx\rho F'(\phi)$ we find 
\be\label{dphi11PW}
\bar{\phi}^{(1)\prime}_1(u\to\infty)\approx-\frac{2iE(\gamma_t\tilde{t}_0)}{\rho} \;.
\ee

Denoting the coefficients in~\eqref{qjsum} as $a\to b$ for $\bar{\phi}^{(2)}$ we find
\be
b_1=\mathcal{O}(\rho^{-3})
\qquad
b_2\approx1+\frac{b_2^{(1)}}{\rho^2}
\qquad
b_3\approx\frac{b_3^{(1)}}{\rho^2}
\ee
and again $b_4=i\rho b_3$. We find $\bar{\phi}^{(2)}_1=\mathcal{O}(\rho^3)$, so this component is negligible. For the other we have
\be
\bar{\phi}^{(2)}_2\approx1+\frac{1}{\rho^2}[b_2^{(1)}+i\phi b_3^{(1)}+\delta z_{[2]}^{(1)}(\phi)] \;.
\ee
Demanding $\bar{\phi}^{(2)\prime}_2\to0$ as $u\to-\infty$ and noting that $\delta z_{[2]}'$ is antisymmetric gives $b_3^{(1)}=-i\delta z_{[2]}^{(1)\prime}(\phi\to\infty)$, so
\be\label{dphi22PW}
\bar{\phi}^{(2)\prime}_2(u\to\infty)\approx\frac{2}{\rho}\delta z_{[2]}^{(1)\prime}(\phi\to\infty) \;.
\ee

Plugging~\eqref{dphi11PW} and~\eqref{dphi22PW} into~\eqref{ddD} gives
\be
\frac{1}{2}D''\approx-\frac{4i}{\rho^2}E(\gamma_t\tilde{t}_0)\delta z_{[2]}^{(1)\prime}(\infty) \;.
\ee
Combining this with~\eqref{XDeltaDeltaPW} gives
\be
X_{\Delta\Delta}\frac{1}{2}D''\approx-\frac{4E}{\Omega}\tilde{f}'(v) \;.
\ee

Now we have everything we need to calculate the prefactor in~\eqref{generalMueller}. With $k_0=\Omega$, $p_0=p'_0\approx m_\LCperp\approx\Omega/2$ we find
\be
{\bf M}\approx\frac{\alpha\{3,0,0,-1\}}{2\pi^{3/2}E\tilde{f}'(v)\Omega^2\lambda}\left|\frac{\beta\hat{X}}{1+\beta\hat{X}}\right|e^{...} \;,
\ee
where $\hat{X}$ and $\beta$ are given by~\eqref{XhatDef} and~\eqref{betaDef}. Next we assume that $\Omega$ is larger than everything else, in particular $\beta\gg1$. Performing the $\Delta$ integral using~\eqref{dDeltaBetaLarge} then just gives a factor of $\lambda\sqrt{\pi}$, and we find
\be
{\bf M}(p_1,p_2,p_3)=\frac{\alpha\{3,0,0,-1\}}{2\pi\tilde{f}'(v)E\Omega^2}e^{...} \;.
\ee
This agrees with the plane-wave result in~\eqref{MuellerVec} (taking into account an extra factor of $\Omega$ coming from the scaling of the momentum variable, $\ud s\approx\ud p_1/\Omega$).

\subsection{Higher orders numerically}

Next we consider higher-order corrections in $1/\Omega\ll1$. For a simple field, such as a Sauter pulse, one may be able to calculate at least some of the coefficients as explicit analytical functions of $\gamma_t$ and $\gamma_z$. However, the calculation might become tedious and the results will probably be long and not particularly illuminating. A faster approach is to choose some particular values of $\gamma_t$ and $\gamma_z$ and then compute the coefficients numerically. The first thing to do is to figure out what structure the expansions of the instanton and $\delta q_{[j]}$ have. We find that the expansions have either only even or only odd powers of $1/\rho$, see~\eqref{tzPWser}, \eqref{delta1ser}, \eqref{delta2ser}, \eqref{delta3ser} and~\eqref{delta4ser}. Next we use Mathematica to analytically expand the equations of motion in power series in $1/\rho^2$. 

Consider first the instanton. For the zeroth order we just take the analytical solution in~\eqref{PWsolLead}. The equation for the next order can be written as
\be\label{ddz1f1f2}
z_{(1)}''=f_1(\phi)+f_2(\phi)\tilde{t}_1 \;,
\ee
where $f_1(\phi)$ and $f_2(\phi)$ are obtained from the zeroth order, which we have already obtained. Integrating the imaginary part of~\eqref{ddz1f1f2} from $\phi=0$ to $\phi=\infty$ and demanding $\text{Im }z_{(1)}'(\infty)=0$ gives
\be
\tilde{t}_1=-i\frac{\int_0^\infty\ud\phi\,\text{Im }f_1(\phi)}{\int_0^\infty\ud\phi\,\text{Re }f_2(\phi)} \;.
\ee
Next we can solve~\eqref{ddz1f1f2} numerically with initial conditions $z_{(1)}(0)=z_{(1)}'(0)=0$. Now that we have $\tilde{t}_0$, $z_{(0)}$, $\tilde{t}_1$ and $z_{(1)}$, the equation for $\tilde{t}_2$ and $z_{(2)}$ is on the form~\eqref{ddz1f1f2}, just with $z_{(1)}\to z_{(2)}$, $\tilde{t}_1\to\tilde{t}_2$ and with different $f_1$ and $f_2$. We can therefore repeat the same steps to obtain $\tilde{t}_2$ and $z_{(2)}$, then $\tilde{t}_3$ and $z_{(3)}$ and so on. Note that we do not need to use the shooting method to obtain this expansion; we only have to solve differential equations with simple initial conditions, so this is numerically fast.   

After we have obtained the expansion of the instanton in~\eqref{tzPWser}, we can now expand the equations in~\eqref{dtdzeq}. The basis solutions in~\eqref{qjsum} already have simple initial conditions, and there are no additional unknown constants to determine. 

\begin{figure}
    \centering
    \includegraphics[width=\linewidth]{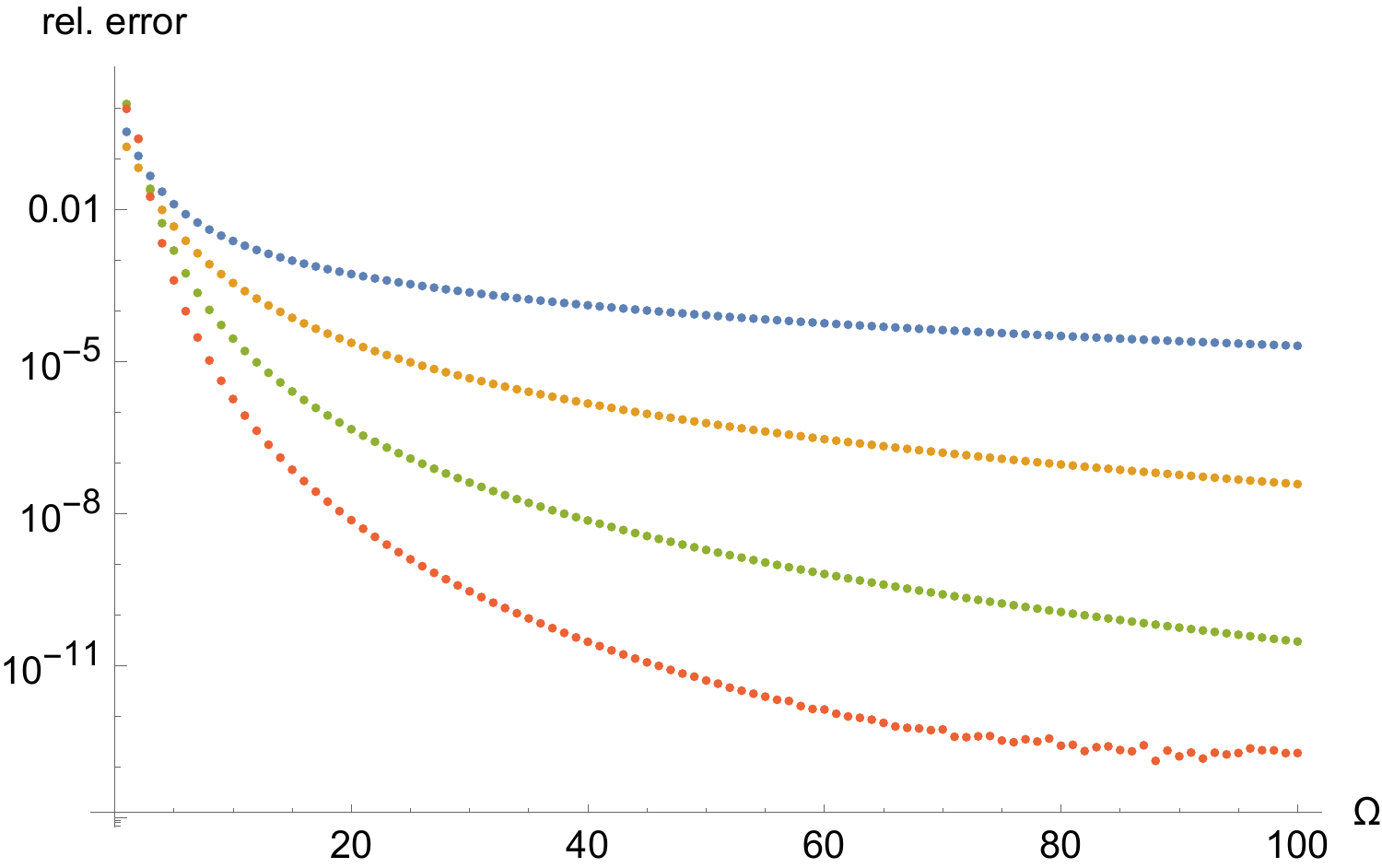}
    \caption{The relative error $|(d_P^{-2})_\text{exact}/(d_P^{-2})_\text{approx.}-1|$ for $\gamma_t=2$ and $\gamma_z=1$, where the ``exact'' result is obtained without making any expansion, and the approximation is obtained by including the first 1, 2, 3 or 4 terms in the expansion in $1/\Omega^2$.}
    \label{fig:dPm2PWrelerror}
\end{figure}

Fig.~\ref{fig:dPm2PWrelerror} shows the relative error of the first four orders of the expansion of $d_P^{-2}$. When we include higher orders and for larger $\Omega$, the relative error becomes very small. To be able to see that each order reduces the error in this regime, one needs to increase the number of digits used in the numerical computation. We have done this using Mathematica by increasing ``WorkingPrecision'' to 30 digits. When including the fourth order we can see in Fig.~\ref{fig:dPm2PWrelerror} that the relative error becomes noisy after $\Omega\gtrsim60$. This is a sign that the error starts to become dominated by the numerical precision rather than the precision of the expansion. When this happens the relative error is $\mathcal{O}(10^{-12})$, which is of course much smaller than what one would need for comparing with experiments, and also much smaller than the corrections to the leading order in the weak field expansion. Our main reason for considering such high precision is to confirm that the expansions of $q_\mu$ and $\delta q_\mu$ indeed have the structure in~\eqref{tzPWser}, \eqref{delta1ser}, \eqref{delta2ser}, \eqref{delta3ser} and~\eqref{delta4ser}, and that the resulting expansions of components of $X$ only have either even or odd powers of $1/\Omega$. This is especially important since we have found that some of the components of $X$ cancel to leading order after performing the $k$ integral.            

\section{Approximation for $\Omega\sim a_0\gg1$}

In this section we consider $\Omega\sim a_0\gg1$. We define
\be
\mu=\gamma p_1=\frac{\Omega}{2a_0}
\ee
and consider $\mu=\mathcal{O}(1)$. We can expand the instanton as
\be
q^\mu(u)=\frac{1}{\gamma}\sum_{n=0}^\infty\gamma^n q_{(n)}^\mu(u) \;,
\ee
and
\be
\delta q^\mu(u)=\sum_{n=0}^\infty\gamma^n \delta q_{(n)}^\mu(u) \;.
\ee
We do not rescale $u$.
To obtain the leading order for $d_P$, $d_\Delta$ and $d_1$ we need to include the first 4 orders of $q^\mu(u)$ and $\delta q^\mu(u)$, i.e. up to $n=3$. 
The analytical expansion of the equations of motion up to the fourth order is done with Mathematica, but is long and not illuminating. With $E_3(t,z)=E F(\gamma t,\gamma z)$ we have to zeroth order
\be\label{q0eq}
t_{(0)}''=F(t_{(0)},z_{(0)})z_{(0)}'
\quad
z_{(0)}''=F(t_{(0)},z_{(0)})t_{(0)}'
\ee
with initial conditions $t_{(0)}(0)=z_{(0)}(0)=z'_{(0)}(0)=0$ and $t'_{(0)}(0)=\mu$. For $\mu=\mathcal{O}(1)$, \eqref{q0eq} cannot be approximated further and so we need to solve it numerically. But we note that $q_{(0)}(u)$ is real if we choose a real $u$ contour, which we assume in the following. The initial conditions for the next order are $t_{(1)}(0)=z_{(1)}'(0)=i$ and $z_{(1)}(0)=t_{(1)}'(0)=0$, so $q_{(1)}(u)$ is imaginary. For $q_{(2)}$ we have $t_{(2)}(0)=z_{(2)}(0)=t_{(2)}'(0)=z_{(2)}'(0)=0$, so $q_{(2)}(u)$ is real. For $q_{(3)}$ we have one nontrivial initial condition. We can write $q_{(3)}(u)=q_{(3p)}(u)+cq_{(3h)}(u)$ as a sum of a particular and a homogeneous solution, where $q_{(3p)}(u)$ is a solution to the same equation as $q_{(3)}(u)$ but with trivial initial conditions $t_{(3p)}(0)=z_{(3p)}(0)=t_{(3p)}'(0)=z_{(3p)}'(0)=0$, while $q_{(3h)}$ is a solution to the same equation as $q_{(1)}$ but with $z_{(3h)}(0)=t_{(3h)}'(0)=z_{(3h)}'(0)=0$ and $t_{(3h)}(0)=i$. The constant $c$ is determined by demanding that $\text{Im}[z'(\infty)]=0$, i.e. $c=-z_{(3p)}'(\infty)/z_{(3h)}'(\infty)$.
Note that all the initial conditions are already determined, i.e. we do not need the shooting method here, which means it is much faster to obtain $q_{(n)}(u)$ compared to $q(u)$.

For $\delta q_{[j](0)}$ we have the same initial conditions as $\delta q_{[j]]}$, while all the higher orders have $\delta q_{(n)}(0)=\delta q_{(n)}'(0)=0$.

\begin{figure}
    \centering
    \includegraphics[width=\linewidth]{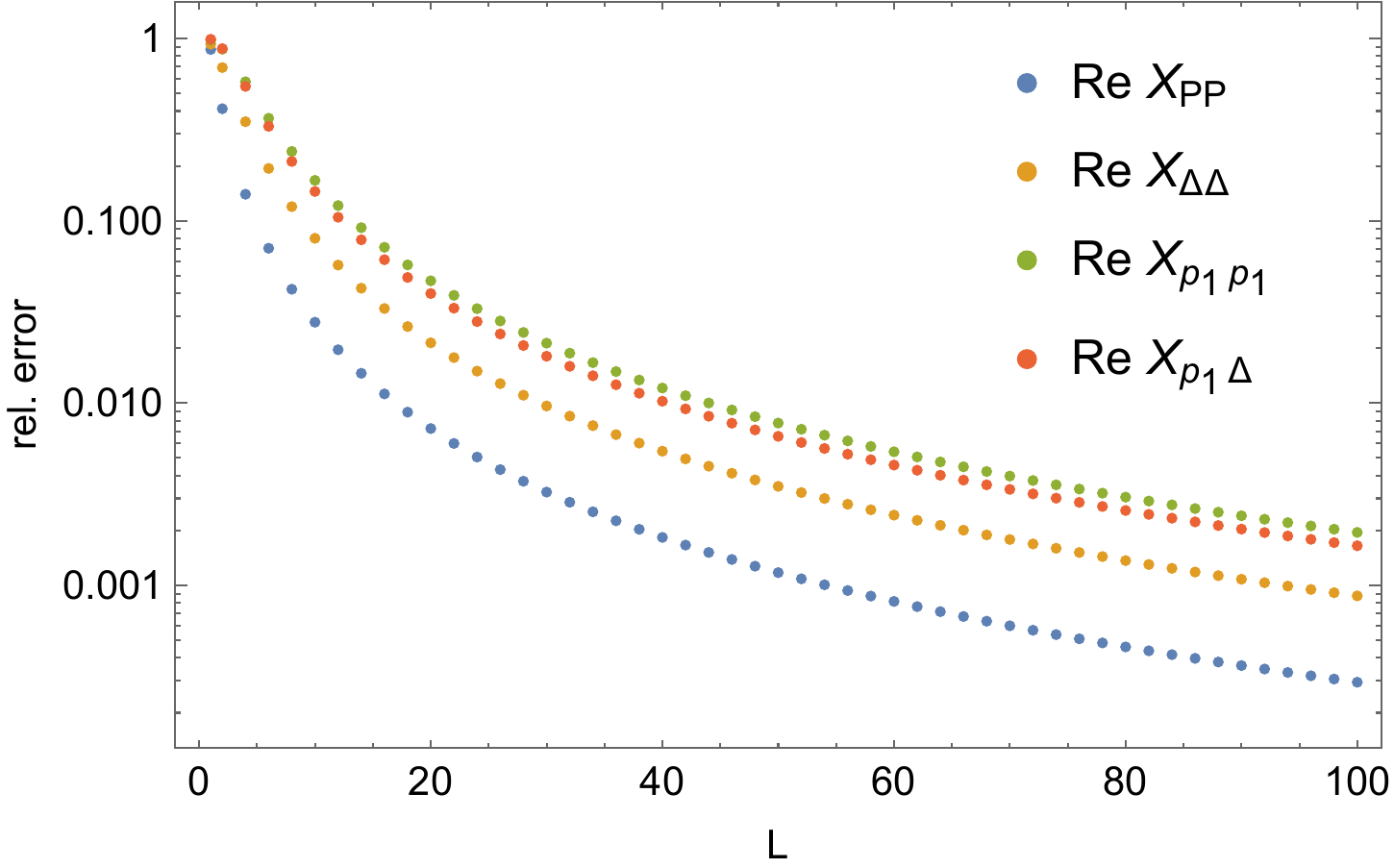}
    \caption{Relative errors $|\text{Re }X_{\text{exact}}/\text{Re }X_{\text{approx}}-1|$ as functions of $L = a_0 = 2\Omega$ to leading order.}
    \label{fig:widthsL}
\end{figure}

We find
\be
X_{kk}\approx ic_{kk}
\qquad
X_{k\Delta}\approx-ic_{k\Delta}
\qquad
X_{kp_1}\approx-ic_{kp_1}
\ee
and
\be
\begin{split}
\text{Re }X_{p_1\Delta}&\approx-c_{p_1\Delta}\gamma^3
\qquad
\text{Re }X_{\Delta\Delta}\approx-c_{\Delta\Delta}\gamma^3 \\
\text{Re }X_{p_1p_1}&\approx-c_{p_1p_1}\gamma^3
\qquad
\text{Re }X_{PP}\approx-c_{PP}\gamma^3 \;,
\end{split}
\ee
where all the constants are real and positive.
For $\mu=1/4$ we have
\be
c_{kk}\approx0.714576
\qquad
c_{k\Delta}\approx1.2035
\qquad
c_{kp_1}\approx0.731135 
\ee
and
\be
\begin{split}
c_{PP}&\approx15.1285
\qquad c_{\Delta\Delta}\approx7.72487 \\
c_{p_1p_1}&\approx88.1843
\qquad c_{p_1\Delta}\approx4.69291 \;.
\end{split}
\ee
Performing the $\delta k$ integral and squaring the amplitude as in~\eqref{widthsDelta1} gives
\be
\begin{split}
&-\gamma^3(c_{PP}\delta P^2+c_{p_1p_1}\delta p_1^2+c_{\Delta\Delta}\delta\Delta^2+2c_{p_1\Delta}\delta p_1\delta\Delta) \\
&-\frac{\lambda^2}{1+c_{kk}^2\lambda^4}(c_{kp_1}\delta p_1+c_{k\Delta}\delta\Delta)^2 \;.
\end{split}
\ee
Note that the cross term $\delta p_1\delta\Delta$ is on the same order of magnitude as the diagonal terms. Thus, this limit is different from the $\Omega\gg1$ limit in the previous section.
A comparison between the ``exact'' (i.e. numerical) and leading order expansions of the widths is shown in Fig.~\ref{fig:widthsL}.


For the exponent~\eqref{expPsiStart}, we find to leading order $e^{-c\gamma}$, and for $\mu=1/4$ we have $c=2\times 8/3$, which is actually what one would expect from the LCF approximation of the plane-wave approximation, i.e. from $e^{-8/(3\chi)}$.
The same number of orders of $q^\mu(u)$ and $\delta q^\mu(u)$ allows us to obtain both the leading and next-to-leading orders for $d_2$,
\be
d_2^{-2}\approx\frac{a}{a_0}+\frac{b}{a_0^3} \;,
\ee
where, for $\mu=1/4$, we find $a=8$ and $b=-8*17/3$. 
And the final ingredient to obtain the full prefactor is the contribution from the path integral,
\be
\frac{1}{2}D''\approx\frac{b_1}{a_0}+\frac{b_3}{a_0^3} \;,
\ee
where $b_1\approx0.381936i$ and $b_3\approx0.66254i$ for $\mu=1/4$.

\section{Perturbative pair production}

While our main focus is on pair production in regimes where one cannot treat the field in perturbation theory, in this section we will calculate the exponent and the widths by treating the field to first order in perturbation theory. We focus on fields with poles on the imaginary axis, such as a Sauter or a Lorentzian pulse. For a pulse shape, $f(\omega t)$, with a pole at $\omega t=i\nu$, where $\nu$ is a constant ($\nu=\pi/2$ for a Sauter pulse), the Fourier transform has an exponential scaling,
\be
f_F(K)\approx\dots\exp\left\{-\frac{\nu}{\omega}|K|\right\} \;,
\ee
where the ellipses denote pre-exponential terms which we will not consider, and where $K$ is the Fourier frequency. We consider fields which are given by a product, $E(t,z)=F_t(t)F_z(z)$, of two pulse shapes, $F_t$ and $F_z$, which both have poles on the imaginary axis. After expressing $E(t,z)$ in terms of its Fourier transform, the space-time integrals give delta functions implying
\be\label{momentumConservation}
k_\mu+K_\mu=p_\mu+p'_\mu \;,
\ee
where $k$, $p$ and $p'$ are again the momenta of the photon, the electron and the positron, and $K$ is the Fourier momentum of the field. The (real) exponential terms come from the wave packet of the photon and the Fourier transform.  On the amplitude level we have
\be\label{perturbativeExpStart}
\exp\left\{-\frac{k_3^2}{2\lambda^2}-\nu_t|K_0|-\nu_z|K_3|\right\} \;,
\ee
where $\nu_t=\frac{\nu}{\omega_t}$, $\nu_z=\frac{\nu}{\omega_z}$, and from~\eqref{momentumConservation}
\be
K_0=p_0+p'_0-k_0
\qquad
K_3=\Delta-k_3 \;.
\ee
We will consider the $z$-independent and $t$-independent limits separately in the next two subsections.

\subsection{$\omega_z\to0$}\label{PerturbativeTimeSec}

Consider the limit where the spatial pulse shape becomes very wide, i.e. $\omega_z\to0$. As $\nu_z\to\infty$ in~\eqref{perturbativeExpStart}, the dominant contribution to the $k_3$ integral comes from $k_3=\Delta$. In other words, the spatial Fourier transform essentially becomes a delta function, which we use to perform the $k_3$ integral. The remaining momentum integral, on the probability level, can be performed using the saddle-point method. We have a saddle point at $p_2=0$, $p_1=p'_1=\rho_1:=\Omega/2=k_1/2$, $P=0$ and $\Delta=2\rho_3$, where $\rho_3$ is either zero or a nontrivial function, depending on whether we are below or above the critical point. 

From the zeroth order in the expansion around the saddle point we find,
\be
\mathbb{P}\sim\exp\left\{-\frac{4\rho_3^2}{\lambda^2}-4\nu_t\left[\sqrt{1+\rho_1^2+\rho_3^2}-\sqrt{\rho_1^2+\rho_3^2}\right]\right\} \;.
\ee

From the first order we find an equation that determines the saddle point for $p_3=p'_3=\rho_3$,
\be\label{firstOrderPer}
\rho_3\left[\frac{1}{\lambda^2}-\frac{\nu_t}{2}\left(\frac{1}{\sqrt{\rho_1^2+\rho_3^2}}-\frac{1}{\sqrt{1+\rho_1^2+\rho_3^2}}\right)\right]=0 \;.
\ee
We always have the trivial solution, $\rho_3=0$. Since the round brackets is a positive and monotonic function, we can also have second, nonzero solution if $\nu\lambda^2$ is larger than some critical value. Just above the the critical point, $\rho_3$ will be nonzero but small, so by setting $\rho_3\to0$ in the square brackets of~\eqref{firstOrderPer} we find and equation for $\lambda_c$,
\be\label{lambdacPerTime}
\frac{1}{\lambda_c^2}=\frac{\nu_t}{2}\left(\frac{1}{\rho_1}-\frac{1}{\sqrt{1+\rho_1^2}}\right) \;.
\ee
This agrees with~\eqref{lambdacX0r} and~\eqref{X0rHighomega}, which we obtained for the $\gamma_t\gg1$ limit of the instanton result.

The nonzero solution of~\eqref{firstOrderPer} can be found analytically as the root of a fourth-order polynomial in $x=\sqrt{\rho_1^2+\rho_3^2}$, but it is a complicated expression. But for $y=(2\nu_t\lambda^2)^{2/3}\gg1$ we can expand it as
\be\label{yexpansion}
x=\frac{\sqrt{y}}{2}\sum_{n=0}^\infty\frac{a_n}{y^n} \;,
\ee
where the first couple of coefficients are $a_0=1$, $a_1=-1$ and $a_2=a_3=1/3$.

From the second order we find the widths,
\be
\begin{split}
\mathbb{P}\propto\exp\Big\{&-d_2^{-2}\delta p_2^2-d_\Delta^{-2}\delta\Delta^2\\
&-\{\delta P,\delta p_1\}\cdot{\bf d}_{1P}^{-2}\cdot\{\delta P,\delta p_1\}\Big\} \;,
\end{split}
\ee
where
\be
d_2^{-2}=\frac{2\nu_t}{\sqrt{1+\rho_1^2+\rho_3^2}} \;,
\ee
\be
d_\Delta^{-2}=\frac{1}{\lambda^2}+\frac{\nu_t}{2}\left[\frac{1+\rho_1^2}{(1+\rho_1^2+\rho_3^2)^{3/2}}-\frac{\rho_1^2}{(\rho_1^2+\rho_3^2)^{3/2}}\right]
\ee
and
\be
{\bf d}_{1P}^{-2}=\frac{2\nu_t}{(1+\rho_1^2+\rho_3^2)^{3/2}}
\begin{pmatrix}
1+\rho_1^2 & \rho_1\rho_3 \\
\rho_1\rho_3 & 1+\rho_3^2 
\end{pmatrix} \;.
\ee
Below the critical point, $\rho_3=0$, the off-diagonal elements vanish and we find agreement with~\eqref{highomegad12P}, \eqref{dDeltaX0r} and~\eqref{X0rHighomega}.
Above the critical point we use~\eqref{firstOrderPer} to rewrite $d_\Delta$ as
\be
d_\Delta^{-2}=\frac{\nu_t\rho_3^2}{2}\left[\frac{1}{(\rho_1^2+\rho_3^2)^{3/2}}-\frac{1}{(1+\rho_1^2+\rho_3^2)^{3/2}}\right] \;,
\ee
from which we see that the nonzero solution of~\eqref{firstOrderPer} is indeed a maximum of the spectrum, in contrast to the $\rho_3=0$ point which becomes a minimum.

\subsection{$\omega_t\to0$}

Next we consider the limit where the temporal pulse shape becomes very wide, i.e. $\omega_t\to0$. As $\nu_t\to\infty$ in~\eqref{perturbativeExpStart}, the dominant contribution to the $k_3$ integral comes from the region around the point where $K_0=0$. We focus on the dependence on $\Delta$ and set $p_2=0$, $p_1=\Omega/2$ and $P=0$ right from the start. The solution to $K_0=0$ is
\be
k_3=\pm2\sqrt{1+\left(\frac{\Delta}{2}\right)^2} \;,
\ee
independent of $\Omega$. We will focus on the $k_3>0$ solution, which will lead to a saddle point for $\Delta$ which is also positive. The only thing that changes for the $k_3<0$ solution is that the sign of $\Delta_s$ changes. We find
\be\label{EzPerExp}
\mathbb{P}\propto\exp\left\{-\frac{4}{\lambda^2}(1+\rho_3^2)-4\nu_z(\sqrt{1+\rho_3^2}-\rho_3)\right\} \;,
\ee
independent of $\Omega=2\rho_1$.
The saddle point for $\rho_3$ is never zero in this case. We can write the saddle point equation as
\be\label{rho3eqSpatial}
\frac{1}{\rho_3}-\frac{1}{\sqrt{1+\rho_3^2}}=\frac{2}{\nu_z\lambda^2} \;.
\ee
We would obtain the same equation if we put $\rho_1\to0$ in the square brackets in~\eqref{firstOrderPer} and replace $\nu_t\to\nu_z$. Thus, we can in some sense view the $\omega_t\to0$ case as equivalent to what we found for $\omega_z\to0$ but by setting $\Omega\to0$. From~\eqref{lambdacPerTime} we see that that this is consistent with the absence of a critical point, or $\lambda_c\to\infty$. We can also reuse the expansion in~\eqref{yexpansion}, by replacing $x\to\rho_3$ and $y\to(2\nu_z\lambda^2)^{2/3}$. 

We can also expand the solution of~\eqref{rho3eqSpatial} in the opposite limit as
\be
\rho_3=r\sum_{n=0}^\infty a_nr^n \;,
\ee
where $r=\nu_z\lambda^2/2$ and the first couple of coefficients are $a_0=1$, $a_1=-1$, $a_2=1$ and $a_3=-1/2$. With this expansion we find to leading order
\be
\mathbb{P}\propto\exp\left\{-\frac{4}{\lambda^2}-4\nu_z\right\} \;.
\ee
Smaller $\nu_z$ means larger $\mathbb{P}$. But smaller $\nu_z$ corresponds to larger $\omega_z$, so in this limit the probability increases if we make the field narrower. For readers who have experience with Schwinger pair production in spatially inhomogeneous fields, this might at first seem counterintuitive, because, in a Schwinger/tunneling regime, a shorter pulse would mean a shorter distance to accelerate the virtual particles, which would make it more difficult for virtual particles to gain enough energy to become real. However, in the perturbative regime, one could have expected larger $\omega_z$ to increase the probability, because of the following argument. We have a field with off-shell Fourier momentum, $K_\mu=\delta _{\mu}^3 K_3$. Such a field could not produce pairs alone at any finite order of perturbation theory. A single on-shell photon can of course not produce any pairs either. But we can produce a pair by absorbing one on-shell photon (with momentum $k_\mu$) and one off-shell Fourier photon from the field. And this is only possible if we absorb a Fourier photon with a nonzero $K_3$ which is large enough to compensate and partially cancel the $k_3$ component of the on-shell photon. Indeed, note that $k_3\gtrsim2$ for $\Delta\gtrsim0$, which means $K_3\sim-k_3\sim-2$, i.e. we need to absorb a Fourier photon with a momentum that is much larger than the momentum of the produced pair. Doing so allows us to absorb larger $k_3$, which means more energy, while still keeping the pair relatively light, which is good because it is easier to produce pairs which have less momentum. Thus, we want to absorb significant momentum from the field, and this becomes easier (less exponentially suppressed) when $\omega_z$ increases, since that makes the Fourier transform wider.

In a recent paper~\cite{Jiang:2023hbo} it was found that the probability of pair production can increase when adding a spatially dependent field to a time-dependent field, $E_t(t)+E_z(z)$, in a perturbative regime and for monochromatic fields. It was explained there that the addition of $E_z(z)$ opens new pair-production channels. We consider instead products of fields, $E_t(t)E_z(z)$, and where both $E_t(t)$ and $E_z(z)$ describe a single pulse without oscillations (and the maximum field strength is constant). However, in the perturbative regime, it seems we are seeing a somewhat similar effect; the spatial inhomogeneity makes additional perturbative production channels available or less suppressed and hence increases the total pair production probability.     

For the width we find
\be
d_\Delta^{-2}=\frac{1}{\lambda^2}+\frac{\nu_z}{2(1+\rho_3^2)^{3/2}} \;.
\ee
The fact that both terms are positive confirms the absence of a critical point.

\section{Conclusions}

In this paper we have shown how to use the worldline instanton formalism for nonlinear Breit-Wheeler pair production in fields that depend on time and one spatial coordinate. We have restricted ourselves to one spatial coordinate for simplicity. We believe that going from 1D fields that only depend on time to the 2D fields considered here is conceptually more nontrivial compared to the generalization from 2D to 4D fields. But already for 2D fields we have several additional nonzero widths terms compared to 1D, and for 4D fields there will be even more terms.   
In~\cite{DegliEsposti:2023qqu} we showed how to use similar methods to obtain the momentum spectrum for Schwinger pair production by e-dipole fields, which is a class of exact solutions to Maxwell's equations that depend on all four space-time coordinates. A natural project for the future is thus to use these methods for nonlinear Breit-Wheeler in such 4D fields.  

Another project for the future is to consider nonzero impact parameter, $b\ne0$, in more detail. As we have hinted in this paper, this gives the instantons a more complicated complex structure. It would also be interesting consider wave packets which are peaked at a nonzero $l_3$, in particular to see what happens with the transition from one spectrum peak (which involves $k_3=0$ for $l_3=0$) to two peaks (which involves $k_3\ne0$ even for $l_3=0$).  

We have found a nontrivial dependence on the width of the photon wave packet. It would be interesting to see how this generalizes when going to 4D fields. In~\cite{DegliEsposti:2021its} we showed how to use worldline instantons for the probability that an electron emits a hard photon in a 1D field. This, nonlinear Compton scattering, is in some sense similar to nonlinear Breit-Wheeler due to crossing symmetry. But it would be interesting to study the role of the electron wave packet, which can lead to qualitative differences both because the mass gives a different dispersion relation and because the electron is accelerated by the field.

\acknowledgements

G. T. is supported by the Swedish Research Council, Contract No. 2020-04327.

\appendix

\section{Wave packet in position space}

In position space we have
\be
f_\mu(x):=\int\frac{\ud^3 k}{(2\pi)^32k_0}\epsilon_\mu(k)f(k)e^{-ikx} \;.
\ee
The motivation for this definition is simply that this is what appears in the probability amplitude after calculating away the mode operators. We can perform the momentum integral analytically for small $\lambda$, assuming that $x^\mu$ is at most $\mathcal{O}(1/\lambda)$. We first change integration variable from
\be
k_j=l_j+\lambda_j\left[\sqrt{2}\kappa_j+i\left(b^j-x^j-\frac{l_j}{l_0}t\right)\lambda_j\right]
\ee
to $\kappa$. We can replace $k_j\to l_j$ in the prefactor. Performing the now Gaussian $\kappa_j$ integral gives (up to an irrelevant constant phase)
\be\label{wavePacketx}
\begin{split}
f_\mu(x)\approx&\sqrt{\frac{\lambda_1\lambda_2\lambda_3}{\pi^{3/2}2l_0}}\epsilon_\mu(l)\\
\times&\exp\left\{-ilx-\sum_{j=1}^3\frac{\lambda_j^2}{2}\left(x^j-b^j+t\frac{l_j}{l_0}\right)^2\right\} \;.
\end{split}
\ee
In the path integral calculation, we have $f_\mu(q(\sigma))$.

As a check we consider a time dependent field. We change variables from ${\bf x}_\LCp$ and ${\bf x}_\LCm$ to ${\bm\varphi}=({\bf x}_\LCp+{\bf x}_\LCm)/2$ and ${\bm\theta}={\bf x}_\LCp-{\bf x}_\LCm$, and shift the path integral variable as $q^j(\tau)\to\varphi^j+q^j(\tau)$ as in~\cite{DegliEsposti:2021its}. The instanton parts do not depend on $\varphi$. In~\cite{DegliEsposti:2021its} we performed the $\varphi$ integral first, and then it gives a momentum conserving delta function, which one can then use to perform the momentum integral ${\bf k}$ from the wave packet. Now, we have instead already performed the ${\bf k}$ integral, and the factor $f_\mu(q(\sigma))$ makes the $\varphi$ integral Gaussian. To remove the linear term in the exponent, we change variable from
\be
\varphi^j=\frac{\sqrt{2}\phi^j}{\lambda_j}+b^j-q^j(\sigma)-t\frac{l_j}{l_0}+\frac{i\Sigma_j}{\lambda_j}
\ee
to $\phi_j$, where 
\be
\Sigma_j=\frac{1}{\lambda_j}(p_j+p'_j-l_j) \;.
\ee
The $\phi$ dependent part of the exponent is now simply $e^{-{\bm\phi}^2}$. We thus find
\be
\begin{split}
\int&\ud^3{\bm\varphi}\; f_\mu e^{i\varphi^j(p_j+p'_j)}=\frac{2\pi^{3/4}}{\sqrt{l_0\lambda_1\lambda_2\lambda_3}}\epsilon_\mu\exp\left\{-ilq-\frac{{\bm\Sigma}^2}{2}\right\} \\
&\times\exp\left\{i\sum_{j=1}^3\lambda_j\left(b^j-q^j-t\frac{l_j}{l_0}\right)\Sigma_j\right\} \;.
\end{split}
\ee
We can neglect the exponent in the second row. Squaring the amplitude gives
\be\label{PtimeDelta}
\begin{split}
&\int\frac{\ud^3 p'}{(2\pi)^3}\frac{4\pi^{3/2}}{l_0\lambda_1\lambda_2\lambda_3}\exp\left\{-\sum_{j=1}^3\frac{(p_j+p'_j-l_j)^2}{\lambda_j^2}\right\}F({\bf p}')\\
&\approx\frac{1}{2l_0}F({\bf p}'={\bf l}-{\bf p}) \;.
\end{split}
\ee
This agrees with Eq.~(A5) in~\cite{DegliEsposti:2021its}, which was obtained without choosing a particular wave packet.
Thus, the widths for ${\bf p}+{\bf p}'$ are $\lambda_j$, same as the width of the wave packet $|f(k)|^2$.

\section{Exponent}\label{app:exponent}

We now compute the exponent at the amplitude level focusing on the contributions from the LSZ, i.e. $\psi$, without the wave packet. To obtain the exponent, we simply evaluate at all the saddle points, namely $\theta^\LCperp = T(p^\LCperp - \sigma k^\LCperp)$ or equivalently
\be
\dot q^\LCperp(1) = -Tp^\LCperp \, \qquad \dot q^\LCperp(0) = Tp'^\LCperp
\ee
for the transverse spatial variables, the Lorentz force equation for the path integral
\be\label{eq:LFE}
\Ddot{q}^\mu = T F^{\mu \nu} \dot q_\nu + Tk^\mu \, \delta(\tau-\sigma)
\ee
and
\be
T^2 = \int_0^1 \ud \tau \, \dot q^2 \qquad k \cdot \dot q(\sigma) = 0
\ee
for $T$ and $\sigma$ respectively, and finally
\be
\dot z(1) = -Tp_3 \, ,\qquad \dot z(0) = Tp'_3 \;,
\ee
which are the same as in~\cite{DegliEsposti:2021its, DegliEsposti:2022yqw}. 

From the Lorentz-force equation we further see that $\dot q^2 = \text{const}$, therefore we can drop the integral above and obtain $T^2 = \dot q^2$. Doing some integration by parts and using the Lorentz-force equation we get
\be\begin{split}\label{ExpSaddle}
    \int_0^1 \ud \tau \, \frac{\dot q^2}{T} + A \dot q &= \left(A + \frac{\dot q}{T} \right) q \biggr|_0^1 - \int_0^1 \ud \tau \, q \left(\frac{\Ddot q}{T} + \dot A \right) \\
&= px_\LCp +p'x_\LCm -\int_0^1 \ud \tau \, q^\mu \partial_\mu A_\nu \dot q^\nu \\
&-\int_0^1 \ud \tau \, Jq \;,
\end{split}
\ee
so the current term as well the free asymptotic states in~\eqref{LSZ3pair} cancel with the terms in~\eqref{ExpSaddle}. Changing variable from $\tau$ to $u = T(\tau - \sigma)$ we find that the final exponent is given by
\be\label{FinalExp}
\psi = i\int \ud u \, q^\mu \partial_\mu A_\nu \frac{\ud q^\nu}{\ud u} \; .
\ee

\begin{figure}
    \centering
    \includegraphics[width=\linewidth]{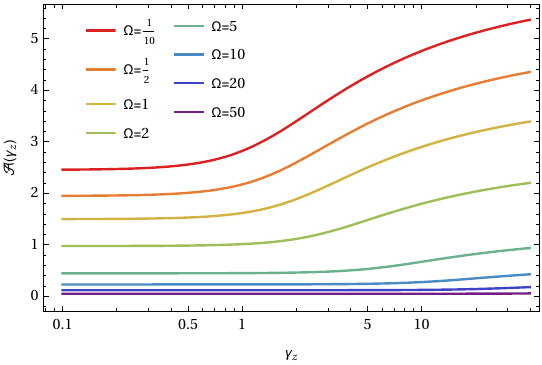}
    \caption{Exponent (without the overall factor of $1/E$) as a function of the size of the field $\gamma_z$ for different values of the photon energy $\Omega$ and $\gamma_t = 1$ in the subcritical regime.}
    \label{fig:exp}
\end{figure}

Since the saddle point value of the $T$-integral is given by (see~\eqref{bfXsaddle})
\be
T_s = \frac{t_\LCp}{p_0} + \frac{t_\LCm}{p'_0} \;,
\ee
i.e. $T\to\infty$ as $t_\LCpm\to\infty$,
we see that the domain of $u$ is from minus to plus infinity. Furthermore, the integrand~\eqref{FinalExp} is analytic up to poles of the field, branch points of the instantons, and the kink at $u = 0$, so we can change the contour of integration as long as we do not cross singular points and keep $u = 0$ fixed. 
\begin{figure*}
    \centering
    \includegraphics[width=.45\linewidth]{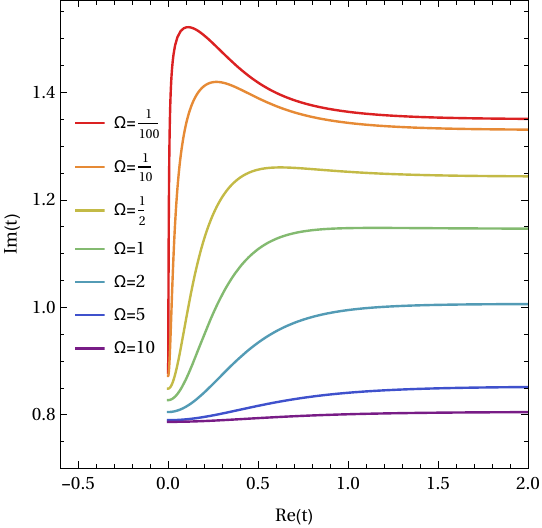}
    \hfill
    \includegraphics[width=.45\linewidth]{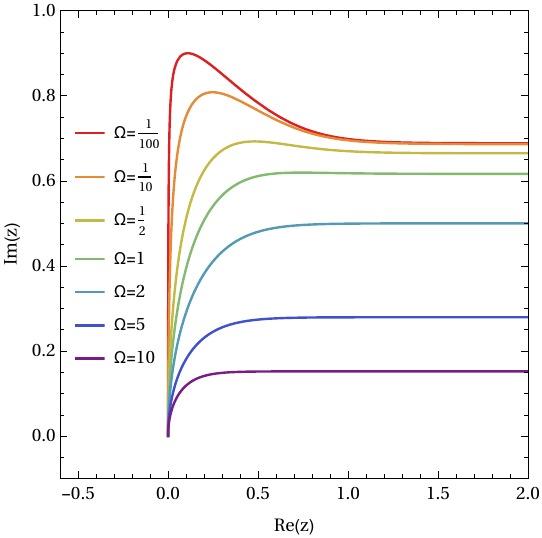}
    \caption{Time and space components of the instantons in the subcritical regime ($\lambda < \lambda_c$) for different photon energies and $\gamma_t = \gamma_z = 1$.}
    \label{fig:instantons}
\end{figure*}

When we make a change of contour, the argument of $\delta (\tau - \sigma) \sim \delta(u)$ becomes complex, so this might look puzzling. However, the delta function merely tells us that the components of the instantons have a kink at $u = 0$ and are therefore not differentiable. Hence, we can simply regard the delta function as effectively splitting the instanton into $q_{(-)}(u)$ when $u<0$ and $q_{(+)}(u)$ when $u>0$ with matching conditions $q_{(-)}(0) = q_{(+)}(0)$ and $q'_{(+)}(0^\LCp) - q'_{(-)}(0^\LCm) = k$. This means that we can make contour deformations in the two regions independently but keeping $u = 0$ fixed. 

We can also take the derivative of the exponent with respect to $\lambda$ and $b$ and take the real part, obtaining
\be\label{eq:DerBL}
\frac{\pa \psi_r}{\pa \lambda} = 2\Re \frac{k^2_s(\lambda,b)}{\lambda^3} \qquad \frac{\pa \psi_r}{\pa b} = -2\Im k_s(\lambda,b) \; .
\ee
From the first equation we see that the exponent is independent of $\lambda$ in the subcritical regime because $k_s = 0$, while in the the supercritical regime it increases so the probability is enhanced. Since the $k_3$-saddle points at $b \neq 0$ are of the form $\{k_s, -k_s^*\}$, from the second equation we see that the exponent is the same for the two peaks even at nonzero impact parameter. 

\section{Path integral}\label{app:PI}

In the $u$-domain, the region where the field is nonzero is some neighborhood of $u = 0$, while outside this neighborhood $E \approx 0$ implies that the instantons are straight lines. This observation allows us to extract the infinite parts arising from the asymptotic regions. 

We start with the path integral and use the Gelfand-Yaglom method~\cite{Dunne:2006st,Dunne:2006ur}. Let $u_0$ and $u_1$ be the initial and final points of the contour, which we eventually take to minus and plus infinity respectively. Expanding the exponent around the instantons $q \to q + \delta q$ up to second order, we obtain
\be\label{pathQuadLambda}
-\frac{i}{2}\int \ud u \, \begin{pmatrix}
	\delta t & \delta z 
\end{pmatrix}
\Lambda
\begin{pmatrix}
	\delta t \\ \delta z
\end{pmatrix} \; .
\ee
with quadratic matrix
\be
\Lambda = \begin{pmatrix} -\partial_u^2+A_{tt}z' & A_{tz}z'+A_t\partial_u \\ A_{tz}z' -\partial_u A_t & \partial_u^2-A_{tz}t' \end{pmatrix} \;,
\ee
with e.g. $A_t=\partial A_3/\partial t$ and similarly for the others. Then, the contribution to the prefactor of the amplitude coming from the path integral is given by
\be
\frac{1}{(2\pi T)^2} \frac{1}{\sqrt{\det \Lambda}} \; .
\ee
To obtain $\det \Lambda$, we find two solutions $\phi^{(1)}$ and $\phi^{(2)}$ of
\be
\Lambda \phi^{(i)} = 0
\ee
with initial conditions
\be
\phi^{(i)}_j(u_0) = 0 \, , \qquad \frac{\ud \phi^{(i)}_j}{\ud u}(u_0) = \frac{1}{T}\delta^i_j
\ee
and then simply evaluate them at the final point to obtain the determinant
\be\label{DetGY}
\det \Lambda = \bigr(\phi^{(1)}_1 \phi^{(2)}_2 - \phi^{(1)}_2 \phi^{(2)}_1\bigr)(u_1) \; .
\ee
However, in the asymptotic limit, the determinant~\eqref{DetGY} is proportional to some power of $t_\LCp$ and $t_\LCm$, therefore in order to evaluate~\eqref{DetGY} we would have to choose some finite (but large) $t_\LCp$ and $t_\LCm$, obtaining some large number which will then cancel against the contributions from other integrals. Therefore, it is much more convenient to extract the divergent asymptotic parts from all the integrals so that they cancel analytically, so that we can take the limit $t_\LCpm \to \infty$ before doing any numerical computations.

The following approach is similar to what we did in~\cite{DegliEsposti:2022yqw} for Schwinger pair production. It turns out that this is actually simpler for nonlinear Breit-Wheeler.
Let $\tilde u_0$ and $\tilde u_1$ be points along the contour, respectively before and after $u = 0$, such that $E \approx 0$ before $\tilde u_0$ and after $\tilde u_1$. Since the solutions are straight lines from $u_0$ to $\tilde u_0$, we have
\be
\phi^{(i)}_j(\tilde u_0) \simeq \frac{1}{T}(\tilde u_0 - u_0) \delta^i_j \; .
\ee
We can rewrite $\tilde u_0 - u_0$ in a more convenient way as follows
\be
\tilde u_0 - u_0 = \int_{u_0}^{\tilde u_0} \ud u = \int_{t_\LCm}^{\tilde t_\LCm}  \frac{\ud t}{t'} \simeq \frac{t_\LCm}{p'_0}
\ee
where the last step follows from the fact that $t' \simeq -p'_0$ in that part of the contour. Similarly one sees that
\be
u_1 - \tilde u_1 \simeq \frac{t_\LCp}{p_0} \; .
\ee
It is now convenient to define rescaled solutions
\be
\bar \phi := \phi \, \frac{T p'_0}{t_\LCm}
\ee
with simple initial conditions now at $\tilde u_0$ given by
\be
\bar \phi^{(i)}_j (\tilde u_0) = \delta^i_j \, , \qquad \frac{\ud \bar \phi^{(i)}_j}{\ud u} 0 \;,
\ee
where for the derivative we have simply taken $T \to \infty$.

Since the $\bar \phi$ solutions are straight lines after $\tilde u_1$, the determinant scales quadratically with $u$
\be
D(u) := \left( \bar \phi^{(1)}_1 \, \bar \phi^{(2)}_2 - \bar \phi^{(1)}_2 \, \bar \phi^{(2)}_1 \right) (u) \simeq \frac{1}{2} D''(\tilde u_1) (u - \tilde u_1)^2
\ee
where
\be\label{ddD}
\frac{1}{2} D''(u) = \left(\bar \phi^{\prime(1)}_1 \, \bar \phi^{\prime(2)}_2 - \bar \phi^{\prime(1)}_2 \, \bar \phi^{\prime(2)}_1 \right) (u) \;.
\ee
Finally, using the relations above, we find
\be\label{DetGYFinal}
\begin{split}
    \det \Lambda &= \left(\frac{t_\LCm}{T p_0'}\right)^2 \frac{1}{2} D''(\tilde u_1) (u_1 - \tilde u_1)^2 \\
    &=\left(\frac{t\LCp t_\LCm}{T p_0 p_0'}\right)^2 \frac{1}{2} D''(\tilde u_1) \;,
\end{split}
\ee
so we have extracted all the asymptotic parts, and the only contribution to find numerically is the finite $\frac{1}{2} D''(\tilde u_1)$.

\section{Ordinary integrals}\label{app:OI}

As to the transverse integrals, we change variables as in~\eqref{varphithetadef}, redefine $q_\LCperp(\tau) \to q_\LCperp(\tau) + \varphi_\LCperp$ and obtain a delta function from the $\varphi_\LCperp$ integral, 
\be
\int\ud^2\varphi\, e^{\dots}=(2\pi)^2 \delta_\LCperp(p+p' - k) \;.
\ee
Next we make a shift in the path-integration variable $q_\LCperp(\tau)\to q_\LCperp(\tau)+\delta q_\LCperp(\tau)$, where afterwards $\delta q_\LCperp(\tau)$ is the new integration variable, while $q_\LCperp(\tau)$ is chosen such that there is no cross term between $\delta q$ and $q$, and $\delta q_\LCperp(0)=\delta q_\LCperp(1)=0$, which means $q(\tau)$ is the solution of the transverse/background-free components of the Lorentz-force equation,
\be
\ddot{q}_\LCperp(\sigma)=Tk_\LCperp\delta(\tau-\sigma) \;,
\ee
with initial conditions $-q_\LCperp(0)=q_\LCperp(1)=\theta_\LCperp/2$. Integrating once gives ($\theta(.)$ with an argument is the step function)
\be
\dot{q}_\LCperp=\theta_\LCperp+Tk_\LCperp[\theta(\tau-\sigma)-(1-\sigma)] \;.
\ee
Plugging this solution into the transverse terms in the exponent gives 
\be
\begin{split}
&i\left\{p_\LCperp x_\LCp^\LCperp+p'_\LCperp x_\LCm^\LCperp-k_\LCperp\dot{q}^\LCperp(\sigma)+\int_0^1\ud\tau\frac{\dot{q}_\LCperp^2}{2T}\right\} \\
&=i\left\{\frac{\theta_\LCperp^2}{2T}+(\sigma k_\LCperp-p_\LCperp)\theta_\LCperp-\frac{T}{2}k_\LCperp^2\sigma(1-\sigma)\right\} \;.
\end{split}
\ee
Performing the $\theta$ integral gives, up to an irrelevant phase,
\be
\int\ud^2\theta_\LCperp\, e^{\dots}=2\pi T\exp\left\{-\frac{iT}{2}[\sigma p_\LCperp^{\prime2}+(1-\sigma)p_\LCperp^2]\right\} \;.
\ee

The integrals over the other ordinary variables, $\bo X := (T,\sigma, z_\LCpm)$, contain divergent contributions which we need to separate analytically, similar to what we did in the previous section for $\det\Lambda$. After the previous integrals have been performed, but before we have performed the integrals over $\bo X$, the exponent is evaluated at the instantons for generic rather than the saddle-point values of $\bo X$, i.e. the instantons at this stage are not yet the ones we focus on in the main part of this paper. But since we can only find the instantons numerically, we do not know their analytical dependence on $\bo X$. However, when we take the first derivatives of the exponent with respect to $\bo X$, the total derivatives are equal to the partial derivatives, i.e. we can treat the previous integration variables as constant, because all the additional terms cancel due to the fact that the previous integration variables are evaluated at their saddle-point values.

We will first introduce two constants, $a_\LCpm$.
$q^{\prime2}$ is a constant of motion, except at the point where the photon is absorbed. We define $a_\LCpm$ as
\be
\begin{split}
&\tau<\sigma: \quad t^{\prime2}-z^{\prime2}=a_\LCm^2 \\
&\tau>\sigma: \quad t^{\prime2}-z^{\prime2}=a_\LCp^2 \;.
\end{split}
\ee

Defining the exponential part as $e^\psi$ we find (cf.~\cite{DegliEsposti:2022yqw})
\be
\frac{\partial\psi}{\partial z_\LCm}=i[p'_3-z'(u_0)]
\qquad
\frac{\partial\psi}{\partial z_\LCp}=i[p_3+z'(u_1)] \;,
\ee
\be
\frac{\partial\psi}{\partial T}=\frac{i}{2}\left\{\sigma a_\LCm^2+(1-\sigma)a_\LCp^2-[1+\sigma p_\LCperp^{\prime2}+(1-\sigma)p_\LCperp^2]\right\}
\ee
and
\be
\frac{\partial\psi}{\partial\sigma}=-ik\dot{q}(\sigma) \;.
\ee
Since $\dot{q}$ is discontinuous at $\tau=\sigma$, it is not immediately obvious that the last derivative is well defined. To show that there is no problem, we note that the discontinuity is given by
\be\label{discqk}
q'_\mu(+0)-q'_\mu(-0)=k_\mu \;,
\ee
so $k\dot{q}(\sigma+0)-\dot{q}(\sigma-0)=Tk^2=0$, since the photon is on shell. Using~\eqref{discqk} again, we therefore have
\be
\begin{split}
k\dot{q}(\sigma)&=\frac{1}{2}k[\dot{q}(\sigma+0)+\dot{q}(\sigma-0)]=\frac{T}{2}[q^{\prime2}(+0)-q^{\prime2}(-0)] \\
&=\frac{T}{2}(a_\LCp^2-a_\LCm^2-p_\LCperp^2+p_\LCperp^{\prime2}) \;.
\end{split}
\ee

Now we have expressed $\partial\psi/\partial{\bf X}$ in terms of four unknowns, $z'(u_0)$, $z'(u_1)$ and $a_\LCpm$. In order to calculate the Hessian matrix we need to differentiate a second time, and for this we first need to express these four unknowns in terms of ${\bf X}$. As mentioned, at finite times this would be complicated, but it simplifies considerably in the asymptotic time, $t_\LCpm\to\infty$, limit.
We have to leading order in $t_\LCpm\to\infty$
\be
\begin{split}
T\sigma &= T\int_0^\sigma\ud\tau=\int_{t_\LCm}^{\tilde t} \frac{\ud t}{t'} \simeq \frac{t_\LCm}{\sqrt{a_\LCm^2 + z'(u_0)^2}} \\
    T(1-\sigma) &=T\int_\sigma^1\ud\tau=\int_{\tilde t}^{t_\LCp} \frac{\ud t}{t'} \simeq \frac{t_\LCp}{\sqrt{a_\LCp^2 + z'(u_1)^2}} \\
    z_\LCm &= \tilde z+\int_{\tilde t}^{t_\LCm} \ud t \frac{z'}{t'} \simeq -\frac{z'(u_0) t_\LCm}{\sqrt{a_\LCm^2 + z'(u_0)^2}} \\
    z_\LCp &= \tilde z+\int_{\tilde t}^{t_\LCp} \ud t \frac{z'}{t'} \simeq \frac{z'(u_1) t_\LCp}{\sqrt{a_\LCp^2 + z'(u_1)^2}} \;,
\end{split}
\ee
which allows us to find the four unknowns as
\be
\begin{split}
    z'(u_0) &= -\frac{z_\LCm}{T \sigma} \qquad z'(u_1) = \frac{z_\LCp}{T(1-\sigma)} \\
    a_\LCm^2 &= \frac{t_\LCm^2 -z_\LCm^2}{T^2 \sigma^2} \qquad \; a_\LCp^2 = \frac{t_\LCp^2 -z_\LCp^2}{T^2 (1-\sigma)^2} \;.
\end{split}
\ee
We also find the saddle-point values of ${\bf X}$,
\be\label{bfXsaddle}
\begin{split}
    z_{\LCm s} &= -\frac{p'_3 t_\LCm}{p'_0} \qquad \; z_{\LCp s} = -\frac{p_3 t_\LCp}{p_0} \\
    T_s &= \frac{t_\LCm}{p'_0} +\frac{t_\LCp}{p_0} \qquad \sigma_s =\frac{t_\LCm}{p'_0}\left(\frac{t_\LCm}{p'_0} +\frac{t_\LCp}{p_0}\right)^{-1} \;.
\end{split}
\ee

Now that we have $\partial\psi/\partial{\bf X}$ expressed explicitly in terms of ${\bf X}$, as well as the saddle-point values, the prefactor follows immediately. Let $\delta \bo X = \bo X - \bo X_s$, then expanding the exponent up to second order gives
\be\label{eq:HessianOrdinary}
\int \ud^4 \bo X \, e^{-\delta \bo X \cdot \bo H \cdot \delta \bo X} = \frac{\pi^2}{\sqrt{\det \bo H}} \;,
\ee
where ${\bf H}_{ij}=(1/2)\partial^2\psi/(\partial X_i\partial X_j)$ is the Hessian matrix.
Calculating ${\bf H}$ and its determinant is conveniently done using Mathematica. The intermediate steps are not particularly illuminating. We just mention that checking that ${\bf H}$ is a symmetric matrix gives us a nontrivial check of the above results for $\partial\psi/\partial{\bf X}$ etc. We finally find a simple result
\be
\det \bo H = \left(\frac{p_0^2 p_0^{\prime 2} T}{4 t_\LCp t_\LCm} \right)^2 \;,
\ee
which has the exact opposite asymptotic contributions with respect to~\eqref{DetGYFinal}, so the factors of $t_\LCpm$ will cancel.

\section{Spin part}\label{app:spin}

After making the shift in~\eqref{replacingAepsilon} and selecting the terms that are linear in $\epsilon$, we obtain two contributions. There is no term coming from making this shift in $\slashed{A}(x^\LCp)$ in~\eqref{propagatorWorldline} because that would correspond to a term where the photon would be absorbed outside the background field. In one part, $T_1$, $\epsilon$ comes from the $A\dot{q}$ term in the exponent of~\eqref{propagatorWorldline}, and in the second part, $T_2$, $\epsilon$ instead comes from the $\sigma^{\mu\nu}F_{\mu\nu}$ term. 

For photons polarized parallel or perpendicular to the field we have
\be\label{eq:SpinVectors}
\epsilon^{(\parallel)}_\mu = \frac{1}{k_0}\left(0, -k_3 ,0,k_1 \right) \, , \qquad \epsilon^{(\perp)}_\mu = (0,0,1,0) \; .
\ee
A general polarization state can be expressed as a superposition,
\be
\epsilon_\mu=\cos\left(\frac{\rho}{2}\right)\epsilon^{(\LCpara)}_\mu+\sin\left(\frac{\rho}{2}\right)e^{i\lambda}\epsilon^{(\LCperp)}_\mu \;,
\ee
where $\rho$ and $\lambda$ are two constants (not to be confused with the momentum and the width of the wave packet). As seen below, the probability can then be expressed in terms of the corresponding Stokes vector,
\be\label{stokesVector}
{\bf N}=\{1,\cos(\lambda)\sin(\rho),\sin(\lambda)\sin(\rho),\cos(\rho)\} \;.
\ee

In the perpendicular case we have $\epsilon^{(\perp)} q'(0) = 0$ and hence no contribution from $T_1$, whereas in the parallel case we find
\be
-i\epsilon^{(\parallel)} q'(0) = 1 \; .
\ee

For $T_1$ we have
\be\label{spinExpT1}
\begin{split}
&\mathcal P \exp\left\{ -i\frac{T}{4} \int_0^1 \ud \tau \, \sigma^{\mu \nu} F_{\mu \nu} \right\} \\
&=\exp\left(\frac{1}{2}\gamma^0\gamma^3\int_{-\infty}^\infty\ud u\,E\right) \;.
\end{split}
\ee
From the Lorentz-force equation we find
\be
\frac{\ud}{\ud u}\ln[\pm t'(u)+z'(u)]=\pm E \;,
\ee
which we use to perform the $u$ integral,
\be
\int_{-\infty}^0\ud u\, E=-\ln\left(\frac{-t'(0^\LCm)+z'(0^\LCm)}{p_0'+p_3'}\right)
\ee
and
\be
\int_0^\infty\ud u\, E=-\ln\left(\frac{t'(0^\LCp)+z'(0^\LCp)}{p_0-p_3}\right) \;.
\ee
Plugging the saddle-point values $p_2=0$ and $p_1=k_1/2$ into~\eqref{dz0Fromk} gives
\be
-t'(0^\LCm)+z'(0^\LCm)=(k_0+k_3)\left(\frac{1}{2}+\frac{i}{k_1}\right)
\ee
and
\be
t'(0^\LCp)+z'(0^\LCp)=(k_0-k_3)\left(\frac{1}{2}+\frac{i}{k_1}\right) \;.
\ee

For $T_2$ we obtain from $\sigma^{\mu\nu}F_{\mu\nu}$ a term proportional to $\slashed{k}\slashed{\epsilon}$ after making the shift~\eqref{replacingAepsilon}. So, instead of~\eqref{spinExpT1}, $T_2$ is proportional to 
\be\label{kepsilonTerm}
\exp\left(\frac{1}{2}\gamma^0\gamma^3\int_0^\infty\ud u\,E\right)\frac{\slashed{k}\slashed{\epsilon}}{2}\exp\left(\frac{1}{2}\gamma^0\gamma^3\int_{-\infty}^0\ud u\,E\right) \;.
\ee
We perform the $u$ integral in the same way as for $T_1$.
For both terms we use $\exp(\gamma^0\gamma^3x)=\cosh(x)+\sinh(x)\gamma^0\gamma^3$.
Thus, we have
\be
\frac{p_0p'_0}{m_\LCperp^2}S:=\left|\bar{u}\gamma^0(\slashed{p}+1)[\epsilon q'(0)\eqref{spinExpT1}+\eqref{kepsilonTerm}]\gamma^0v\right|^2 \;.
\ee

The rest of the calculation of $S$ involves linear algebra of multiplication of Dirac matrices and spinors, $u$ and $v$. Although this calculation can of course be done without choosing a particular representation of the Dirac matrices, it is nevertheless quite convenient to do so, and then use Mathematica to do the matrix multiplications. For the class of fields we consider here, we have found it convenient to use the spin basis in Eq.~(8) in~\cite{DegliEsposti:2021its} for $u$ and $v$.   

We finally find
\be
S=N^\LCp_iN^\LCm_j N^\gamma_z M_{ijk} \;,
\ee
where ${\bf N}^\gamma$ is the Stokes vector for the photon~\eqref{stokesVector}, and ${\bf N}^\LCpm$ are the Stokes vectors for the positron and electron spins, the indices take four values\footnote{Note that the Stokes vectors are not Lorentz vectors.}, $i=1,2,3,4$, and $M_{ijk}$ is a $4\times4\times4$ Mueller matrix (cf.~\cite{Dinu:2019pau} for nonlinear Breit-Wheeler in a plane wave). Summing over the final spins and including a factor of $1/2^2$ (coming from the factor of $\ud T/2$ in~\eqref{propagatorWorldline}, i.e. we are summing not averaging) gives
\be
\frac{1}{4}\sum_\text{spins} S={\bf N}^\gamma\cdot{\bf m} \;,
\ee
where
\be
{\bf m}=\{1+3p_1^2,0,0,1-p_1^2\} \;,
\ee
which is the same as in Eq.~(154) in~\cite{DegliEsposti:2021its} or Eq.~(44) in~\cite{Brodin:2022dkd}.
We plan to study the other components of $M_{ijk}$ elsewhere.

\section{Widths for plane waves}

For large enough $\Omega$ we expect to see convergence to the results obtained by replacing the field $E_3(t,z)$ with an appropriate plane wave. To show that this is the case, we first have to derive the plane-wave results. In this limit the results are conveniently expressed in terms of $\chi=\sqrt{-(F^{\mu\nu}k_\nu)^2}$. 
Eqs.~(35) to (40) in~\cite{Dinu:2019pau} give, for an arbitrary plane wave, the probability integrated over the momentum components, $p_{(\LCperp)}$, perpendicular to the plane wave propagation direction (not to be confused with $p_\LCperp$ which we use in this paper for the components perpendicular to the field direction). From those equations we can obtain a saddle-point approximation for the lightfront longitudinal width and compare that with our instanton results. We also want to compare with the widths in the $p_{(\LCperp)}$ directions, which we cannot derive from the results presented in~\cite{Dinu:2019pau}. Starting with the Volkov solutions, we find after some straightforward simplification an integrand with an exponential part that can be expressed as
\be
\exp\left\{\frac{ir\theta}{2b_0}(M^2+Q_{(\LCperp)}^2)\right\} \;,
\ee
where
\be\label{Qdef}
Q_{(\LCperp)}=p_{(\LCperp)}-\frac{kp}{kl}l_{(\LCperp)}-\langle a_{(\LCperp)}\rangle \;,
\ee
$p$ in this subsection is actually the canonical momentum,
$\varphi=(\phi_2+\phi_1)/2$ and $\phi_2-\phi_1$ are two lightfront time variables. 
There is no approximation at this stage. The $p_{(\LCperp)}$ integral is exactly Gaussian and can therefore be performed analytically without approximation (see e.g.~\cite{Baier:1998vh,Dinu:2013hsd}). We assume $l_2=l_3=0$. Generalizing the following results to $l_{2,3}\ne0$ is straightforward. We perform the lightfront-time integrals with the saddle-point approximation. We consider a linearly polarized, symmetric plane wave with a single (dominant) maximum, $a_\mu(\phi)=\delta_{\mu3}a_0 f(\phi)$ with $f(-\phi)=-f(\phi)$, cf. the saddle-point treatment of nonlinear Compton scattering in~\cite{Dinu:2018efz}. We have a saddle point at $\varphi\sim0$ and $\theta=2iv$, where $v$ is given by~\eqref{vDef}.
For the lightfront longitudinal momentum we have a saddle point at $s=1/2$. The $p_{(\LCperp)}$ integrals are already written as a quadratic term in the exponent, but due to $\langle a_3\rangle$ in~\eqref{Qdef}, there is a cross term between $\varphi$ and $p_3$,
\be
\begin{split}
&\exp\left\{-\frac{4a_0}{\chi}\left[a_0\tilde{f}'(v)-\frac{1}{v}\right]\varphi^2-\frac{4a_0 v}{\chi}\left[p_3-\frac{\varphi}{u}\right]^2\right\}=\\
&\exp\left\{-\frac{4a_0^2\tilde{f}'}{\chi}\left[\varphi-\frac{p_3}{a_0\tilde{f}'}\right]^2-\frac{4a_0v}{\chi}\left[1-\frac{1}{a_0v\tilde{f}'}\right]p_3^2\right\}\;,
\end{split}
\ee
so the saddle point is $\varphi_s=p_3/(a_0\tilde{f}')$ (this has been obtained assuming $p_3=\mathcal{O}(\sqrt{\chi})$). In terms of $\delta\varphi=\varphi-\varphi_s$ and $\delta\theta=\theta-2iv$ we have   
\be\label{phiQuad}
\exp\left\{-\frac{a_0^2\tilde{f}'(v)}{\chi}(4\delta\varphi^2+\delta\theta^2)\right\} \;,
\ee
and there are no other cross terms. We define a Stokes vector ${\bf N}$ for the photon polarization as in~\cite{DegliEsposti:2021its}. ${\bf N}=\{1,0,0,\pm1\}$ correspond to a basis of two linearly polarized photon states. 
We find $\mathbb{P}={\bf M}\cdot{\bf N}$, where
\be\label{MuellerVec}
\begin{split}
{\bf M}=&\frac{\alpha}{2\pi\chi\tilde{f}'(v)}\{3,0,0,-1\}\\
\times&\exp\left\{-\mathcal{A}_{PW}-d_s^{-2}\delta s^2-d_2^{-2}p_2^2-d_3^{-2}p_3^2\right\} \;,
\end{split}
\ee
where 
\be\label{ds}
\mathcal{A}_{PW}=\frac{4a_0}{\chi}(v-a_0^2\mathcal{J}) 
\qquad
d_s^{-2}=4\mathcal{A}_{PW} \;,
\ee
\be\label{d23PW}
d_2^{-2}=\frac{4a_0v}{\chi} 
\qquad
d_3^{-2}=\frac{4a_0v}{\chi}\left(1-\frac{1}{a_0v\tilde{f}'(v)}\right) \;,
\ee
and
\be
\mathcal{J}=\int_0^v\ud x\tilde{f}^2(x) \;.
\ee
By performing the Gaussian integrals over $\delta s$ and $p_{2,3}$ we find agreement with~\cite{DegliEsposti:2021its,Brodin:2022dkd}. For a Sauter pulse, $f(\phi)=\tanh(\phi)$, we have $v=\text{arccot}(a_0)$,
\be
\mathcal{A}_{PW}=\frac{4a_0}{\chi}[(1+a_0^2)\text{arccot}(a_0)-a_0] \;,
\ee
\be
d_2^{-2}=\frac{4a_0}{\chi}\text{arccot}(a_0) \;,
\ee
and
\be
d_3^{-2}=\frac{4a_0}{\chi}\left(\text{arccot}(a_0)-\frac{a_0}{1+a_0^2}\right)
\;.
\ee

To compare with the high energy limit of our instanton results, we have $s=kp/kl\approx p_1/l_1=p_1/\Omega$, so we should compare $d_1$ with $\Omega d_s$.

\section{Next-to-leading exponent correction}

Let $\mathcal A(\gamma_z)$ be the exponent at the probability level as a function of $\gamma_z$ with all other parameters fixed. The leading order result of the time-dependent limit is $\mathcal A(0)$, and in general $\mathcal A(\gamma_z)$ can be written as power series
\be
\mathcal A (\gamma_z) = \mathcal A(0) + \frac{1}{2}\mathcal A''(0) \gamma_z^2 + \mathcal O(\gamma_z^4) \; .
\ee
To find the next-to-leading correction $\mathcal A''(0)$, it is convenient to use
\be
\frac{1}{2}\mathcal A''(0) = \lim_{\gamma_z \to 0} \frac{1}{2\gamma_z} \frac{\ud \mathcal A}{\ud \gamma_z}
\ee
with
\be
\mathcal A(\gamma_z) = 2 \Im \left[px_\LCp + p'x_\LCm -\frac{T}{2} -\int_0^1 \ud\tau \frac{\dot q^2}{2T} + A \dot q + J q\right]
\ee
because, although the saddle points are all implicitly functions of $\gamma_z$, contributions from the derivatives of the saddle points cancel and we only need to consider the explicit dependence on $\gamma_z$ in the field $A(t,z)=\frac{1}{\gamma_t} f(\gamma_t t) F_z(\gamma_z z)$. After taking the derivative we perform some integrations by parts in order to obtain an integrable function when $\gamma_z \to 0$ and expand the instantons as
\be
q(u) = q_{(0)}(u) + \gamma_z^2 \, q_{(1)}(u) + \mathcal O(\gamma_z^4) \; .
\ee 
With the field normalized as in~\eqref{eq:FieldNorm}, at the end we are left with
\be\label{ddAgammaz2}
\frac{1}{2}\mathcal A''(0) = -\frac{2}{3} \Im \int \ud u \, z^3_{(0)} t'_{(0)} E (t_{(0)}) \; .
\ee
\begin{figure}
    \centering
    \includegraphics[width=\linewidth]{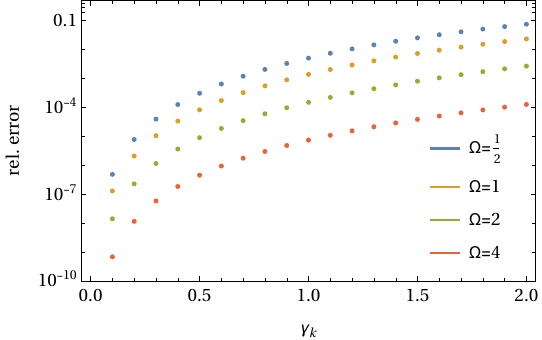}
    \caption{Relative error~\eqref{eq:experr} as a function of $\gamma_k$ for a few values of $\Omega$ and $\gamma_t = 1$.}
    \label{fig:exp_err}
\end{figure}
We can see the relative error
\be\label{eq:experr}
\left|\frac{\mathcal A(\gamma_k)}{\mathcal A(0) + \frac{1}{2}\mathcal A''(0)\gamma_k^2} - 1 \right|
\ee
in Fig.~\ref{fig:exp_err}.

At $\gamma_t = 0$ the instantons for a constant electric field are given by
\be
\begin{split}
    t_{(0)}(u) &= i\cosh(u) +\frac{\Omega}{2}\sinh u 
    \\
    z_{(0)}(u) &= i \sinh(u) +\Omega \sinh^2 \frac{u}{2} \;,
\end{split}
\ee
which can be obtained e.g. from the $\gamma_t\to0$ limit of the analytical solutions for a time-dependent Sauter pulse in~\cite{DegliEsposti:2021its}. We can separate out the imaginary part in~\eqref{ddAgammaz2} conveniently by choosing a contour which starts at $u=0$, follows the imaginary axis until $u=-i\text{arccot}(\Omega/2)$, and then goes parallel to the real axis to $-i\text{arccot}(\Omega/2)+\infty$. Only the part of the contour that follows the imaginary axis contributes to the imaginary part of~\eqref{ddAgammaz2}. We find
\be
\begin{split}
&\mathcal{A}\approx\frac{2}{E}\left([1+\rho^2]\text{arccot}(\rho)-\rho\right) \\
&+\frac{\gamma_z^2}{E}\left\{\frac{1}{2}(1+6\rho^2+5\rho^4)\text{arccot}(\rho)-\frac{\rho}{6}(13+15\rho^2)\right\} \;,
\end{split}
\ee
where $\rho=\Omega/2$. The zeroth order is the same as in~\cite{Dunne:2009gi}. 
In the limit $\Omega \to 0$ this reduces to
\be
\mathcal A \approx \frac{\pi}{E} \left( 1 + \frac{\gamma_k^2}{4} \right) \; .
\ee
Both $\mathcal{O}(\gamma_z^0)$ and $\mathcal{O}(\gamma_z^2)$ are monotonically decreasing function of $\Omega$, and for $\Omega\gg1$ we have
\be
\mathcal A \approx\frac{4}{3E\rho}\left(1-\frac{1}{5\rho^2}+\frac{2\gamma_k^2}{35\rho^2}\right) \;.
\ee

\section{Time dependent field of $d_1$, $d_2$ and $d_P$}\label{timed12Psection}

For large $\rho=\Omega/2$ a natural scaling is $u=\mathcal{O}(1/\Omega)$ and $z=\mathcal{O}(1/\Omega)$, so to leading order we can neglect $z$ in $E(t,z)$. This is because at large $\Omega$ we still have only $z'=\mathcal{O}(1)$, while $t'=\mathcal{O}(\Omega)$, so the instanton will leave the field region before it has time to move significantly in the $z$ direction and therefore does not feel the $z$ dependence. We can therefore solve (most of) the equations of motion with a purely time dependent field, $E(t)=A'(t)$. We will therefore first derive exact results for $E(t)$, and afterwards take the high-energy limit of those results. 

We have
\be\label{tztime}
z'=A(t)
\qquad
t'=\sqrt{m_\LCperp^2+A^2} \;,
\ee
so we can change variable from $u$ to $t$.
We have $t\approx\tilde{t}+\rho u$, where $A(\tilde{t})=i$, which can be written in the same way as in the plane wave case, $A(t)=a_0 f(\omega t)$, so $\tilde{t}=iv/\omega$, where $v$ is given by~\eqref{vDef}. We have
\be
u=\int_{\tilde{t}}^t\frac{\ud\bar{t}}{t'} \;.
\ee

From~\eqref{d2k0} we immediately find
\be\label{d2time}
d_2^{-2}\approx2\text{Im}\left[\frac{t}{p_0}-\int_{\tilde{t}}^t\frac{\ud\bar{t}}{t'}\right] 
=2\text{Im}\int_0^{\tilde{t}}\frac{\ud\bar{t}}{t'} \;,
\ee
where the second expression is obtained by choosing a real $t$. 

Integrating~\eqref{dtdzeq} (using $t'\delta t'-z'\delta z'=d=$const.) gives
\be
\delta z'=E\delta t+c
\qquad
\delta t'=\frac{d+cz'+Ez'\delta t}{t'} \;,
\ee
and integrating the second equation gives
\be
\delta t(u)=t'(u)\left(\frac{\delta t(0)}{\rho}+\int_{\tilde{t}}^{t(u)}\ud\bar{t}\frac{d+cA}{(m_\LCperp^2+A^2)^{3/2}}\right) \;.
\ee

For $\delta q_{p_1}$, the boundary/initial conditions in~\eqref{deltaqp1Der} imply $c_{p_1}=\delta t_{p_1}(0)=0$ and $d_{p_1}=\rho$, so
\be
\delta t_{p_1}(u)=\rho t'(u)\int_{\tilde{t}}^{t(u)}\frac{\ud\bar{t}}{(m_\LCperp^2+A^2)^{3/2}} \;.
\ee
Inserting this into~\eqref{Xp1p1k0} gives
\be\label{d1time}
\begin{split}
d_1^{-2}&\approx2\text{Im}\left[\frac{1+P^2}{p_0^3}t-\int_{\tilde{t}}^t\ud\bar{t}\frac{1+A^2}{(m_\LCperp^2+A^2)^{3/2}}\right] \\
&=2\text{Im}\int_0^{\tilde{t}}\ud\bar{t}\frac{1+A^2}{(m_\LCperp^2+A^2)^{3/2}}\;,
\end{split}
\ee
where $P=z'(\infty)=A(\infty)$ and the second row is obtained by choosing a real $t$.

For $\delta q_P$, the conditions in~\eqref{deltaqPDer} imply $c_P=1$, $d_P=0$ and $\delta t_P(0)=-1/E(\tilde{t})$, so
\be
\delta t_P=t'\left(-\frac{1}{\rho E(\tilde{t})}+\int_{\tilde{t}}^t\ud\bar{t}\frac{A}{(m_\LCperp^2+A^2)^{3/2}}\right)
\ee
and
\be
\delta z_P=\frac{z'}{t'}\delta t_P+\frac{i}{\rho E(\tilde{t})}+m_\LCperp^2\int_{\tilde{t}}^t\frac{\ud\bar{t}}{(m_\LCperp^2+A^2)^{3/2}} \;.
\ee
Inserting this into~\eqref{XPPk0} gives
\be\label{d3time}
\begin{split}
d_P^{-2}&=2\text{Im}\left[\frac{m_\LCperp^2}{p_0^3}t-\frac{i}{E(\tilde{t})\rho}-m_\LCperp^2\int_{\tilde{t}}^t\frac{\ud\bar{t}}{t^{\prime3}}\right] \\
&=2\text{Im}\left[-\frac{i}{E(\tilde{t})\rho}+m_\LCperp^2\int_0^{\tilde{t}}\frac{\ud\bar{t}}{t^{\prime3}}\right] \;.
\end{split}
\ee

We have checked that~\eqref{d2time}, \eqref{d1time}, and~\eqref{d3time} agree with what one finds starting with Eq.~(148) in~\cite{DegliEsposti:2021its}, which in turn can be obtained immediately from the WKB approximations of the wave functions.

\eqref{d2time}, \eqref{d1time} and~\eqref{d1time} are exact for a time dependent field. Expanding these results to leading order in $1/\Omega\ll1$ gives high-energy approximations that are still valid for fields that also depend on $z$.
For $1/\Omega\ll1$, \eqref{d2time}, \eqref{d1time} and~\eqref{d1time} reduce to 
\be\label{timed1HighEnergy}
d_1^{-2}\approx\frac{16a_0}{E\Omega^3}\left(v-a_0^2\mathcal{J}\right) 
\ee
and
\be\label{highEnergydPz}
d_2^{-2}\approx\frac{4a_0}{E\Omega}v 
\qquad
d_P^{-2}\approx\frac{4a_0}{E\Omega}\left(v-\frac{1}{a_0\tilde{f}'(v)}\right) \;,
\ee
which agree with the plane-wave widths in~\eqref{d23PW} and~\eqref{ds}. 

If we instead take $\Omega=\mathcal{O}(1)$ and $\gamma\gg1$, and consider field shapes with poles on the imaginary axis, then we can approximate $m_\LCperp^2+A^2\to m_\LCperp^2$ in the above integrals, since $A$ is only $\mathcal{O}(1)$ very close to the pole. The integrals become trivial and give a factor of $\tilde{t}$. The $E(\tilde{t})$ term in~\eqref{d3time} is negligible to leading order. Using $\tilde{t}\approx t_\text{pole}=i\nu/\omega$, where $\nu$ is a constant ($\nu=\pi/2$ for a Sauter pulse ), \eqref{d2time}, \eqref{d1time} and~\eqref{d3time} become
\be\label{highomegad12P}
d_2^{-2}=\frac{2\nu}{\omega m_\LCperp} 
\qquad
d_1^{-2}=\frac{2\nu}{\omega m_\LCperp^3} 
\qquad
d_P^{-2}=\frac{2\nu}{\omega m_\LCperp}
\;.
\ee
While this limit is probably less relevant from an experimental point of view, it is useful because it allows us to compare with the results obtained by treating the field to leading order in perturbation theory, see Sec.~\ref{PerturbativeTimeSec}, i.e. without using worldline instantons or WKB.

The results for the widths $d_1$, $d_2$ and $d_P$ in this section do not depend on $\lambda$ and hence do not show any sign of any critical point. In the next two sections we will study criticality in $E(t)$ fields by consider $d_\Delta(\lambda)$.

\section{Time dependent limit of $d_\Delta$}

\subsection{Time-dependent criticality}

If we start with a time-dependent field, the integral over $k_3$ is trivial due to the delta function from the spatial integrals $\delta(\Delta - k_3)$, so the exponent is simply evaluated at $k_3 = \Delta$. As a consequence, since the widths are calculated from total derivatives with respect to the momenta, only $d_\Delta^{-2}$ is affected by the wave packet, while the other widths are
\be
d_P^{-2} = -\frac{1}{2}R_{PP} \, \, \qquad d_1^{-2} = -\frac{1}{2}R_{p_1 p_1} \; ,
\ee
which we calculated in the previous section.
Neglecting the labels of the other momenta for simplicity, the exponent becomes
\be
\psi_r(k_3 = \Delta, \Delta) \;,
\ee
so when we take the second (total) derivative with respect to $\Delta$ we get, using the chain rule with the trivial $\ud k_3 /\ud \Delta = 1$,
\be
-\frac{1}{\lambda^2} + X_0 \, , \qquad X_0 := X_{\Delta \Delta} + 2X_{k \Delta} + X_{kk} \; .
\ee
If we evaluate it at $k_3 = \Delta = 0$ we find, using~\eqref{Xkk0}, \eqref{XDeltaDeltak0} and both forms of\eqref{XkDelta0} 
\be\label{X0def}
X_0 = i\left( \delta z_\nabla(\infty) - \frac{P}{p_0}\delta t_\nabla(\infty) +\frac{m_\LCperp^2 t(\infty)}{2p_0^3} -\frac{t(0)}{k_0} \right) \;,
\ee
where 
\be
\delta q_\nabla^\mu(u)=\delta q_\Delta^\mu(u)-\delta q_\Delta^\mu(0)+\delta q_k^\mu(u)-\delta q_k^\mu(0) \;.
\ee
For the $\gamma_z=0$ case we consider here, we have $\delta t_k(0)=\delta t_\Delta(0)=0$ and $\delta z$ only appears in~\eqref{dtdzeq} via $\delta z'$ and $\delta z''$. So $\delta q_\nabla$ is actually a solution to~\eqref{dtdzeq} despite the two constant terms $\delta q_\Delta^\mu(0)$ and $\delta q_k^\mu(0)$. $\delta q_\nabla$ can therefore be expressed as a superposition~\eqref{qjsum} of the basis solutions. 
From the initial conditions at $u=0+$, we find (for $u>0$)
\be
\delta q_\nabla^\mu(u)=-\frac{i}{2\rho}\delta q_{[3]}(u)-\frac{1}{2}\delta q_{[4]}(u) \;.
\ee

In general, even for $\gamma_z=0$, the saddle point is not necessarily $\Delta_s = 0$, since the width in the $\Delta$ direction
\be\label{dDeltaX0r}
d_\Delta^{-2} = \frac{1}{\lambda^2} - X_{0r} \, , \qquad X_{0r} := \Re \, X_0
\ee
vanishes at some finite $\lambda_c$,
\be\label{lambdacX0r}
\lambda_c = \frac{1}{\sqrt{X_{0r}}} \;,
\ee
which is the value after which $\Delta = 0$ becomes a local minimum.

Above the critical point $R_{p_1 P}$ is nonzero as well, therefore we have mixing between $p_1$ and $P$ as in the spacetime case
\be
\bo d^{-2}_{1P} = -\frac{1}{2}
\begin{pmatrix}
    R_{PP} & R_{p_1 P} \\
    R_{p_1 P} & R_{p_1 p_1}
\end{pmatrix}
\ee
but there is still no mixing with $\Delta$. While the latter is explicitly a function of $\lambda$, the $p_1$ and $P$ components depend on $\lambda$ only implicitly through the instantons.

The $z$ component satisfies
\be
\begin{split}
    z'(u) &= - \frac{1}{\gamma_t} -p_3 + A(t) \qquad u > 0 \\
    z'(u) &= - \frac{1}{\gamma_t} +p'_3 + A(t)  \qquad u < 0 \;.
\end{split}
\ee
So $z'(0^\LCp) - z'(0^\LCm) = -k_3$ is consistent with $\Delta = p_3 + p'_3 = k_3$.

From symmetry we have $-p_3-1/\gamma_t = -p'_3+1/\gamma_t = -
k_3/2$, so $P = 1/\gamma_t$.
The turning point $\tilde t = t(0)$ can now be found from the condition $z'(0^\LCp) = A(\tilde t) - k_3/2$, so for example for a Sauter pulse we find
\be
\tilde t = \frac{i}{\omega} \arctan\left[\gamma_t\sqrt{1+\frac{k_3^2}{\Omega^2}} \right] \;,
\ee
which agrees with Eq.~(47) in~\cite{Brodin:2022dkd}.

However, the value of $k_3$ itself is not trivial and follows the saddle point equation for $\Delta$
\be
\begin{split}
    \Re \, \biggr\{& i\biggr( z(\infty) + z(-\infty) +\frac{p_3}{p_0}t(\infty) + \frac{p'_3}{p'_0}t(-\infty) \\ &-2z(0) -2\frac{k_3}{k_0}t(0)\biggr) \biggr\} -\frac{2k_3}{\lambda^2} = 0 \; .
\end{split}
\ee
Since the instanton equations are invariant under a shift of $z(0)$ we can choose $z(0) = 0$, however, it is simpler to choose $z(0)$ as in~\eqref{z0Saddle} so that the last terms cancel and the equation reduces to
\be
    \Im \, \biggr\{ z(\infty) + z(-\infty) +\frac{p_3}{p_0}t(\infty) + \frac{p'_3}{p'_0}t(-\infty) \biggr\} =0 \; .
\ee

\subsection{$\gamma_z \to 0$ limit of spacetime field}

In the previous section we started with $\gamma_z=0$. In this section we will start with a nonzero $\gamma_z$ and consider the limit of $\gamma_z\ll1$.
Consider~\eqref{widthsDelta1}. In the limit $\gamma_z\ll1$ we find
\be
\begin{split}
X_{\Delta\Delta}&=\frac{X_{\Delta\Delta}^{(-2)}}{\gamma_z^2}+X_{\Delta\Delta}^{(0)}+\mathcal{O}(\gamma_z^2) \\
X_{k\Delta}&=\frac{X_{k\Delta}^{(-2)}}{\gamma_z^2}+X_{k\Delta}^{(0)}+\mathcal{O}(\gamma_z^2) \\
X_{kk}&=\frac{X_{kk}^{(-2)}}{\gamma_z^2}+X_{kk}^{(0)}+\mathcal{O}(\gamma_z^2) \;,
\end{split}
\ee
while the other $X$'s are $\mathcal{O}(\gamma_z^0)$. Thus,
\be
X_{p_1p_1}+\frac{X_{kp_1}^2\lambda^2}{1-X_{kk}\lambda^2}=X_{p_1p_1}+\mathcal{O}(\gamma_z^2) \;.
\ee
We will show below that the coefficient in front of $\delta\Delta^2$ is also $\mathcal{O}(\gamma_z^0)$, while the coefficient of the cross term, $\delta\Delta\delta p_1$ can be neglected. Thus, \eqref{d1time} already gives all the information of the width for $\delta p_1$, and is in particular not affected by $\delta\Delta$ or $\lambda$. 

The four independent solutions in~\eqref{qjsum} all start at $\delta q_{[j]}(u)=\mathcal{O}(\gamma_z^0)$. For $\delta q_{[1]}$, $\delta q_{[3]}$ and $\delta q_{[4]}$ this $\mathcal{O}(\gamma_z^0)$ part is the only part we need to obtain the $X$'s to leading order. But for $\delta q_{[2]}$ we have
\be\label{eq:tDepBelow}
\begin{split}
\delta z_{[2]}(u)&=1+\gamma_z^2\delta z_{[2]}^{(2)}(u)+\mathcal{O}(\gamma_z^4) \\
\delta t_{[2]}(u)&=\gamma_z^2\delta t_{[2]}^{(2)}(u)+\mathcal{O}(\gamma_z^4) \;,
\end{split}
\ee
i.e. the $\mathcal{O}(\gamma_z^0)$ part is constant, so, since we have a factor of $1/\delta z_{[2]}'$ in~\eqref{deltaqDeltaP21}, \eqref{deltaqp1From32} and~\eqref{deltakFrom234}, we need to calculate the $\mathcal{O}(\gamma_z^2)$ terms. To solve~\eqref{dtdzeq} we assume the electric field is given by a product, $E(t,z)=E(t)F(\gamma_z z)$, and that $F$ has a maximum at $z=0$. Without loss of generality, we normalize $F$ and $\gamma_z$ such that $F(0)=1$ and $F''(0)=-2$. The factor of $2$ is arbitrary but convenient since for a Sauter pulse we then have $F(u)=\text{sech}^2u$ rather than e.g. $\text{sech}^2(2u)$ or $\text{sech}^2(u/2)$. Expanding the second equation in~\eqref{dtdzeq} to $\mathcal{O}(\gamma_z^2)$, using $E(t)t'=z''$ and integrating gives
\be
\delta z_{[2]}^{(2)\prime}(u)=E(t[u])\delta t_{[2]}^{(2)}(u)-2\int_0^u\!\ud v\, z(v)z''(v) \;.
\ee
Thus,
\be
\frac{\delta z_{[2]}(u)}{\delta z_{[2]}'(\infty)}=-\frac{2a}{\gamma_z^2}+\mathcal{O}(\gamma_z^0)
\qquad
\frac{\delta t_{[2]}(u)}{\delta z_{[2]}'(\infty)}=\mathcal{O}(\gamma_z^0) \;,
\ee
where
\be
a=\frac{1}{4}\left(\int_0^\infty\ud u\, zz''\right)^{-1} 
\ee
is a constant. Inserting this into~\eqref{deltaqDeltaP21} and~\eqref{deltakFrom234} gives 
\be
\delta z_\Delta(u)=\frac{a}{\gamma_z^2}+\mathcal{O}(\gamma_z^0)
\qquad
\delta z_k(u)=-\frac{a}{\gamma_z^2}+\mathcal{O}(\gamma_z^0) \;,
\ee
and inserting these into~\eqref{Xkk0}, \eqref{XkDelta0} and~\eqref{XDeltaDeltak0} gives
\be\label{Xminus2gammaz}
X_{kk}^{(-2)}=X_{\Delta\Delta}^{(-2)}=-X_{k\Delta}^{(-2)}=\frac{ia}{\gamma_z^2} \;.
\ee
Using~\eqref{Xminus2gammaz} we find
\be\label{dDeltagammaz0}
X_{\Delta\Delta}+\frac{X_{k\Delta}^2\lambda^2}{1-X_{kk}\lambda^2}=-\frac{1}{\lambda^2}+X_0+\mathcal{O}(\gamma_z^2) \;,
\ee
where $X_0$ is given by~\eqref{X0def}.
Note that the $\mathcal{O}(1/\gamma_z^2)$ terms have cancelled thanks to~\eqref{Xminus2gammaz}. 
At this point we can see that we have managed to rewrite $X_0$ in term of quantities which we can expand to $\mathcal{O}(\gamma_z^0)$ without needing any information on $F(\gamma_z z)$.
Thus, the results will be the same as in the previous section where we started with $\gamma_z=0$. In particular, $d_\Delta$ is given by~\eqref{dDeltaX0r}.

Before we turn to the calculation of $X_{0r}$ in~\eqref{dDeltaX0r}, we will show that the cross term in~\eqref{widthsDelta1} is negligible. We have
\be
\begin{split}
&X_{\Delta p_1}+\frac{X_{k\Delta}X_{kp_1}\lambda^2}{1-X_{kk}\lambda^2}=X_{\Delta p_1}+X_{kp_1}+\mathcal{O}(\gamma_z^2) \\
&=i\left[\delta z_{1'}(\infty)-\frac{P}{p_0}\delta t_{1'}(\infty)+\frac{\rho P}{p_0^3}t(\infty)\right] \;,
\end{split}
\ee
where
\be
\delta q_{1'}^\mu(u)=\delta q_{p_1}^\mu(u)-\delta q_{p_1}^\mu(0) \;.
\ee
From the initial conditions at $u=0+$, we find (for $u>0$) $\delta q_{1'}(u)=\delta q_{[3]}(u)$. Using~\eqref{t3gammaz0} and~\eqref{z3gammaz0} we find
\be
\text{Re}\left(X_{\Delta p_1}+X_{kp_1}\right)=\mathcal{O}(\gamma_z^2) \;,
\ee
so the cross term is indeed negligible.

A similar calculation as in Sec.~\ref{timed12Psection} gives to $\mathcal{O}(\gamma_z^0)$
\be\label{t3gammaz0}
\delta t_{[3]}(u)\approx\rho t'(u)\int_{\tilde{t}}^{t(u)}\frac{\ud\bar{t}}{(m_\LCperp^2+A^2)^{3/2}} \;,
\ee
\be\label{z3gammaz0}
\delta z_{[3]}(u)\approx\rho\int_{\tilde{t}}^t\ud\bar{t}\frac{A(t)-A(\bar{t})}{(m_\LCperp^2+A^2[\bar{t}])^{3/2}} \;,
\ee
\be
\delta t_{[4]}(u)\approx t'\int_{\tilde{t}}^t\ud\bar{t}\frac{A-i}{(m_\LCperp^2+A^2)^{3/2}} 
\ee
and
\be
\delta z_{[4]}\approx\rho\int_{\tilde{t}}^t\ud\bar{t}\frac{m_\LCperp^2+A(t)A(\bar{t})-i[A(t)-A(\bar{t})]}{(m_\LCperp^2+A^2[\bar{t}])^{3/2}} \;.
\ee
Thus,
\be\label{ReX0}
X_{0r}:=\text{Re }X_0=\frac{1}{2}\text{Im}\left[\frac{\tilde{t}}{\rho}-m_\LCperp^2\int_0^{\tilde{t}}\frac{\ud\bar{t}}{(m_\LCperp^2+A^2)^{3/2}}\right] \;.
\ee

In the high-energy limit, $\Omega\gg1$, we find
\be\label{X0rlargeOmega}
\lim_{\Omega\gg1} X_{0r}=\frac{2a_0}{E\Omega^3}\left(v-3a_0^2\mathcal{J}\right) \;.
\ee
For $a_0=1/\gamma_t\gg1$, we can expand the above $\Omega\gg1$ approximation as
\be\label{X0rlargeOmegaThenLargea0}
\lim_{a_0\gg1}\left[\lim_{\Omega\gg1}E\Omega^3X_{0r}\right]=\frac{8}{15a_0^2}+\frac{2(f^{(5)}(0)-40)}{105a_0^4}+\mathcal{O}(a_0^{-6}) \;,
\ee
where $f^{(5)}(0)=d_u^5 f(u)|_{u=0}$ and $A(t)=f(\omega t)/\gamma_t$, and we have normalized $f$ and $\omega$ such that $f'(0)=1$ and $f'''(0)=-2$. For fields with a pole on the imaginary axis at $v=v_p$ (e.g. for a Sauter pulse we have $v_p=\pi/2$) we find 
\be\label{X0rHighOmegaLowomega}
\lim_{a_0\ll1}\left[\lim_{\Omega\gg1}X_{0r}\right]=\frac{2a_0 v_p}{E\Omega^3} \;.
\ee

In the LCF limit we find
\be\label{X0rLCF}
\lim_{\gamma_t\ll1}X_{0r}=\frac{\gamma_t^2}{6\rho E}\left(2+3\rho^2-3\rho[1+\rho^2]\text{arccot }\rho\right) \;.
\ee
The high-frequency limit of this LCF limit is
\be
\lim_{\Omega\gg1}\left[\lim_{\gamma_t\ll1}X_{0r}\right]=\frac{\gamma_t^2}{15\rho^3 E} \;,
\ee
which agrees with the first term in~\eqref{X0rlargeOmegaThenLargea0}. Thus, we find the same result regardless of whether we first take the large-$\Omega$ limit and then the large-$a_0$ limit or vice versa. For low frequencies we instead find
\be\label{X0rLCFlowOmega}
\lim_{\Omega\ll1}\left[\lim_{\gamma_t\ll1}X_{0r}\right]=\frac{\gamma_t^2}{3\rho E} \;.
\ee

If we take the low-frequency limit first, then the dominant contribution to the integral in~\eqref{ReX0} comes from $\bar{t}$ close $\tilde{t}$. We find
\be\label{X0rlowOmega}
\lim_{\Omega\ll1}X_{0r}=\frac{1}{2\rho E}\left(\frac{v}{\gamma_t}-\frac{1}{\tilde{f}'(v)}\right) \;.
\ee
(Curiously, this is the same function as in the $\Omega\gg1$ (opposite limit) approximation of $d_P^{-2}$ in~\eqref{highEnergydPz}.) 
\be
\lim_{\gamma_t\ll1}\left[\lim_{\Omega\ll1}X_{0r}\right]=\frac{\gamma_t^2}{3\rho E} \;,
\ee
which agrees with~\eqref{X0rLCFlowOmega}, so the two limits $\gamma_t\to0$ and $\Omega\to0$ commute. 

For fields with a pole on the imaginary axis and for $\gamma\gg1$ we can neglect the $A^2$ in the integral in~\eqref{ReX0} because it is only $\mathcal{O}(1)$ in very short interval close to $\tilde{t}$, so we find
\be\label{X0rHighomega}
\lim_{\gamma_t\gg1} X_{0r}=\frac{v_p}{2\gamma_t E}\left(\frac{1}{\rho}-\frac{1}{m_\LCperp}\right) \;.
\ee
\eqref{X0rHighomega} reduces to~\eqref{X0rHighOmegaLowomega} in the limit $\Omega\gg1$. We also find
\be
\lim_{\Omega\ll1}\left[\lim_{\gamma_t\gg1} X_{0r}\right]=\frac{v_p}{\gamma_t\Omega E}=\lim_{\gamma_t\gg1}\left[\lim_{\Omega\ll1} X_{0r}\right] \;,
\ee
where $\lim_{\Omega\ll1} X_{0r}$ is given by~\eqref{X0rlowOmega}.

Finally, the width in $\delta\Delta$ is obtained from~\eqref{dDeltaX0r} and~\eqref{ReX0},
and the critical point by~\eqref{lambdacX0r}.
For $\gamma_t=\Omega=1$, $X_{0r}\approx0.11/E$ and hence $\lambda_c\approx3.0\sqrt{E}$.
For $\Omega\gg1$, \eqref{X0rlargeOmega} gives $\lambda_c\propto\sqrt{E}\Omega^{3/2}$, which agrees with~\eqref{lambdacLargeOmega}. In the LCF limit, $\gamma_t\ll1$, \eqref{X0rLCF} gives $\lambda_x\propto\sqrt{E}/\gamma_t$.
Thus, the two most popular limits, large $\Omega$ and large $a_0$, both lead to a higher value of $\lambda_c$, i.e. make it more difficult to observe the critical point. From~\eqref{X0rLCFlowOmega} we see that $\lambda_c\propto\sqrt{E\Omega}$ for $\Omega\ll1$, so $\lambda_c$ continues to decrease as $\Omega$ decreases in the $\Omega\ll1$ regime, but not as fast as in the $\Omega\gg1$ regime. Similarly, from~\eqref{X0rHighomega} we see $\lambda_c\propto\sqrt{E\gamma_t}$ for $\gamma_t\gg1$. All these limits show that one should choose smaller $\Omega$ and $\gamma_t$ if one wants to have a lower critical point $\lambda_c$.

\section{$E(z)$}

In this section we consider a field that only depends on $z$. We are in particular interested in the $\gamma_z\gg1$ limit in order to compare with the perturbative results, and we will therefore focus on parts of the momentum space where $|P|$ is smaller than $|\Delta/2|$. This means both the electron and the positron end up travelling in either $z\to+\infty$ or $z\to-\infty$ direction, in contrast to the focus of the rest of this paper, or the case without the photon, where the field accelerates the electron and positron in opposite directions. We focus on the saddle point with $\Delta<0$, which means both the electron and the positron travel to $z\to\infty$. (For $\Delta>0$ they would instead travel to $z\to-\infty$.) Since the field does not depend on time, we have a delta function which implies $k_0=p_0+p'_0$, which, upon setting $p_2=0$ and $p_1=\Omega/2$, gives 
\be
k_3=\pm2\sqrt{(p_0+p'_0)^2-k_1^2} \;.
\ee
We will choose $k_3=-2\sqrt{...}$ so that the instanton goes to $z\to\infty$. Let $E(z)=A'(z)$, then $t''=Ez'$ can be integrated to give
\be
u>0: t'=c_\LCp+A(z)
\qquad
u<0: t'=c_\LCm+A(z) \;,
\ee
where $c_\LCpm$ are two constants. With the asymptotic energy conditions, $t'(+\infty)=p_0$ and $t'(-\infty)=-p'_0$, and $A(+\infty)=1/\gamma_z$, we find
\be
c_\LCp=p_0-\frac{1}{\gamma_z}
\qquad
c_\LCm=-p'_0-\frac{1}{\gamma_z} \;.
\ee
Since both ends of the instanton are at $z\to+\infty$, we could have chosen $A$ such that $A(+\infty)=0$, but since we are mostly considering symmetric fields we choose instead $A$ to be antisymmetric. From~\eqref{dt0Fromk} we find a condition for $z(0)=\tilde{z}$,
\be\label{ftildez}
f(\omega_z\tilde{z})=1-i\frac{k_3}{k_1}\gamma_z \;,
\ee
where $A(z)=f(\omega_zz)/\gamma_z$ ($\gamma_z=\omega_z/E$). From $t^{\prime2}-z^{\prime2}=m_\LCperp^2$ we find
\be
\begin{split}
u<0: z'&=-\sqrt{(c_\LCm+A)^2-m_\LCperp^2} \\
u>0: z'&=\sqrt{(c_\LCp+A)^2-m_\LCperp^2} \;.
\end{split}
\ee
To evaluate the exponent~\eqref{expPsiStart}, we can choose $A_0(z)=-A(z)$,
\be
\psi=-i\int\ud u\, zE(z)t' \;.
\ee
Next we change integration variable from proper time $u$ to $z$,
\be
2\text{Re }\psi=2\text{Im}\left[\int_{\tilde{z}}^{z_1}\ud z\, zE\frac{t'}{z'}\Big|_{c_\LCp}-\int_{\tilde{z}}^{z_1}\ud z\, zE\frac{t'}{z'}\Big|_{c_\LCm}\right] \;,
\ee
where $z_1$ is some arbitrary point on the positive real axis. Using partial integration we find
\be
\begin{split}
&2\text{Re }\psi=2\text{Im}\bigg[k_3\tilde{z}\\
&-\int_{\tilde{z}}^{z_1}\left(\sqrt{(c_\LCm+A)^2-m_\LCperp^2}+\sqrt{(c_\LCp+A)^2-m_\LCperp^2}\right)\bigg] \;.
\end{split}
\ee

In the $\gamma_z\gg1$ limit and for fields with poles on the imaginary axis, $\tilde{z}$ will approach the pole on the positive imaginary axis which is closest to the origin, but $A(\tilde{z})$ does not diverge, it just approaches a constant as seen in~\eqref{ftildez}. In fact, $A$ is only $\mathcal{O}(1)$ in a very small region close to $\tilde{z}$, whereas for most of the integration contour $A$ is negligible. We can therefore neglect $A$ in the integrand and we find a trivial integral. 
As we have seen in the rest of this paper, when we are not considering the perturbative limit, it is natural to think of $\hat{\lambda}:=\lambda/\sqrt{E}$ as an $E$ independent parameter, in the same way as we usually think of $\gamma=\omega_z/E$ as $E$ independent. But in the perturbative limit we can instead think of $\gamma$ as $\mathcal{O}(1/E)$ and $\hat{\lambda}$ as $\mathcal{O}(1/\sqrt{E})$. In this limit we have found that $P_s$ does indeed decrease as $1/\gamma$, while $\Delta_s$ converges to an $\mathcal{O}(E^0)$ function of $\gamma_z/\hat{\lambda}^2=\omega_z/\lambda^2$.  
We thus find (recall $\Delta<0$)
\be
\begin{split}
\lim_{\gamma_z\gg1}2\text{Re }\psi&=2\text{Im }\tilde{z}(k_3-\Delta)\approx 2\nu_z(k_3-\Delta)\\
&=-2\nu_z|k_3-\Delta| \;,
\end{split}
\ee
where $\nu_z=\nu/\omega_z$ and the pole closes to the real axis (with positive imaginary part) is $z=i\nu_z$. For a Sauter pulse we have $\nu=\pi/2$. Adding the $-k_3^2/\lambda^2$ term in~\eqref{expPsiStart} we find agreement with~\eqref{EzPerExp}, which we obtained by treating the field in lowest order of perturbation theory. (In~\eqref{EzPerExp} we considered the other saddle point, where $\Delta>0$, while here we focused on $\Delta<0$ so that the instanton goes to $z\to+\infty$, but in both approaches we find both saddle points and they are simply related by changing signs.)

\end{document}